\documentclass[10pt,a4paper]{article}
\usepackage{lineno,hyperref}
\modulolinenumbers[5]
\usepackage{float}
\usepackage{geometry}
\usepackage[utf8]{inputenc}
\usepackage{indentfirst}
\usepackage{graphicx}
\usepackage{algorithm}
\usepackage{algorithmicx}
\usepackage{algpseudocode}
\usepackage{amsmath}
\usepackage{amsfonts}
\usepackage{booktabs}
\usepackage{multirow}
\usepackage{threeparttable}
\usepackage[figuresright]{rotating}

\usepackage{tikz}
\usetikzlibrary{arrows,graphs}
\usetikzlibrary{chains, decorations.pathreplacing, positioning} 
\usetikzlibrary{positioning} 
\usetikzlibrary{shapes.geometric}
\usepackage{pgf}
\usetikzlibrary{arrows,automata}
\usepackage{verbatim}
\usepackage{chngcntr}
\usepackage[position=top]{subfig}
\usepackage{natbib}
\usepackage{amsmath, amsthm}
\usepackage{longtable}

\usepackage{smartdiagram}
\usepackage{url}
\usepackage{rotating}

\usepackage{pgfplots}
\pgfplotsset{compat=1.18}

\usetikzlibrary{decorations.pathreplacing}

\usetikzlibrary{arrows.meta, positioning}

\title{\bfseries New Formulations and Discretization Insights for the Electric Autonomous Dial-a-Ride Problem}

\author{
	Boshuai Zhao$^{*}$\\
	{\small Université Paris-Saclay, CentraleSupélec, Laboratoire Génie Industriel, France}\\
	{\small \texttt{boshuai.zhao@centralesupelec.fr; zhaoboshuai1995@163.com}}
	\and
	Adam Abdin\\
		{\small Université Paris-Saclay, CentraleSupélec, Laboratoire Génie Industriel}\\
	{\small \texttt{adam.abdin@centralesupelec.fr}}
	\and
	Jakob Puchinger\\
	{\small EM Normandie Business School, Métis Lab, France}\\
	{\small Université Paris-Saclay, CentraleSupélec, Laboratoire Génie Industriel}\\
	{\small \texttt{jakob.puchinger@centralesupelec.fr}}
}

\begin{document}
	\maketitle

		\textbf{Abstract:} The Electric Autonomous Dial-a-Ride Problem (E-ADARP) involves routing and scheduling electric autonomous vehicles under battery capacity and partial recharging constraints, aiming to minimize total travel cost and excess ride time. In practice, operational data for time and state-of-charge (SoC) are often available only at a coarse granularity. This raises a natural question: can discretization be exploited to improve computational performance by enabling alternative formulation structures? To investigate this question, we develop three formulations reflecting different levels of discretization. The first is an improved event-based formulation (IEBF) with arc-flow SoC variables for the continuous-parameter E-ADARP, serving as a strengthened baseline. The latter two are fragment-based formulations designed for discretized inputs. The second is a time-space fragment-based formulation with continuous SoC arc-flow variables (TSFFCS), which discretizes time while keeping SoC continuous. The third is a battery-time-space fragment-based formulation (BTSFF), which discretizes both time and SoC. Here, an event denotes a tuple consisting of a location and a set of onboard customers, while a fragment denotes a partial path. Computational results show that IEBF improves upon the existing event-based formulation for the original E-ADARP. Under discretized settings, TSFFCS tends to outperform IEBF, particularly when recharging is frequent and time discretization is relatively coarse, indicating that time discretization can improve computational performance across a wide range of settings. In contrast, BTSFF rarely outperforms TSFFCS unless the number of reachable SoC levels is limited, suggesting that explicit SoC discretization is beneficial only in relatively restricted settings.
			

		\section{Background}\label{ch3:background}
		
		The Dial-a-Ride Problem (DARP) involves designing vehicle routes and schedules to transport customers from their origins to destinations while minimizing travel cost. The Electric Autonomous Dial-a-Ride Problem (E-ADARP) extends DARP by considering fleets of electric, self-driving vehicles. Here, the autonomous feature mainly implies that driver-related cost and constraints are not modeled explicitly.
		The E-ADARP setting introduces additional operational constraints due to limited battery capacity, while eliminating those associated with human drivers (e.g., maximum driving time). The need for recharging further increases complexity, requiring decisions about when, where, and how much to recharge--decisions that must balance energy needs with charging time. Additionally, the objective function not only minimizes travel cost but also accounts for excess user ride time. These additional constraints and the objective structure further increase computational complexity.
		
		Most existing studies assume highly precise or continuous parameters for time and SoC, particularly in applications such as food delivery (e.g., Meituan, Uber Eats, Deliveroo) and ride-hailing (e.g., DiDi, Uber). In many practical settings, however, operational data are available only at a coarse granularity, which leads to discretized representations of time and SoC. For instance, services such as non-emergency patient transport, campus shuttle operations, and tourist transportation often adopt scheduling intervals of 5 or 10 minutes due to lower operational urgency or practical limitations. Similarly, short-horizon applications such as drone delivery typically operate within a 2-4 hour window; even with minute-level granularity, this results in a limited number of time steps and SoC levels.
		
		These observations suggest that discrete representations of time and SoC can be a natural fit in many operational contexts. This raises a central question: can discretization be exploited to improve computational performance for E-ADARP, and if so, how does it reshape the relative effectiveness of different formulations and inform formulation selection? To study this setting, we consider a discretized representation of E-ADARP, referred to as D-E-ADARP.

		Beyond the motivation from discretization, the study of D-E-ADARP is driven by recent algorithmic advances. Specifically, \citet{Rist2021} introduce a fragment-based formulation for the DARP that significantly outperforms existing methods--including branch-and-cut \citep{doi:10.1287/opre.1060.0283}, branch-and-price \citep{doi:10.1287/trsc.1090.0272,doi:10.1287/opre.1100.0881}, and the more recent event-based approach \citep{Gaul2024}--on standard benchmark instances. Here, a ``fragment'' denotes a partial path with empty load only at the start and end nodes, while an ``event'' represents a tuple consisting of a location and a set of onboard customers.
		However, partial recharging and continuous SoC in E-ADARP hinder the direct application of fragment-based formulations, as the resulting time schedules become difficult to handle. \citet{Su2024} incorporate fragments into a branch-and-price (B\&P) framework but only use them as building blocks for column generation in their subproblem. In contrast, D-E-ADARP, with discretized time and SoC indices, facilitates the direct use of fragment-based methods.
		
		Rather than advocating a universally superior formulation, this study investigates how discretization reshapes the computational effectiveness of different exact formulations for E-ADARP. To address this question, we construct a formulation comparison framework that allows us to examine the impact of discretization in both time and SoC. In particular, we develop a formulation for E-ADARP and two formulations for D-E-ADARP. The formulation for E-ADARP is an improved event-based formulation (IEBF) that serves as a strengthened baseline. The two formulations for D-E-ADARP are fragment-based: a time-space fragment-based formulation with continuous SoC arc-flow variables (TSFFCS) and a battery-time-space fragment-based formulation (BTSFF). TSFFCS relies on a network augmented with discrete time while maintaining continuous SoC variables, whereas BTSFF relies on a network augmented with both discrete time and SoC dimensions. The comparison between TSFFCS and IEBF isolates the role of time discretization, whereas the comparison between TSFFCS and BTSFF isolates the role of SoC discretization.

		The contributions of this paper are threefold.
		First, we propose an improved event-based formulation for the E-ADARP by incorporating modeling techniques inspired by the electric vehicle routing problem literature. In particular, we employ arc-flow SoC variables and unified node-charging station-node arcs instead of the node-based SoC variables and separate charging arcs used in \citet{Stallhofer2025}. Second, we develop a formulation comparison framework that enables a controlled investigation of discretization effects. Third, using this framework, we derive formulation-selection insights and clarify the computational benefits of time and SoC discretization. Our results show that time discretization can improve computational performance across a relatively broad range of settings, whereas explicit SoC discretization is beneficial only in restricted regimes where the number of reachable SoC states remains small. Outside these regimes, the event-based formulation remains a robust choice.
		
		Practically, our work provides both improved and alternative solution methods for electric dial-a-ride and related transportation applications, and offers guidance on the computational value of discretized time and SoC parameters. Academically, we develop improved formulations and analyze their performance under different discretization settings through computational experiments, yielding insights for formulation design in E-ADARP. These insights may also inform formulation design in other routing problems involving time and energy constraints.
		
		The remainder of this article is organized as follows. Section~\ref{ch4:liter} presents a comprehensive review of the relevant literature. Section~\ref{DARPSV} presents the problem definition and the proposed formulations, including IEBF, TSFFCS, and BTSFF. In Section~\ref{numericaldiscussion}, we present and discuss the computational results. Finally, Section~\ref{Conclusions} concludes the paper.

		\section{Literature review}	\label{ch4:liter}
		In this section, we review existing studies on DARP and E-ADARP and discuss works that employ time and SoC discretization in vehicle routing problems.
		
		\subsection{Electric autonomous dial-a-ride problem}\label{literEDARPS}
		
		Before introducing E-ADARP, we first review the research on DARP. Traditional methods, such as branch-and-cut \citep{doi:10.1287/opre.1060.0283} and branch-and-price \citep{doi:10.1287/trsc.1090.0272,doi:10.1287/opre.1100.0881}, have been widely used to solve DARP. 
		
		One recent line of research centers on fragment-based methods. \citet{Alyasiry2019} propose and apply this method to the Pickup and Delivery Problem (PDP) with Time Windows under a Last-In-First-Out (LIFO) loading constraint. PDP is a generalization of the DARP, in which multiple items from a single pickup node may be destined for different delivery nodes, and no maximum ride time constraint is imposed. In this context, fragments are defined as partial paths in which only the start and end nodes have an empty load, and the sequence adheres to the LIFO rule. Example fragment sequences include $p1p2p3d3d2d1$ and $p1p2d2p3d3d1$, where ``$p$'' denotes a pickup node, ``$d$'' denotes a delivery node, and the number indicates the customer index. This notation is adopted throughout the remainder of the paper. By enumerating combinations of such fragments, the authors construct a main formulation over a time-space network in which the travel times of fragments/arcs are rounded down. While solving the main formulation, a callback mechanism is invoked to identify and eliminate time-infeasible schedules that may arise from this rounding. Their computational results demonstrate the effectiveness of the approach. This fragment concept has also been extended beyond the strict LIFO setting. Recent studies, such as \citet{Rist2021}, \citet{Rist2022}, and \citet{Zhang2022}, demonstrate that fragment-based methods can significantly enhance computational performance in solving DARP and its variants. In these works, fragments are not restricted to LIFO. For example, a sequence like $p1p2d1p3d2d3$ still satisfies the general fragment definition where only the start and end nodes have an empty load. An important advantage of the fragment-based approach is that key constraints--such as pairing, precedence, capacity, time windows, and maximum ride time--are embedded during fragment generation. As a result, many DARP constraints are implicitly satisfied, reducing the burden on the main formulation.
		
		Another recent line of research on the DARP introduces an event-based formulation \citep{Gaul2022, Gaul2024}. This formulation resembles the classical arc-based one of \citet{doi:10.1287/opre.1060.0283}, but instead relies on a network in which nodes represent events and arcs capture feasible transitions between them. This structure inherently satisfies pairing, precedence, and capacity constraints, thereby reducing the number of explicit constraints required in the model and enhancing computational performance. 

		We now turn to the development of exact algorithms for E-ADARP. In addition to the standard DARP constraints, E-ADARP has to track the SoC to determine when, where, and how much to recharge, which complicates both the routing and scheduling decisions. \citet{Bongiovanni2019} are the first to solve E-ADARP exactly, employing a branch-and-cut approach based on an arc-based formulation. Subsequently, \citet{Su2024} propose a B\&P algorithm that significantly improves computational efficiency compared to \citet{Bongiovanni2019}. As mentioned earlier, their approach leverages fragments as the basis for generating paths in the subproblem. More recently, \citet{Stallhofer2025} apply an event-based formulation to E-ADARP, extending the works of \citet{Gaul2022, Gaul2024}. With the support of additional tailored cuts, their method achieves superior performance over \citet{Su2024} on most reported benchmark instances when each recharging station is allowed to be visited at most once. However, no additional tests are provided for cases with higher visit limits.
		
		Recent studies have explored event-based formulations for E-ADARP and incorporated fragment ideas within B\&P frameworks. The best-performing approaches are the event-based method of \citet{Stallhofer2025} and the B\&P algorithm of \citet{Su2024}, with the former showing greater advantage. However, it remains unclear how discretization changes the relative strength of event-based versus fragment-based formulations in E-ADARP	. Given that fragment-based approaches outperform event-based ones in DARP on their benchmark instances, this gap motivates us to explore their direct application to (D-)E-ADARP. Table~\ref{tab:methodstreams} summarizes the fragment- and event-based approaches across related DARP settings.

		\begin{table}[h]
			\caption{Fragment- and event-based approaches across related DARP settings}
			\footnotesize
			\centering
			\setlength{\tabcolsep}{4pt}
			\renewcommand{\arraystretch}{1}
			\begin{threeparttable}
				\begin{tabular}{@{}l p{3.9cm} p{3.0cm} p{3.5cm}@{}}
					\toprule
					Problem / Method & Fragment-based & Event-based & Traditional \\ \midrule
					
					DARP 
					& Direct use \citep{Alyasiry2019,Rist2021}
					& \citet{Gaul2022,Gaul2024} 
					& B\&C \citep{doi:10.1287/opre.1060.0283}, B\&P \citep{doi:10.1287/trsc.1090.0272,doi:10.1287/opre.1100.0881} \\\midrule
					
					E-ADARP 
					& Indirect use \citep{Su2024}
					& \citet{Stallhofer2025} 
					& B\&C \citep{Bongiovanni2019} \\\midrule
					
					D-E-ADARP (this paper)
					& Direct use enabled by discretization 
					& Improved event-based formulation (IEBF) 
					& -- \\
					
					\bottomrule
				\end{tabular}
				\begin{tablenotes}
					\scriptsize
					\item [1] B\&C: branch-and-cut; direct use: fragments are used in the main formulation; indirect use: fragments are not used directly in the main formulation.
				\end{tablenotes}
			\end{threeparttable}
			\label{tab:methodstreams}
		\end{table}
		
		\subsection{Time and SoC discretization in routing problems}\label{literDRP}
		
		Another relevant body of work addresses the use of time and SoC discretization in vehicle routing problems. 
		
		Many studies have adopted discrete time settings for modeling and optimization. For example, \citet{SANTINI2018607} design a B\&P framework to efficiently solve a feeder network design problem, which is similar to PDP. Based on real-world scenarios, they use a discrete time representation. Their computational experiments include two instances: the Baltic scenario, with a total planning horizon of one week and a time unit of two hours, and the Western African (WAF) scenario, with a four-week horizon and eight-hour time units, both corresponding to 84 time indices. In addition, \citet{Mahmoudi2016} apply a state-space-time formulation to the PDP with time window constraints, motivated by ride-sharing applications. Their notion of ``state'' captures onboard customer information, which is conceptually similar to the ``event'' concept proposed by \citet{Gaul2022}. Notably, they adopt a discrete time setting with one-minute intervals over a four-hour planning horizon, resulting in 240 time indices. 
		
		Several studies on the Electric Vehicle Routing Problem (EVRP) adopt discretized SoC settings. \citet{Fernandez2022} investigate a variant of the EVRP in the context of arc routing with speed-dependent energy consumption. They demonstrate that a discretization granularity of 1000 kWs for battery capacities ranging from 25 to 75 kWh (approximately 1.3\%-4\% of battery capacity) offers a good trade-off between solution quality and computational tractability, and is sufficient for most case studies with negligible impact on solution quality. \citet{Bruni2025} examine the EVRP with non-linear charging functions. They state that accurately estimating SoC per arc is intricate due to the presence of many uncertainties. Hence, they consider using a fine-grained discretization of 5\% of battery capacity to be a reasonable choice for their study. 
		
		There are also related studies on discretized modeling in electric vehicle routing. For example, \citet{LIU2025104233} develop a state-space-time model that jointly captures spatial, temporal, and battery dimensions. However, such studies mainly use discretization as a modeling tool and are not specifically tailored to E-ADARP. Unlike existing studies where discretization is primarily used as a modeling device, in our setting discretization fundamentally changes the structure of the formulation, as it enables the construction of fragment-based representations that are not available under continuous settings. Although discretized representations are widely used, there is a lack of systematic analysis of how discretization affects formulation performance in E-ADARP.

		\section{Problem statement and new formulations for (D-)E-ADARP}\label{DARPSV}
		
		(D-)E-ADARP concerns the design of vehicle routes and schedules for serving a set of customers while incorporating recharging decisions. This section first presents a formal problem description of (D-)E-ADARP (Section~\ref{problemDef}).	We then introduce three formulations that reflect different modeling choices regarding time and state-of-charge (SoC) discretization. We first present IEBF, a strengthened event-based formulation for the original continuous-parameter E-ADARP. Next, we introduce TSFFCS, which discretizes time while keeping SoC continuous, followed by BTSFF, which discretizes both time and SoC. Together, these formulations enable a controlled comparison framework to examine how discretization reshapes formulation structure and computational performance. In particular, comparing IEBF with TSFFCS isolates the impact of time discretization, whereas comparing TSFFCS with BTSFF isolates the impact of SoC discretization.
		
		Unlike the existing event-based formulation, IEBF incorporates two modeling techniques: the use of SoC arc-flow variables and a unified node-station-node arc structure. These techniques are also adopted in TSFFCS, which makes IEBF an appropriate baseline for comparison.
		
		\subsection{Problem description}\label{problemDef}
		
		We first introduce the (original) E-ADARP problem description and then extend the description to D-E-ADARP. E-ADARP involves a total number of customers $n$, with each customer~$i$ having a unique pickup and delivery location pair represented as $(i, i+n)$, where $i$ represents the pickup location and $i+n$ the corresponding delivery location. The pickup and delivery locations are represented by the sets $P = \{1, 2, \dots, n\}$ and $D = \{n+1, n+2, \dots, 2n\}$, respectively. The set of charging stations is denoted by $S$. The location set $N = P \cup D \cup S \cup \{0, 2n+1\}$ includes all pickup and delivery locations, charging station locations, as well as the origin depot $0$ and the destination depot $2n+1$. Each location~$i \in P \cup D$ is associated with a time window $(et_i, lt_i)$, defining the earliest arrival time and latest departure time. Additionally, $t_{\min}$ and $t_{\max}$ denote the earliest departure time from the origin depot and the latest arrival time at the destination depot, respectively. The arc set is denoted by $A$, including all arcs $(i,j) \in N \times N$ with $i \ne j$. For each arc $(i, j) \in A$, the minimum travel cost, travel time, and SoC consumption are represented by $C_{ij}$, $C^t_{ij}$, and $C^b_{ij}$, respectively (the same notation applies to the node arcs introduced below).
		
		The set of vehicles is denoted by $V$, and the parameter $Q$ represents the vehicle load capacity. The load quantity change at location $i \in N$ is expressed as $q_i$, with $q_0 = q_{2n+1} = 0$ and $q_i = -q_{n+i}$ ($q_i \in \mathbb{Z}_{+}$) for $i \in P$. This value represents the customer demand: at pickup location $i \in P$, the vehicle increases its load by $q_i$, while at delivery location $i+n \in D$, it decreases the load by $q_i$.
		
		Let $b_{\max}$ denote the vehicle's maximum effective battery size. The terms $b_{\min}$ and $\gamma b_{\max}$ represent the minimum required SoC during the trip and before returning to the depot, respectively, where $\gamma$ denotes the corresponding ratio. The SoC charging and discharging rates per unit time are denoted by $\alpha$ and $\beta$, respectively.
		
		The problem has the following assumptions: first, the vehicle speed remains constant, ensuring proportionality between cost, time, distance, and SoC consumption. Second, recharging is only permitted when there are no customers in the vehicle. Third, partial recharging is allowed, where the replenished amount of SoC is proportional to the recharging time.

		The objective of E-ADARP is to minimize a weighted cost while ensuring that no more than $|V|$ vehicles are used to serve all pickup and delivery tasks. The weighted cost consists of the total travel cost incurred by all vehicles and the total customer ride time cost, weighted by $\lambda$ and $1 - \lambda$, respectively. The ride time cost is defined as the difference between each customer's actual ride time and the corresponding direct travel time. Each vehicle starts its route at the origin depot~$0$ and ends at the destination depot~$2n+1$. Before its SoC drops to $b_{\min}$, each vehicle must visit a charging station $s \in S$ for recharging. Prior to reaching the destination, the vehicle's SoC cannot be lower than $\gamma b_{\max}$. Each pickup location $i\in P$ must be visited by the same vehicle as its corresponding delivery location~$i+n\in D$ (pairing), with the pickup location visited before the delivery location (precedence). Moreover, each customer~$i\in P$ has a maximum ride time $R_i$, each location~$i\in N$ has its time window, and each vehicle is limited to a capacity of at most $Q$.
		
		As a notational simplification, we incorporate the service time at each location into time-related parameters and omit it explicitly throughout the paper. Following \citet{Rist2021}, the service time at location~$i$ is embedded in both the travel time of arcs departing from $i$ and the maximum ride time of customer~$i$. Hence, after simplification, an arc's travel time is not necessarily proportional to its travel cost, distance, or SoC consumption.

		Moreover, this study considers only the setting where the maximum number of allowable visits to each charging station, denoted by $N_{\text{max}}$, is unlimited. This assumption is also considered in the extended scenario of \citet{Su2024}. In existing arc- and event-based formulations, a key role of setting values on $N_{\text{max}}$ is to enforce network flow through charging stations by replicating each station $N_{\text{max}}$ times, which may substantially increase the model size. In contrast, our formulations handle the unlimited recharging case without introducing additional station replications. Since our focus is on comparing discretization strategies rather than restricting station visits, we consider the unlimited setting across the three formulations. Nevertheless, for benchmarking purposes, we additionally implement an IEBF variant with limited station visits to compare with existing benchmarks.
		
		Based on the above description of E-ADARP, we define its discretized counterpart, referred to as D-E-ADARP. The key distinction lies in the use of discretized input parameters: time and SoC values are quantized into predefined units. Consequently, $C^t_{ij}$ and $C^b_{ij}$ on all arcs are restricted to integer multiples of these units. The resulting finite sets of time indices and SoCs are denoted by $T$ and $B$, respectively, for which $et_i, lt_i \in T$, and $b_{\max}, b_{\min}, \gamma b_{\max} \in B$. The time horizon set $T = \{t_{\min}, t_1, t_2, \dots, t_{\max}\}$ consists of a sequence of increasingly ordered time indices, while the SoC horizon set $B = \{b_{\min}, b_1, b_2, \dots, b_{\max}\}$ comprises a sequence of increasingly ordered SoCs.

		\subsection{Improved event-based formulation for E-ADARP}\label{IEBF}
		
		In this subsection, we present an improved event-based formulation for E-ADARP, which serves as the continuous-parameter baseline in our study. 
		The formulation extends the event-based model of \citet{Stallhofer2025} through two refinements: SoC is tracked using continuous arc-flow variables, and recharging is represented by unified node-station-node event arcs. Such modeling choices have been shown to strengthen formulations for electric vehicle routing problems \citep{WuYaman2025EVRP}.
		
		
		IEBF builds on an \textit{event-based network} $G_E=(V_E,A_E)$ with event set~$V_E$ and event arc set~$A_E$. 
		Each \textit{event} is a tuple $u=(i,U)$, where $i\in N$ is a physical location and $U\subseteq P$ is the set of onboard customers immediately after serving~$i$. 
		For simplicity, when $i\in P$, customer $i$ is not included in~$U$. 
		Capacity feasibility requires $\sum_{j\in U\cup\{i\}} q_j\le Q$ if $i\in P$, and $\sum_{j\in U} q_j\le Q$ if $i\in D$. For notational convenience, an empty-load event $(i,\emptyset)$ may be referred to simply by its physical location $i$. 
		The origin and destination depot events are denoted by $eO^+$ and $eO^-$, respectively; more generally, in multiple-depot settings, $eOv^+$ and $eOv^-$ denote the sets of origin and destination depot events (equivalently, locations), respectively.
		
		An \textit{event arc} connects two events in $V_E$ and represents a feasible transition between vehicle states. 
		The arc set $A_E$ is partitioned into the set of normal event arcs $A_E^n$ and the set of charging event arcs $A_E^c$. 
		A normal event arc $((i,U),(j,U'))\in A_E^n$ corresponds to direct travel from location $i$ to $j$ without recharging, and the onboard sets are updated according to the standard DARP rules. 
		Arcs leaving an origin depot event $o\in eOv^+$ connect only to events $(i,\emptyset)$ with $i\in P$, and only events $(i,\emptyset)$ with $i\in D$ connect to a destination depot event $d\in eOv^-$. 
		A \textit{charging event arc} $((i,\emptyset),k,(j,\emptyset))\in A_E^c$ connects two empty-load events via a charging station $k\in S$; it represents traveling from a delivery or an origin-depot location~$i\in D\cup eOv^+$ to a station $k$ and then to a pickup or a destination location~$j\in P\cup eOv^-$. 
		Let $\mathcal{A}^n$ and $\mathcal{A}^c$ denote the sets of associated physical normal arcs $(i,j)$ and physical charging arcs $(i,k,j)$, respectively. Events and normal event arcs are generated by enumeration following \citet{Gaul2024}; charging event arcs are generated by combining feasible event arcs from empty-load delivery (or origin-depot) events to empty-load pickup (or destination-depot) events with charging stations.
		
		The event-based network inherently enforces pairing, precedence, and capacity, since events explicitly include the information of onboard customers and event-arc transitions ensure load transfer.
		Figure~\ref{fig:event_network} illustrates an event-based network for DARP; in E-ADARP, additional charging event arcs are added.

		\begin{figure}[h]
			\centering
			\scalebox{1}[0.9]{
				\begin{tikzpicture}[>=latex, every node/.style={font=\small}]
					\tikzset{
						node/.style={draw, ellipse, inner sep=2pt},
						base/.style={semithick, ->},
					}
					\node[node] (Oplus) at (2,0) {$eO^+$};
					\node[node] (p2) at (4,0) {$p2$};
					\node[node] (p3set) at (6,0) {$(p3,\{2\})$};
					\node[node] (p1) at (6.5,1.2) {$(p1,\{2,3\})$};
					\node[node] (p3only) at (5.2,-1.0) {$p3$};
					\node[node] (d32) at (9.5,1.2) {$(d3,\{1,2\})$};
					\node[node] (d1) at (9.5,-0.5) {$(d1,\{2\})$};
					\node[node] (d3) at (11,0.3) {$d3$};
					\node[node] (d2set) at (6.5,-1.6) {$(d2,\{3\})$};
					\node[node] (d2) at (9.5,-1.5) {$d2$};
					\node[node] (Ominus) at (13,0) {$eO^-$};
					\draw[base] (Oplus) -- (p2);
					\draw[base] (p2) -- (p3set);
					\draw[base] (p3set) -- (p1);
					\draw[base] (p1) -- (d32);
					\draw[base] (d32) -- (d1);
					\draw[base] (d1) -- (d2);
					\draw[base] (p3set) -- (d2set);
					\draw[base] (d2set) .. controls (8.3,0.6) and (10.0,0.5) .. (d3);
					\draw[base] (d2) -- (Ominus);
					\draw[base] (d3) -- (Ominus);
					\draw[base] (p2) -- (d2);
					\draw[base] (Oplus) -- (p3only);
					\draw[base] (d2) .. controls (6,-1) .. (p3only);
					\draw[base] (p3only) .. controls (7.5,-0.4) and (9.8,0.1) .. (d3);
					\draw[base] (d3) .. controls (10,0.8) and (6.5,0.9) .. (p2);
				\end{tikzpicture}
			}
			\caption{An event-based network for DARP}
			\label{fig:event_network}
		\end{figure}

		Let $\delta^+(u)$ and $\delta^-(u)$ denote the sets of outgoing and incoming event arcs of event $u\in V_E$, respectively, and let $A_E^+(i)$ denote the set of event arcs leaving from physical location~$i$. 
		Each event arc $e \in A_E$ inherits the travel cost and travel time of its underlying physical location arc, denoted by $C_e$ and $C_e^t$, respectively (previously, same notations are also defined for location arcs). Let $A_E^n(i,j)$ denote the set of normal event arcs whose underlying physical movement is $(i,j)$, and let $A_E^c(i,k,j)$ denote the set of charging event arcs whose underlying physical movement is $(i,k,j)$.

		\begin{table}[h]
			\footnotesize
			\centering
			\caption{Decision variables of IEBF}
			\begin{tabular}{@{}ll@{}}
				\toprule
				Variables & Definition \\
				\midrule
				$z_e$ & $=1$ if event arc $e\in A_E$ is traversed; $=0$ otherwise \\
				$\tau_i$ & service start time at physical location $i\in P\cup D\cup eOv^+ \cup eOv^-$ \\
				$W_{ij}$ & entry SoC of physical normal arc $(i,j)\in \mathcal{A}^n$ \\
				$W_{ikj}$ & entry SoC of physical charging arc $(i,k,j)\in \mathcal{A}^c$ \\
				$E_{ikj}$ & recharging time at station $k$ on charging arc $(i,k,j)\in \mathcal{A}^c$ \\
				$r_i$ & excess ride time of customer $i\in P$ \\
				\bottomrule
			\end{tabular}\label{IEBFvar}
		\end{table}
		
		The decision variables of IEBF are defined in Table~\ref{IEBFvar}. The formulation IEBF is shown as follows.
		
		\allowdisplaybreaks
		{\footnotesize
			\setlength{\jot}{1pt}
			\begin{align}
				\min \quad 
				& \lambda \sum_{e\in A_E} C_e z_e + (1-\lambda) \sum_{i\in P} r_i
				\label{IEBFobj}
			\end{align}
			\begin{align}
				\text{s.t.}\quad
				\sum_{e\in\delta^-(u)} z_e &= \sum_{e\in\delta^+(u)} z_e && \forall\, u\in V_E\setminus (eOv^+ \cup eOv^-)
				\label{IEBFflow} 
				\\
				\sum_{e\in A_E^+(i)} z_e &= 1 && \forall\, i\in P
				\label{IEBFcover} 
				\\
				\tau_j + M^1_{ij}\Big(1-\sum_{e \in A_E^n(i,j)} z_e\Big) &\ge \tau_i + C^t_{ij} && \forall\, (i,j)\in \mathcal{A}^n
				\label{IEBFtimeN} 
				\\
				\tau_j + M^2_{ikj}\Big(1-\sum_{e \in A_E^c(i,k,j)} z_e\Big) &\ge \tau_i + C^t_{ik}+C^t_{kj} + E_{ikj} && \forall\, (i,k,j)\in \mathcal{A}^c
				\label{IEBFtimeC} 
				\\
				\tau_{i+n}-\tau_i&\le R_i && \forall\, i\in P
				\label{IEBFRride} 
				\\
				\tau_{i+n}-\tau_i - C^t_{i(i+n)}&\le r_i && \forall\, i\in P
				\label{IEBFrride} 
				\\
				b_{\max} \sum_{e \in A_E^n(o,j)} z_e &= W_{oj} && \forall\, (o,j)\in \mathcal{A}^n,\; o\in eOv^+
				\label{IEBFstartN} 
				\\
				b_{\max} \sum_{e \in A_E^c(o,k,j)} z_e &= W_{okj} && \forall\, (o,k,j)\in \mathcal{A}^c,\; o\in eOv^+
				\label{IEBFstartC} 
				\\
				(C^b_{id}+\gamma b_{\max}) \sum_{e \in A_E^n(i,d)} z_e &\le W_{id} && \forall\, (i,d)\in \mathcal{A}^n,\; d\in eOv^-
				\label{IEBFendN} 
				\\
				(C^b_{ik}+C^b_{kd}+\gamma b_{\max}) \sum_{e \in A_E^c(i,k,d)} z_e &\le W_{ikd} + \alpha E_{ikd} && \forall\, (i,k,d)\in \mathcal{A}^c,\; d\in eOv^-
				\label{IEBFendC} 
				\\
				W_{ij} &\le b_{\max} \sum_{e \in A_E^n(i,j)} z_e && \forall\, (i,j)\in \mathcal{A}^n
				\label{IEBFactn} 
				\\
				W_{ikj} &\le b_{\max} \sum_{e \in A_E^c(i,k,j)} z_e && \forall\, (i,k,j)\in \mathcal{A}^c
				\label{IEBFactc} 
				\\
				0 \le E_{ikj} &\le b_{\max}/\alpha \sum_{e \in A_E^c(i,k,j)} z_e && \forall\, (i,k,j)\in \mathcal{A}^c
				\label{IEBFactE}  
				\\
				(C^b_{ij}+b_{\min}) \sum_{e \in A_E^n(i,j)} z_e &\le W_{ij} && \forall\, (i,j)\in \mathcal{A}^n
				\label{IEBFfeasN} 
				\\
				(C^b_{ik}+b_{\min}) \sum_{e \in A_E^c(i,k,j)} z_e &\le W_{ikj} && \forall\, (i,k,j)\in \mathcal{A}^c
				\label{IEBFreachK} 
				\\
				W_{ikj} + \alpha E_{ikj} - C^b_{ik} \sum_{e \in A_E^c(i,k,j)} z_e &\le b_{\max} && \forall\, (i,k,j)\in \mathcal{A}^c
				\label{IEBFcap} 
				\\
				\sum_{(h,k,i)\in \mathcal{A}^c} \Big( W_{hki} - (C^b_{hk}+C^b_{ki}) \sum_{e \in A_E^c(h,k,i)} z_e + \alpha E_{hki} \Big)
				\nonumber
				\\
				\quad + \sum_{(h,i)\in \mathcal{A}^n} \Big( W_{hi} - C^b_{hi} \sum_{e \in A_E^n(h,i)} z_e \Big)
				\ge \sum_{(i,j)\in \mathcal{A}^n} W_{ij} &+ \sum_{(i,k,j)\in \mathcal{A}^c} W_{ikj} && \forall\, i\in P\cup D
				\label{IEBFbatflow} 
				\\
				z_e &\in \{0,1\} && \forall\, e \in A_E
				\label{IEBFdomZ} 
				\\
				et_i \le \tau_i &\le lt_i  && \forall\, i \in P\cup D
				\label{IEBFdomtau} 
				\\
				0 \le W_{ij} &\le b_{\max} && \forall\, (i,j) \in \mathcal{A}^n
				\label{IEBFdomWn} 
				\\
				0 \le W_{ikj} &\le b_{\max} && \forall\, (i,k,j) \in \mathcal{A}^c
				\label{IEBFdomWc} 
				\\
				r_{i} &\ge 0 && \forall\, i \in P
				\label{IEBFdomr}
			\end{align}
		}
		
		The objective~(\ref{IEBFobj}) minimizes a weighted sum of the total travel cost over all event arcs and the total excess ride time over all customers. 
		Constraints~(\ref{IEBFflow}) enforce flow conservation at each event. 
		Constraints~(\ref{IEBFcover}) ensure that each pickup location is visited exactly once; pairing and precedence are implicitly enforced by the event representation.
		Constraints~(\ref{IEBFtimeN}) and~(\ref{IEBFtimeC}) propagate service start times along normal physical arcs $(i,j)\in \mathcal{A}^n$ and charging physical arcs $(i,k,j)\in \mathcal{A}^c$, respectively. 
		For charging arcs, the temporal propagation includes the recharging duration $E_{ikj}$. 
		We use $M^1_{ij}=\max\{0,\; lt_i + C^t_{ij} - et_j\}$ for normal arcs and $M^2_{ikj}=\max\{0,\; lt_i + C^t_{ik}+C^t_{kj}+b_{\max}/\alpha  - et_j\}$ for charging arcs. 
		Constraints~(\ref{IEBFRride}) set the ride time limit $R_i$ of each customer~$i$, while constraints~(\ref{IEBFrride}) define the excess ride time of each customer. Up to this point, all constraints are about network flow, visiting customers, and time increments. 
		
		In the following, we introduce the SoC-related constraints~(\ref{IEBFstartN})--(\ref{IEBFbatflow}), which ensure SoC initialization, feasibility along normal and charging arcs, consistency of recharging decisions, and flow conservation. In contrast to the node-based SoC formulation in \citet{Stallhofer2025}, the SoC level is tracked through arc-flow variables associated with physical arcs.
		Constraints~(\ref{IEBFstartN}) and~(\ref{IEBFstartC}) initialize the SoC to full battery capacity when leaving an origin depot on normal arcs $(o,j)\in \mathcal{A}^n$ and charging arcs $(o,k,j)\in \mathcal{A}^c$, respectively. 
		Constraints~(\ref{IEBFendN}) and~(\ref{IEBFendC}) enforce a minimum SoC level $\gamma b_{\max}$ upon arrival at a destination depot along normal arcs $(i,d)\in \mathcal{A}^n$ and charging arcs $(i,k,d)\in \mathcal{A}^c$. 
		Constraints~(\ref{IEBFactn}) and~(\ref{IEBFactc}) activate the SoC variables only when the corresponding normal arc $(i,j)\in \mathcal{A}^n$ or charging arc $(i,k,j)\in \mathcal{A}^c$ is selected. 
		Constraints~(\ref{IEBFactE}) activate and set the limit for the recharging time variable on the charging arc $(i,k,j)\in \mathcal{A}^c$. 
		Constraints~(\ref{IEBFfeasN}) enforce a minimum SoC requirement, ensuring that sufficient SoC is available to traverse a normal arc $(i,j)\in \mathcal{A}^n$.
		Constraints~(\ref{IEBFreachK}) and~(\ref{IEBFcap}) enable SoC feasibility before and after recharging. For each charging arc $(i,k,j)\in \mathcal{A}^c$, constraints~(\ref{IEBFreachK}) guarantee that sufficient SoC is available to traverse the first leg $(i,k)$, while constraints~(\ref{IEBFcap}) specify that the SoC after recharging at station $k$ on charging arcs $(i,k,j)\in \mathcal{A}^c$ does not exceed the capacity $b_{\max}$.  
		Constraints~(\ref{IEBFbatflow}) enforce SoC flow conservation at each physical pickup and delivery location by balancing incoming and outgoing SoC flows over both normal arcs $(h,i)\in \mathcal{A}^n$ and charging arcs $(h,k,i)\in \mathcal{A}^c$, accounting for both SoC consumption and recharging on charging arcs. 
		Constraints~(\ref{IEBFdomZ})-(\ref{IEBFdomr}) define the domains of the variables $z$, $\tau$, $W$, and $r$.
		
		Under the setting where the maximum number of visits to each charging station is limited (i.e., $N_{\max}=1$), an additional constraint is required: $\sum_{e\in A_E^c(k)} z_e \le N_{\max}$ for all $k\in S$, where $A_E^c(k)$ denotes the set of charging event arcs whose underlying station is $k$.
		
		\subsection{Fragment-based formulations for D-E-ADARP}\label{BTSFF+TSFFCS}

		In this subsection, we introduce two fragment-based formulations for the discretized problem D-E-ADARP. We first describe the fragment-based network for DARP and the fragment generation procedure adapted for E-ADARP in Section~\ref{ch4:Preliminaries}. We then present the time-space fragment-based formulation with continuous SoC arc-flow variables (TSFFCS) in Section~\ref{TSFFCS}, followed by the battery-time-space fragment-based formulation (BTSFF) in Section~\ref{BTSFF}.
		
		\subsubsection{Fragment-based network for DARP and fragment generation for E-ADARP}\label{ch4:Preliminaries}
		
		This subsection introduces the fragment-based network for DARP together with the fragment generation and cost computation procedures for E-ADARP.
		
		The definitions largely follow \citet{Rist2021}. A \textit{DARP route} is a route $(0, i_1, i_2, \dots, i_L, 2n+1)$ for which at least one feasible schedule exists, satisfying the pairing, precedence, capacity, maximum ride time (on customers), and time window constraints of E-ADARP. SoC-related constraints are not considered at this stage. The sequence of locations $(i_1, i_2, \ldots, i_L)$, excluding depot locations, is called a \textit{DARP route path}. Any segment of this sequence is referred to as a route path. A \textit{node} is defined as a pair consisting of a location and a set of loads in the previous studies \citep{Rist2021,ZHAO2025}. Since we only use the fragments that have empty loads at their start and end nodes, all nodes in this study correspond directly to locations. Nevertheless, we retain the term ``node'' for consistency with previous studies. A \textit{fragment} refers to a partial DARP route path where only the start and end nodes have an empty load. For instance, a valid fragment could have a sequence of locations $(p1,p2,d2,d1)$ or $(p1,p2,d1,p3,d2,d3)$. An invalid case is $(p1, d1, p2, p3, d2, d3)$ as this violates the empty load condition after location $d1$ and should be decomposed into $(p1, d1)$ and $(p2, p3, d2, d3)$, ensuring that a fragment starts from a pickup node and ends at a delivery node. Here, pickup and delivery nodes are simply pickup and delivery locations. Next, we refer to the \textit{node arc} as the connection between fragments, specifically linking a delivery node to a pickup node or the destination depot, or connecting the origin depot to a pickup node.
		
		Building a fragment-based network involves first enumerating all feasible fragments that connect their corresponding pickup nodes to their respective delivery nodes. The network is then established by linking the origin depot with all pickup nodes, each pickup node with all delivery nodes, and all delivery nodes with the destination depot, thereby generating all the necessary node arcs. For fragment enumeration, each fragment must satisfy the DARP's constraints along its path, including the time window, maximum ride time, vehicle capacity, pairing, and precedence constraints. 	Fig.~\ref{ch4:fig3} presents a fragment-based network for DARP. Here, ellipses represent nodes, solid lines (with ellipses on both sides) indicate fragments (the information is described above or to the right), and dashed lines with arrows signify node arcs. $O^+$ and $O^-$ denote the origin and destination depots, respectively. An inherent characteristic of a feasible path is its typical initiation with a node arc extending from the origin depot to a pickup node, succeeded by a fragment from the pickup node to a delivery node. Subsequently, a node arc transition occurs from a delivery node to a pickup node or the destination depot. This pattern persists in an alternating sequence. As illustrated in Fig.~\ref{ch4:fig3}, dashed lines and solid lines alternate throughout any feasible path.

		\begin{figure}[H]
			\centering
			\resizebox{0.8\textwidth}{!}{
				\begin{tikzpicture}
					\node[draw,ellipse] (0+) at (-8,1) {$O^+$};
					\node[draw, ellipse] (0-) at (5,1) {$O^-$};
					\node[draw, ellipse] (p2) at (-0.5,1) {p2};
					\node[draw, ellipse] (p3) at (-3,3) {p3};
					\node[draw, ellipse] (p1) at (-6,1) {p1};
					\node[draw, ellipse] (d2) at (2.5,1) {d2};
					\node[draw, ellipse] (d3) at (3,3) {d3};
					\node[draw, ellipse] (d1) at (-2,1) {d1};
					
					\draw[line width=0.75pt, ->,dashed,>=latex] (0+) to[out=40, in=-180]  (p3);
					\draw[line width=0.75pt, ->,dashed,>=latex] (0+) ->  (p1);
					\draw[line width=0.75pt, ->,dashed,>=latex] (d2) ->  (0-);
					\draw[line width=0.75pt, ->,dashed,>=latex] (d3) ->  (0-);
					\draw[line width=0.75pt, ->,dashed,>=latex] (d2)  to[out=140, in=-10]  (p3);
					\draw[line width=0.75pt, ->,dashed,>=latex] (d1) -> (p3);
					\draw[line width=0.75pt, ->,dashed,>=latex] (d1) -> (p2);
					
					\draw[line width=0.75pt, ->,>=latex] (p2) -- node[above] {(p2,d2)} (d2);
					\draw[line width=0.75pt, ->,>=latex] (p3) -- node[above] {(p3,d3)} (d3);
					\draw[line width=0.75pt, ->,>=latex] (p1) -- node[above] {(p1,d1)} (d1);
					\draw[line width=0.75pt, ->,>=latex] (p1) to[out=40, in=140] node[above] {(p1,p4,d4,d1)} (d1);
					\draw[line width=0.75pt, ->,>=latex] (p2)  to[out=40, in=-165] node[right]{(p2,p3,d2,d3)} (d3);
			\end{tikzpicture}}
			\caption{A fragment-based network for DARP}\label{ch4:fig3}
		\end{figure}

		Based on the preceding discussion, we next describe the fragment generation process for E-ADARP and the computation of the weighted costs (travel and ride-time costs). The fragment generation procedure follows that of DARP but additionally incorporates the SoC-related constraints. Then, we obtain the set $F$ containing all feasible fragments, and the set $A^n$ containing all node arcs that do not involve charging stations. The arcs in $A^n$ are referred to as \textit{normal node arcs}. We can also compute a tightened time window for each fragment $f\in F$ and define its feasible departure time index set $DT_f$. For the weighted costs, we focus primarily on fragments, as they capture all pickup-to-delivery movements of customers, exclusively involving the ride-time costs. As shown by \citet{Su2024}, each fragment has a predetermined minimum ride-time cost that is precomputable and independent of its position in a route. The weighted cost for a fragment~$f\in F$ is then defined as $\lambda$ times its travel cost plus $(1-\lambda)$ times this precomputed minimum ride-time cost. The weighted cost for a node arc~$a\in A^n$ equals $\lambda$ times its travel cost.

		\subsubsection{Time-space fragment-based formulation with SoC arc-flow variables for D-E-ADARP}\label{TSFFCS}
		
		TSFFCS represents the discrete-time and continuous-SoC formulation in our study.
		It allows us to assess the role of time discretization by comparing it with the continuous-parameter formulation IEBF, and the role of SoC discretization by comparing it with the fully discretized formulation BTSFF.
		Here, SoC consumption is not necessarily discretized, whereas the recharging time, and consequently the recharged amount, are discretized. 
		TSFFCS also employs time-space charging arcs to couple the discretized recharging duration with the continuous SoC change.

		The time-space fragment network is denoted by $G=(tsN, tsF, tsA)$, where $tsN$, $tsF$, and $tsA$ denote the sets of time-space node, time-space fragment, and time-expanded node arc, respectively. A \textit{time-space node} $h=(i,t)\in tsN$ corresponds to a physical location~$i\in N$ and a time index~$t\in T$. $tsN$ includes the time-space origin depot $tsO^+=(0,t_{\min})$, the time-space destination depot $tsO^-=(2n+1,t_{\max})$, and all time-space pickup and delivery nodes associated with pickup and delivery locations. Let $P_{ts}$ and $D_{ts}$ denote the sets of pickup and delivery time-space nodes.
		
		A \textit{time-space fragment} $(f,t)\in tsF$ corresponds to a physical fragment~$f$ with a feasible departure time~$t$. Since the earliest arrival time at the end of $f$ can be determined by $(f,t)$, we use $(f,t)$ to uniquely identify a time-space fragment. The set $tsF$ is generated by combining each physical fragment with all feasible departure times.

		A \textit{time-expanded node arc} $(a,t,t') \in tsA$ represents movement along arc $a$ starting at time $t$ and ending at time $t'$. It can be one of three types: a \textit{time-space normal node arc} when $a=(i,i')$ and $t'=t+C_a^t$, a \textit{time-space idle arc} when $a=(i,i)$, or a \textit{time-space charging arc} when $a=(i,k,i')$. Here, $i$ and $i'$ are the physical start and end locations, and $k$ denotes a charging station. 
		Let $tsA^{n}\subseteq tsA$ and $tsA^{c}\subseteq tsA$ denote the sets of time-space normal node arcs and time-space charging arcs, respectively. 
		A time-space normal node arc represents physical travel and thus incurs the same travel cost and SoC consumption as its physical counterpart.
		A time-space idle arc indicates waiting without spatial movement.
		For a time-space charging arc $(a=(i,k,i'),t,t') \in tsA^{c}$, the time difference $t'-t$ incorporates both travel and recharging durations. 
		
		We define $\zeta_{att'}$ as the SoC change along the arc $(a,t,t')$, defined as the consumed SoC minus the recharged SoC: 
		$\zeta_{att'} = C^b_{ik}+C^b_{ki'}-\alpha\bigl(t'-t-C^t_{ik}-C^t_{ki'}\bigr)$, 
		where $a=(i,k,i')$. Here, $C^b_{ik}+C^b_{ki'}$ represents the consumed SoC, and $\alpha\bigl(t'-t-C^t_{ik}-C^t_{ki'}\bigr)$ represents the recharged SoC, where the recharging time equals the time difference ($t'-t$) minus the travel time ($C^t_{ik}+C^t_{ki'}$). 
		The notation $(a, t, t')$ is used for time-expanded node arcs rather than the notation $(a, t)$ because, for a time-space charging arc, the earliest arrival time $t'$ cannot be uniquely deduced from $a$ and $t$ alone. The set $tsA$ is constructed by combining physical node arcs and nodes with feasible time indices and, when applicable, with charging stations. For each $a=(i,k,i')$, we let $s(a)$ and $e(a)$ denote the physical subarcs $(i,k)$ and $(k,i')$, respectively.

		Thereafter, we let $C_f^b$ (which previously also denotes SoC consumption on location arcs) denote the SoC consumption incurred on a physical fragment or a normal node arc $f \in F \cup A^n$. Notably, the SoC change along a time-space charging arc is represented by $\zeta$. Additionally, we use $C^{ts}_f$ to denote the weighted cost associated with the physical counterpart~$f$ of a time-space fragment or a time-expanded node arc. 
		
		For each time-space node~$h \in tsN$, let $tsF(h)^+$ ($tsF(h)^-$) denote the set of time-space fragments starting (ending) at $h$, and $tsA(h)^+$ ($tsA(h)^-$) the set of time-expanded node arcs starting (ending) at $h$. For each pickup location~$i \in P$, let $tsF_i$ denote the set of time-space fragments serving~$i$. For each physical location~$i \in P \cup D$, let $tsF_i^+$ ($tsF_i^-$) denote the set of time-space fragments whose starting (ending) location is $i$, $tsA_i^+$ ($tsA_i^-$) the set of time-expanded node arcs whose starting (ending) location is $i$, and $F_i^+$ ($F_i^-$) and $A_i^+$ ($A_i^-$) the sets of physical fragments and normal node arcs starting (ending) at $i$, respectively.

		\begin{figure}[h]
			\centering
			
			\begin{minipage}[t]{0.495\textwidth}
				\centering
				\begin{tikzpicture}
					\begin{axis}[
						width=7.55cm,
						height=5.9cm,
						xmin=-0.4, xmax=13.4,
						ymin=-9.8, ymax=0.8,
						xtick={0,1,...,13},
						xticklabels={$t_0$,$t_1$,$t_2$,$t_3$,$t_4$,$t_5$,$t_6$,$t_7$,$t_8$,$t_9$,$t_{10}$,$t_{11}$,$t_{12}$,$t_{13}$},
						ytick={0,-1,-2,-3,-4,-5,-6,-7,-8,-9},
						yticklabels={$L_0$,$L_1$,$L_2$,$L_3$,$L_4$,$L_5$,$L_6$,$L_7$,$L_8$,$L_9$},
						xlabel={Time},
						ylabel={Space},
						xmajorgrids,
						ymajorgrids,
						x grid style={draw=gray!20},
						y grid style={draw=gray!20,dashed},
						clip=false,
						enlargelimits=false,
						tick label style={font=\small},
						label style={font=\small},
						every axis plot/.append style={line cap=round},
						]
						
						\addplot[only marks, mark=*, mark size=2.1pt, black] coordinates {(0,0)};
						\node[font=\small, right=1pt] at (axis cs:0,0) {$O^+$};
						
						\addplot[only marks, mark=*, mark size=2.1pt, black] coordinates {(13,-9)};
						\node[font=\small, left=1pt] at (axis cs:13,-9) {$O^-$};
						
						\addplot[only marks, mark=*, mark size=1.9pt, blue] coordinates {(1,-1)};
						\node[font=\small, below left=0 pt] at (axis cs:1,-1) {$p_1$};
						
						\addplot[only marks, mark=*, mark size=1.9pt, blue] coordinates {(3,-2)};
						\node[font=\small, below=1pt] at (axis cs:3,-2) {$d_2$};
						
						\addplot[only marks, mark=*, mark size=1.9pt, blue] coordinates {(4,-2)};
						\node[font=\small, below ] at (axis cs:4,-2) {$d_2$};
						
						\addplot[only marks, mark=*, mark size=1.9pt, blue] coordinates {(5,-3)};
						\node[font=\small, below] at (axis cs:5,-3) {$p_4$};
						
						\addplot[only marks, mark=*, mark size=1.9pt, blue] coordinates {(7,-4.1)};
						\node[font=\small, below=1pt] at (axis cs:7,-4.1) {$d_4$};
						
						\addplot[only marks, mark=*, mark size=1.9pt, blue] coordinates {(8,-5.2)};
						\node[font=\small, below=1pt] at (axis cs:8,-5.2) {$s_1$};
						
						\addplot[only marks, mark=*, mark size=1.9pt, blue] coordinates {(10,-5.2)};
						\node[font=\small, above=1pt] at (axis cs:10,-5.2) {$s_1$};
						
						\addplot[only marks, mark=*, mark size=1.9pt, blue] coordinates {(11,-7.2)};
						\node[font=\small, above right=0pt] at (axis cs:11,-7.2) {$p_7$};
						
						\addplot[only marks, mark=*, mark size=1.9pt, blue] coordinates {(12,-8.1)};
						\node[font=\small, right=1pt] at (axis cs:12,-8.1) {$d_7$};
						
						\addplot[dashed, black!75, line width=1.0pt] coordinates {(0,0) (1,-1)};
						\addplot[dashed, black!75, line width=1.0pt] coordinates {(4,-2) (5,-3)};
						\addplot[dashed, black!75, line width=1.0pt] coordinates {(12,-8.1) (13,-9)};
						
						\addplot[black!75, line width=1.3pt, smooth]
						coordinates {(1,-1) (2.1,-1.45) (3,-2)};
						\node[font=\small, above=-0.5pt] at (axis cs:3.15,-1.3) {$p_1$-$p_2$-$d_1$-$d_2$};
						
						\addplot[black!75, line width=1.3pt, smooth]
						coordinates {(5,-3) (6,-3.45) (7,-4.1)};
						\node[font=\small, above right=-2pt] at (axis cs:6,-3.35) {$p_4$-$d_4$};
						
						\addplot[black!75, line width=1.3pt, smooth]
						coordinates {(11,-7.2) (11.45,-7.55) (12,-8.1)};
						\node[font=\small, above right=-3 pt] at (axis cs:11.42,-7.52) {$p_7$-$d_7$};
						
						\addplot[dashed, draw=gray!60, line width=0.8pt]
						coordinates {(3,-2) (4,-2)};
						
						\addplot[dashed, draw=gray!65, line width=1.6pt]
						coordinates {(7,-4.1) (11,-7.2)};
						
						\addplot[draw=gray!55, line width=0.85pt] coordinates {(7,-4.1) (8,-5.2)};
						\addplot[draw=gray!55, line width=0.85pt] coordinates {(8,-5.2) (10,-5.2)};
						\addplot[draw=gray!55, line width=0.85pt] coordinates {(10,-5.2) (11,-7.2)};
						
					\end{axis}
				\end{tikzpicture}
				
				\vspace{1mm}
				{\small (a) Time-space view for TSFFCS}
			\end{minipage}
			\hfill
			\begin{minipage}[t]{0.495\textwidth}
				\centering
				\begin{tikzpicture}
					\begin{axis}[
						width=7.55cm,
						height=5.9cm,
						xmin=-0.4, xmax=13.4,
						ymin=0, ymax=105,
						xtick={0,1,...,13},
						xticklabels={$t_0$,$t_1$,$t_2$,$t_3$,$t_4$,$t_5$,$t_6$,$t_7$,$t_8$,$t_9$,$t_{10}$,$t_{11}$,$t_{12}$,$t_{13}$},
						ytick={0,10,...,100},
						xlabel={Time},
						ylabel={SoC (\%)},
						xmajorgrids,
						ymajorgrids,
						x grid style={draw=gray!20},
						y grid style={draw=gray!20,dashed},
						clip=false,
						enlargelimits=false,
						tick label style={font=\small},
						label style={font=\small},
						every axis plot/.append style={line cap=round},
						]
						
						\addplot[only marks, mark=*, mark size=2.1pt, black] coordinates {(0,100)};
						\node[font=\small, right=1pt] at (axis cs:0,100) {$O^+$};
						
						\addplot[only marks, mark=*, mark size=2.1pt, black] coordinates {(13,10)};
						\node[font=\small, left=1pt] at (axis cs:13,10) {$O^-$};
						
						\addplot[only marks, mark=*, mark size=1.9pt, blue] coordinates {(1,90)};
						\node[font=\small, below left= 0pt] at (axis cs:1,90) {$p_1$};
						
						\addplot[only marks, mark=*, mark size=1.9pt, blue] coordinates {(3,70)};
						\node[font=\small, below=1pt] at (axis cs:3,70) {$d_2$};
						
						\addplot[only marks, mark=*, mark size=1.9pt, blue] coordinates {(4,70)};
						\node[font=\small, above=1pt] at (axis cs:4,70) {$d_2$};
						
						\addplot[only marks, mark=*, mark size=1.9pt, blue] coordinates {(5,60)};
						\node[font=\small, below left=0pt] at (axis cs:5,60) {$p_4$};
						
						\addplot[only marks, mark=*, mark size=1.9pt, blue] coordinates {(7,40)};
						\node[font=\small, below=1pt] at (axis cs:7,40) {$d_4$};
						
						\addplot[only marks, mark=*, mark size=1.9pt, blue] coordinates {(8,30)};
						\node[font=\small, below=1pt] at (axis cs:8,30) {$s_1$};
						
						\addplot[only marks, mark=*, mark size=1.9pt, blue] coordinates {(10,70)};
						\node[font=\small, above=1pt] at (axis cs:10,70) {$s_1$};
						
						\addplot[only marks, mark=*, mark size=1.9pt, blue] coordinates {(11,60)};
						\node[font=\small, above right=0pt] at (axis cs:11,60) {$p_7$};
						
						\addplot[only marks, mark=*, mark size=1.9pt, blue] coordinates {(12,50)};
						\node[font=\small, right=1pt] at (axis cs:12,50) {$d_7$};
						
						\addplot[dashed, black!75, line width=1.0pt] coordinates {(0,100) (1,90)};
						\addplot[dashed, black!75, line width=1.0pt] coordinates {(4,70) (5,60)};
						\addplot[dashed, black!75, line width=1.0pt] coordinates {(12,50) (13,10)};
						
						\addplot[black!75, line width=1.3pt, smooth]
						coordinates {(1,90) (2.1,80) (3,70)};
						\node[font=\small, above= 2pt] at (axis cs:3.15,84) {$p_1$-$p_2$-$d_1$-$d_2$};
						
						\addplot[black!75, line width=1.3pt, smooth]
						coordinates {(5,60) (6,50) (7,40)};
						\node[font=\small, above right=-5pt] at (axis cs:6,54) {$p_4$-$d_4$};
						
						\addplot[black!75, line width=1.3pt]
						coordinates {(11,60) (12,50)};
						\node[font=\small, above right=-1pt] at (axis cs:11.45,55.5) {$p_7$-$d_7$};
						
						\addplot[dashed, draw=gray!60, line width=0.8pt]
						coordinates {(3,70) (4,70)};
						
						\addplot[dashed, draw=gray!65, line width=1.6pt]
						coordinates {(7,40) (11,60)};
						
						\addplot[draw=gray!55, line width=0.85pt] coordinates {(7,40) (8,30)};
						\addplot[draw=gray!55, line width=0.85pt] coordinates {(8,30) (10,70)};
						\addplot[draw=gray!55, line width=0.85pt] coordinates {(10,70) (11,60)};
						
					\end{axis}
				\end{tikzpicture}
				\vspace{1mm}
				{\small (b) SoC-time view for BTSFF}
			\end{minipage}
			
			\caption{Time-space and SoC-time representations of the same feasible path}\label{fig:two_views_path}
		\end{figure}
		
		Figure~\ref{fig:two_views_path} presents the same feasible path under two aligned representations, for illustrating TSFFCS and BTSFF, respectively. $O^+$ and $O^-$ denote the origin and destination depots, respectively. The left panel of Figure~\ref{fig:two_views_path} illustrates the time-space fragment-based representation of a feasible path in TSFFCS. Here, the horizontal axis represents time and the vertical axis represents physical locations. Dark solid lines denote time-space fragments, dark dashed lines denote normal node arcs, the gray dashed line denotes an idle arc, and the thick gray dashed line denotes a time-space charging arc. The light gray solid path further shows the detailed route of the time-space charging arc via station $s_1$. 
		
		\begin{table}[h]
			\footnotesize
			\centering
			\caption{Decision variables of TSFFCS}
			\begin{tabular}{@{}ll@{}}
				\toprule
				Variables & Definition \\\midrule
				$x_{ft}$ & = 1, if a time-space fragment $(f,t) \in tsF$ is traversed; = 0, otherwise \\
				$y_{att'}$ & = 1, if a time-expanded node arc $(a,t,t') \in tsA$ is traversed; = 0, otherwise \\
				$w_u$ & entry SoC associated with a physical fragment or a physical normal node arc~$u \in F \cup A^n$ \\
				$w_{att'}$ & entry SoC associated with a time-space charging node arc~$(a,t,t') \in tsA^c$ \\ 
				\bottomrule
			\end{tabular}\label{table:ch3notation3.24}
		\end{table}

		The decision variables of TSFFCS are summarized in Table~\ref{table:ch3notation3.24}. The problem can then be formulated as follows.

		{\footnotesize
			\setlength{\jot}{1pt}
			
			\begin{align}
				\min \quad
				\sum_{(f,t)\in tsF} C^{ts}_f\, x_{ft}
				+ \sum_{(a,t,t')\in tsA} C^{ts}_a\, y_{att'}
				\label{ch4:TSFFobj}
			\end{align}
			
			\text{s.t.}
			\begin{align}
				\sum_{(f,t)\in tsF(h)^-} x_{ft}
				+ \sum_{(a,t,t')\in tsA(h)^-} y_{att'}
				= \sum_{(a,t,t')\in tsA(h)^+} y_{att'}+&\sum_{(f,t)\in tsF(h)^+} x_{ft}
				&& h\in P_{ts}\cup D_{ts}
				\label{ch4:TSFFflow}
				\\
				\sum_{(f,t)\in tsF_i} x_{ft}
				&= 1
				&& i\in P
				\label{ch4:TSFFcover}
				\\
				\sum_{(a,t,t')\in tsA(tsO^+)^+} y_{att'}
				&\le |V|
				\label{ch4:TSFFvehicle}
				\\
				\sum_{f\in F_j^-} w_f - \sum_{(f,t)\in tsF_j^-} C_{f}^b x_{ft} 
				\ge \sum_{(a,t,t')\in tsA_j^+ \cap tsA^c} w_{att'}  + & \sum_{a\in A_j^+} w_a
				&& j\in D
				\label{ch4:TSFFbatteryD}
				\\
				\sum_{(a,t,t')\in tsA_j^-\cap tsA^c} w_{att'} + \sum_{a\in A_j^-} w_a 
				- \sum_{(a,t,t')\in tsA_j^-\cap tsA^c} \zeta_{att'} y_{att'} \notag \\
				\qquad
				- \sum_{(a,t,t')\in tsA_j^-\cap tsA^n} C_{a}^b y_{att'}
				&\ge \sum_{f\in F_j^+} w_f
				&& j\in P
				\label{ch4:TSFFbatteryP}
				\\
				\sum_{(a,t,t')\in tsA_j^-\cap tsA^c} {C_{e(a)}^b}  y_{att'}
				+ \sum_{f\in F_j^+} w_f
				&\le b_{\max}
				&& j\in P
				\label{ch4:TSFFfromstation}
				\\
				(b_{\min}+{C_{s(a)}^b}) y_{att'}
				&\le  w_{att'}
				&& (a,t,t')\in tsA^c
				\label{ch4:TSFFtoStation}
				\\
				(\gamma b_{\max} + C_{a}^b)\, y_{att'}
				&\le w_a 
				&& (a,t,t')\in tsA_{2n+1}^-\cap tsA^n
				\label{ch4:TSFFendArc}
				\\
				(\gamma b_{\max} + \zeta_{att'}) y_{att'}
				&\le w_{att'}
				&& (a,t,t')\in tsA_{2n+1}^-\cap tsA^c
				\label{ch4:TSFFchargingendArc}
				\\
				w_f
				&\le b_{\max} \sum_{t\in T} x_{ft}
				&& f\in F
				\label{ch4:TSFFwf}
				\\
				w_a &\le b_{\max} \sum_{t,t' \in T} y_{a t t'} && a \in A^n \label{ch4:TSFFwa}
				\\
				w_{att'}
				&\le b_{\max} y_{att'}
				&& (a,t,t')\in tsA^c
				\label{ch4:TSFFchargingwa}\\
				w_a& = b_{\max} \sum_{t,t' \in T} y_{a t t'}&& a\in A_{0}^+\label{ch4:TSFForiginArc}\\
				w_{att'} &= b_{\max} y_{att'} && (a,t,t')\in tsA_{0}^+\cap tsA^c \label{ch4:TSFForiginChargingArc}\\
				x_{ft}
				&\in \{0,1\}
				&& (f,t)\in tsF
				\label{ch4:TSFFx}
				\\
				y_{att'}
				&\in \{0,1\}
				&& (a,t,t')\in tsA
				\label{ch4:TSFFy}
				\\
				0 \le w_u
				&\le b_{\max}
				&& u\in F\cup A^n
				\label{ch4:TSFFw}\\
				0 \le w_{att'}
				&\le b_{\max}
				&& (a,t,t')\in tsA^c
				\label{ch4:TSFFwc}
			\end{align}
		}
		
		The objective~(\ref{ch4:TSFFobj}) minimizes the total weighted cost of all traversed time-space fragments and time-expanded node arcs. 
		Constraints~(\ref{ch4:TSFFflow}) enforce the network flow constraints at each time-space node. 
		Constraints~(\ref{ch4:TSFFcover}) guarantee that each pickup request is covered exactly once by a time-space fragment. 
		Constraints~(\ref{ch4:TSFFvehicle}) limit the number of vehicles departing from the origin depot to at most $|V|$. Up to this point, the formulation coincides with the time-space fragment-based one presented in \citet{Alyasiry2019}.
		
		The subsequent constraints are specifically tailored for SoC considerations.
		Constraints~(\ref{ch4:TSFFbatteryD}) and~(\ref{ch4:TSFFbatteryP}) propagate the SoC continuously along time-space fragments and time-expanded node arcs. 
		Constraints~(\ref{ch4:TSFFbatteryD}) ensure that, at each delivery location, the entry SoC of fragments ending at that location, after subtracting their SoC consumption, is no less than the entry SoC assigned to the outgoing physical normal or time-space charging arcs. 
		Constraints~(\ref{ch4:TSFFbatteryP}) enforce the symmetric propagation at pickup nodes: the entry SoC of normal or time-space charging arcs ending at the location, after accounting for SoC consumption or charging gains, is no less than the entry SoC of the fragments departing from that location. Notably, the notation $\zeta_{att'}$ equals the consumed SoC minus the recharged SoC.
		Together, these constraints ensure continuous SoC consistency across fragment-arc transitions. 
		
		Additional constraints impose upper and lower bounds on SoC and specify the initial SoC.
		Constraints~(\ref{ch4:TSFFfromstation}) and~(\ref{ch4:TSFFtoStation}) enforce the maximum recharging amount and the minimum SoC on time-space charging arcs, as detailed after the general introduction of all these constraints. 
		Constraints~(\ref{ch4:TSFFendArc}) and~(\ref{ch4:TSFFchargingendArc}) impose minimum SoC requirements for the normal and time-space charging arcs entering the destination depot. 
		Constraints~(\ref{ch4:TSFFwf})-(\ref{ch4:TSFFchargingwa}) link the continuous SoC variables $w$ to the binary selection variables $x$ and $y$, ensuring that SoC variables are active only when the corresponding time-space fragments or time-expanded node arcs are selected. 
		Constraints~(\ref{ch4:TSFForiginArc}) and (\ref{ch4:TSFForiginChargingArc}) initialize the SoCs on normal arcs and time-space charging arcs departing from the origin depot, respectively.
		Constraints~(\ref{ch4:TSFFx})-(\ref{ch4:TSFFwc}) define the domains of the decision variables $x$, $y$, and $w$.
		
		In particular, constraints~(\ref{ch4:TSFFfromstation}) ensure that, at each pickup location, the SoC consumption from a charging station to that location plus the entry SoC of the outgoing fragments do not exceed the battery capacity~$b_{\max}$. 
		We aggregate the entry SoC of all time-space charging arcs ending at the same pickup node, which yields a tighter formulation than handling each charging time-space node arc separately. 
		In the disaggregated form, the constraint would be: $C_{s(a)}^b y_{att'} + \sum_{(f,t)\in tsF^+_{j: j=el(a)}} w_f \le b_{\max}$ for each $(a,t,t')\in tsA^c$, where $el(a)$ denotes the end location of $a$. 
		Constraints~(\ref{ch4:TSFFtoStation}) ensure that, for each charging time-space node arc, the entry SoC assigned to that arc is at least the SoC consumption from its start location to the charging station plus the required minimum SoC level. The constraints are kept in a disaggregated form, as they are tighter than an aggregated form. The key difference between constraints~(\ref{ch4:TSFFfromstation}) and (\ref{ch4:TSFFtoStation}) lies in the domain of the $w$ variable: the former uses the physical fragment-based variable to model entry SoC, while the latter uses the time-space arc-based one.

		Consistent with \citet{Bongiovanni2019} and \citet{Su2024}, when vehicles have distinct origin depots (equal in number to the fleet size) and multiple potential destination depots (possibly exceeding the fleet size), the sets of time-space origin and destination depots are denoted by $tsOv^+$ and $tsDv^-$, respectively. Then, constraints~(\ref{ch4:vehicle2tso}) and~(\ref{ch4:vehicle2tsd}) replace constraints~(\ref{ch4:TSFFvehicle}). All other constraints remain unchanged, with their domains extended accordingly. This separate presentation aligns the formulation above with existing fragment-based approaches.

		\begin{align}
			\sum_{(a,t,t')\in tsA{(i)}^+} y_{att'} &= 1  && i\in tsOv^+ \label{ch4:vehicle2tso}\\
			\sum_{(a,t,t')\in tsA{(i)}^-} y_{att'} &\le 1 && i\in tsDv^- \label{ch4:vehicle2tsd}
		\end{align}

		\subsubsection{Battery-time-space fragment-based formulation for D-E-ADARP}\label{BTSFF}

		BTSFF serves as the fully discretized formulation in our study: it discretizes both time and SoC and therefore allows us to evaluate the computational impact of explicit SoC discretization within the fragment-based framework.
		
		Before presenting BTSFF, we first introduce its network for D-E-ADARP. The battery-time-space (BTS)~fragment-based network~$G(N_N, F_N, A_N)$ for D-E-ADARP comprises the BTS~node set~$N_N$, the BTS~fragment set~$F_N$, and the battery-time-expanded node arc set~$A_N$. The BTS network extends the (physical) fragment-based network (see Section~\ref{ch4:Preliminaries}) by incorporating time and SoC dimensions.
		
		The \textit{BTS~node}~$h \in N_N$ is a tuple consisting of a location~$i\in N$, a time index~$t\in T$, and a SoC index~$b\in B$. Specifically, $h$ can be expressed as $(i, t, b)$. The BTS node set $N_N$ includes the BTS origin depot $\mathcal{O}^+ = (0, t_{\min}, b_{\max})$, the BTS destination depot $\mathcal{O}^- = (2n + 1, t_{\max}, \gamma b_{\max})$, the BTS pickup nodes, and the BTS delivery nodes. The BTS pickup and delivery nodes correspond to the pickup and delivery locations, with the respective sets denoted by $P_N$ and $D_N$. All BTS nodes can be generated by assigning each location $i$ with all its feasible time indices $t \in \{tn \in T : et_i \le tn \le lt_i\}$ and all its feasible SoC indices $b \in B$. Here, the origin depot has ${t_{\min}}$ and $b_{\max}$ as its only feasible time and SoC indices, respectively, while the destination depot has ${t_{\max}}$ and $\gamma b_{\max}$ as its only feasible time and SoC indices. 
		
		The \textit{BTS fragment} represents a three-dimensional movement, capturing SoC consumption, time progression (including waiting time induced by time-window feasibility along its physical fragment), and physical travel. A BTS fragment $g$ can be expressed as $(h, f, h')$, where $f$ represents a physical fragment, $h = (i, t, b)$ is a BTS start node, and $h' = (i', t', b')$ is a BTS end node. $F_N$ is constructed by combining all fragments with various feasible departure times and initial SoCs. A fragment with physical start node $i$, physical end node $i'$, departing time $t \in DT_f$, and initial SoC $b \in B$ (where each fragment may have a subset of SoC indices) can be represented in BTS form as $((i, t, b), f, (i', t', b'))$.  Here, $t'$ is the earliest arrival time at $i'$ when a vehicle departs from node $i$ at time $t$, follows $f$, and satisfies all the time window and maximum ride time constraints. $b'$ is equal to $b$ minus the total SoC consumption of all arcs within $f$. During creation, the condition $b' \ge b_{\min}$ must hold.
		
		The \textit{battery-time-expanded node arc} can be classified into three types, namely, the \textit{BTS node arc}, the \textit{idle node arc}, and the \textit{charging node arc}. Its set is denoted by $A_N$. A battery-time-expanded node arc can be generally represented as $((i,t,b), \phi, (i',t', b'))$, where $\phi$ in this context might denote an arc, a location, or a partial path. 
		
		The BTS node arc represents movement in terms of SoC consumption, time progression, and physical travel between two physical nodes (locations). The BTS node arc follows the same structure as the BTS fragment, and its set is constructed analogously to $F_N$ based on all physical node arcs. Here, for a BTS node arc ending at $\mathcal{O}^-$, if the resulting SoC satisfies $b' \ge \gamma b_{\max}$, we set $b'=\gamma b_{\max}$.
		
		The idle node arc reflects movement solely in the time dimension while maintaining the same SoC and physical location. Its set is generated by assigning each location~$i$ with all feasible departure times $t \in T$ and initial SoCs $b \in B$. An idle node arc typically has the format $((i, t, b), i, (i, t', b))$, where $t' \in T$ is the immediate successor of $t \in T$.
		
		The charging node arc represents movement from a customer location to a charging station and then to another customer location. The charging node arc set is constructed by combining all node arcs with feasible departure time, initial SoCs, final SoCs, and charging stations. If a node arc $(i,i')$ passes through a charging station $s \in S$ and departs at time $t \in T$ with an initial SoC $b \in B$ and a final SoC $b' \in B$, the corresponding charging node arc is represented as $((i,t,b), (i,s,i'), (i',t',b'))$. The earliest arrival time $t'$ is then given by $t' = t + C^t_{is} + C^t_{si'} + \frac{(b'-b) + (C^b_{is}+C^b_{si'})}{\alpha}$. Here, $C^b_{is}$ and $C^b_{si'}$ denote the SoC consumption from $i$ to $s$ and from $s$ to $i'$, respectively, while $\tfrac{(b'-b)}{\alpha}$ and $\tfrac{(C^b_{is}+C^b_{si'})}{\alpha}$ represent the recharging time required to increase the SoC by $b'-b$ and the additional recharging time needed to compensate for the SoC consumed during the trip.
		During the generation of the charging node arc, the conditions $b - C^b_{is} \geq b_{\min}$ and $b' + C^b_{si'} \leq b_{\max}$ must hold.
		
		Building on these definitions, the weighted costs $C^{bts}_g$ for $g\in F_N$ and $C^{bts}_c$ for $c\in A_N$ are derived directly from their physical counterparts.

		The right panel of Figure~\ref{fig:two_views_path}  illustrates the relationship between SoC and time along a feasible path. Here, the horizontal axis represents time, while the vertical axis indicates SoC. The dark solid lines represent BTS fragments corresponding to the trips from $p1$ to $d2$, $p4$ to $d4$, and $p7$ to $d7$; if a fragment contains intermediate locations, its route path is labeled above the line (e.g., $p1 \to p2 \to d1 \to d2$). The dark dashed lines denote BTS node arcs, such as the connections from $O^+$ to $p1$, from $d2$ to $p4$, and from $d7$ to $O^-$. Both BTS fragments and BTS node arcs result in a decrease in SoC while the time index increases. The (normal-thickness) gray dashed line corresponds to an idle node arc, indicating the passage of time without changes in both location and SoC (e.g., between two $d2$ nodes). Finally, the ultra-thick dashed gray line corresponds to a charging node arc, representing the movement from $d4$ to $p7$ with an intermediate stop at station $s1$. The detailed route path along this arc is shown as a light gray solid line. Along this arc, the SoC first decreases during travel from $d4$ to $s1$ (the left one), increases while recharging at $s1$, and then decreases again from $s1$ (the right one) to $p7$, while the time index increases monotonically due to both travel and charging durations.

		We will now formalize the foregoing concepts. We use the notation $F_N(h)^+$ for the set of BTS fragments starting at BTS node $h\in P_N$, $F_N(h)^-$ for those ending at BTS node $h\in D_N$, and $F_N^i$ for the set of BTS fragments traversing location $i\in P$. Additionally, $A_N(h)^+$ and $A_N(h)^-$ represent the set of battery-time-expanded node arcs starting from BTS node $h\in N_N\setminus \{\mathcal{O}^-\}$ and the set of battery-time-expanded node arcs ending at BTS node $h\in N_N\setminus \{\mathcal{O}^+\}$, respectively.

		\begin{table}[h]
			\footnotesize
			\centering
			\caption{Decision variables of BTSFF}
				\begin{tabular}{@{}ll@{}}
					\toprule
					Decision variables & Definition \\\midrule
					$X_{g}$ & = 1, if the BTS fragment  $g\in F_N$ is traversed; = 0, otherwise \\
					$Y_{c}$ & = 1, if the battery-time-expanded node arc  $c\in A_N$ is traversed; = 0, otherwise\\\bottomrule
				\end{tabular}
			\label{table:ch3notation3.23}
		\end{table}

		The decision variables of BTSFF are summarized in Table~\ref{table:ch3notation3.23}. The problem can then be formulated as follows.
		
		\allowdisplaybreaks
		\begin{align}
			\min&& \sum_{g\in F_N} C^{bts}_g X_{g}+ \sum_{c\in A_N} C^{bts}_c Y_{c}& &&\label{ch4:MDARPobj2}\\
			s.t.
			&&\sum_{g\in F_N(h)^-} X_{g}+\sum_{c\in A_N(h)^-} Y_{c}  &=\sum_{g\in F_N(h)^+} X_{g}+\sum_{c\in A_N(h)^+}  Y_{c}      &&h\in P_N\cup D_N  \label{ch4:FragmentArc2}\\
			&&\sum_{g\in F_N^i}  X_{g}                  &= 1   &&i\in P\label{ch4:cover3}\\
			&&\sum_{c\in A_N(\mathcal{O}^+)^+} Y_{c}                  &\le  |V|   &&\label{ch4:vehicle2}\\
			&&X_{g}					&\in\{0, 1\}		&& g{\in}F_N\label{ch4:DARdomain31}\\
			&&Y_{c}					&\in \{0,1\}	&&c{\in}A_N\label{ch4:DARdomain32}
		\end{align}

		The objective function~(\ref{ch4:MDARPobj2}) minimizes the total weighted cost of traversed BTS fragments and battery-time-expanded node arcs. 
		Constraints~(\ref{ch4:FragmentArc2}) describe the network flow for the BTS fragments and battery-time-expanded node arcs.
		This can be divided into two scenarios: network flow at a BTS pickup node belonging to $P_N$ or at a BTS delivery node belonging to $D_N$, since a BTS fragment can only start at a BTS pickup node and end at a BTS delivery node.
		Constraints~(\ref{ch4:cover3}) ensure that each pickup location is visited exactly once. Since pairing constraints are already embedded within each fragment, it is not necessary to separately ensure that the delivery location is visited. Constraints~(\ref{ch4:vehicle2}) state that at most $|V|$ vehicles are used. This is achieved by restricting the number of vehicles traversing the battery-time-expanded node arcs that originate from $\mathcal{O}^+$.
		Constraints~(\ref{ch4:DARdomain31}) and (\ref{ch4:DARdomain32}) specify the domains of $X$ and $Y$, respectively.
		
		Consistent with \citet{Bongiovanni2019} and \citet{Su2024}, when vehicles have distinct origin nodes (equal in number to the fleet size) and multiple potential destination nodes (possibly exceeding the fleet size), the corresponding BTS origin and destination sets are denoted by $Ov^+$ and $Ov^-$, respectively. Constraints~(\ref{ch4:vehicle2o}) and~(\ref{ch4:vehicle2d}) then replace constraints~(\ref{ch4:vehicle2}).
		
		\begin{align}
			\sum_{c\in A_N(h)^+} Y_{c}                  &=  1  &&h\in Ov^+ &&\label{ch4:vehicle2o}\\
			\sum_{c\in A_N(h)^-} Y_{c}                  &\le  1  &&h\in Ov^- &&\label{ch4:vehicle2d}
		\end{align}	
		

		\section{Computational results}\label{numericaldiscussion}
		
		The computational study examines how discretization affects the relative effectiveness of different formulations for the E-ADARP. In particular, we investigate whether discretizing time and/or SoC can simplify the problem and change the preferred formulation. To this end, we compare BTSFF, TSFFCS, and IEBF, where the comparison between BTSFF and TSFFCS highlights the role of SoC discretization, while the comparison between TSFFCS and IEBF reveals the role of time discretization.

		\subsection{Test instances and experiments}\label{IG}
		
		To evaluate the computational efficiency of different formulations, we employ the benchmark instances commonly used in E-ADARP studies \citep{Bongiovanni2019, Su2024, Stallhofer2025}. The instances are categorized into three types: a, r, and u. Type-a and type-r are both adapted from classical DARP datasets \citep{doi:10.1287/opre.1060.0283, doi:10.1287/trsc.1090.0272}, differing primarily in problem scale. Each instance assumes unit demand and a vehicle capacity of three. Type-a instances (14 in total) involve 2-5 vehicles, 16-50 user requests, and an operating horizon of around 10 hours, whereas type-r instances (10 in total) comprise 5-8 vehicles, 60-96 requests, and a 12-hour horizon. All vehicles start with a full battery of 16.5~kWh, the maximum battery capacity, corresponding to a driving range of 5 hours. They must retain at least 10\% of full capacity at all times (thus yielding an effective capacity of 14.85~kWh). Both charging and discharging rates (previously denoted $\alpha$ and $\beta$) are set to 0.055~kWh/min, corresponding to a total driving time of 270 minutes. Type-u instances (14 in total) are generated from an Uber dataset by \citet{Bongiovanni2019}. Similar to type-a, they involve 2-5 vehicles, 16-50 requests, and a 3.5-hour horizon. The effective battery capacity is set to 3.5~kWh, with charging and discharging rates of 0.055 and 0.0715~kWh/min, respectively, corresponding to approximately 50 minutes of driving time. Type-a and type-r instances have a uniform maximum ride time (which has incorporated the service time during preprocessing) of 33 minutes, and type-u has 8.5 minutes. Across all instance types, a minimum SoC level before returning to the depot, $\gamma b_{\max}$, where $\gamma \in \{10\%, 40\%, 70\%\}$, is imposed. The parameter $\lambda$ is set to 0.75. Public access to these instances is provided by \citet{doi:10.1287/opre.1060.0283} and \citet{Bongiovanni2019}.  

		\begin{table}[h]
			\centering
			\caption{Overview of the experimental design}
			\label{table:overview}
			\scriptsize
			\begin{threeparttable}
				\begin{tabular}{p{1.9 cm} p{5.2 cm} p{3.0 cm} p{0.8 cm} p{2.1 cm}}
					\toprule
					\multicolumn{1}{l}{Problem} &
					\multicolumn{1}{l}{Instance parameters} &
					\multicolumn{1}{l}{Compared formulations} &
					\multicolumn{1}{l}{$N_{\max}$} &
					\multicolumn{1}{c}{$\gamma$} \\
					\midrule
					E-ADARP &
					Continuous: type-a and type-u &
					REBF\tnote{1}, IEBF &
					$\{1,\infty\}$ &
					\multirow{2}{*}{$\{10\%,40\%,70\%\}$} \\
					
					D-E-ADARP &
					Discrete\tnote{2}. Type-a: $\{5,2,1\}$; type-u: $\{2,1\}$  &
					IEBF, TSFFCS, BTSFF &
					$\infty$ & \\
					
					\bottomrule
				\end{tabular}
				\begin{tablenotes}
					\scriptsize
					\item[1] REBF: replicated event-based formulation of \citet{Stallhofer2025}.
					\item[2] D-E-ADARP instances are derived from E-ADARP instances through time and SoC discretization (via rounding parameter values); numbers denote the corresponding discretization units (minutes). For type-a, a 5-min unit yields 120 time indices and 60 SoC levels, while for type-u, a 2-min unit yields 100 time indices and 25 SoC levels.
				\end{tablenotes}
				
			\end{threeparttable}
		\end{table}
		
		An overview of the experimental design is presented in Table~\ref{table:overview}.
		For E-ADARP, we primarily compare our IEBF with a replicated event-based formulation (REBF) based on \citet{Stallhofer2025}. REBF uses the cuts~1, 4, 6, and~7 in \citet{Stallhofer2025}. Cut~1 is designed for the DARP setting (and is thus also used in IEBF), whereas the remaining cuts (4, 6, and~7) address SoC constraints. Our replication proves reliable: for the original E-ADARP with $N_{\max}=1$ (see Appendix~A), the results match or surpass those originally reported, especially on medium and large instances, likely due to our use of Gurobi instead of CPLEX, which is used in \citet{Stallhofer2025}. For the setting of $N_{\max}=\infty$ (unlimited visits), REBF, unlike IEBF, requires station replication to represent multiple visits. In type-a and type-u instances, stations are typically visited at most 2-4 times. We therefore set $N_{\max}=5$ to represent the case of $N_{\max}=\infty$.
		Additionally, we include the B\&P results with the settings of $N_{\max}=\infty$ and $N_{\max}=1$ reported by \citet{Su2024} in Appendix~A. For simplicity, experiments are performed on type-a and type-u instances with $\gamma \in \{10\%,40\%,70\%\}$, where $\gamma=70\%$ represents the most challenging setting due to a higher frequency of recharging.

		For D-E-ADARP, we compare several formulations, namely IEBF, TSFFCS, and BTSFF, under the setting $N_{\max}=\infty$, which allows us to focus on discretization effects. The instances are discretized by rounding all travel times, SoC consumptions, time windows, and maximum ride times to the nearest unit of a chosen granularity, derived from the original continuous-parameter benchmark data. Such discretization may change the instances, in particular the resulting number of feasible fragments or events, since time-window interactions and ride-time constraints can be affected by the rounding. The discretized instances are not intended to approximate the continuous problem exactly; rather, they allow us to study how different time and SoC granularities influence the size of the generated networks and the computational performance of the formulations. By construction, the discrete- and continuous-parameter formulations yield the same objective values when applied to the same discretized instances. Here, time and SoC discretizations are defined via equivalent driving-time units. We then select discretization levels that yield comparable numbers of time indices and SoC levels across instance types. For type-a instances (10-hour horizon, 5-hour battery size), we set a 5-minute time step as a default choice, yielding 120 time indices and 60 SoCs (with each unit taking up 1.85\% of battery size). Finer levels, 2 and 1 minutes, are also tested to assess discretization effects. For type-u instances (200-minute horizon, 50-minute battery size), we set a 2-minute time step as a default choice, yielding 100 time indices and 25 SoC levels (each corresponding to 4\% of battery capacity). A finer level of 1 minute is further tested to assess discretization effects. Type-r instances are excluded from the experiments because they share the same structural characteristics as type-a instances, while the use of coarse rounding parameters leads to a very large number of generated fragments. As a result, the computational performance of fragment-based formulations becomes dominated by fragment enumeration, which makes formulation comparisons less informative.

		In evaluating computational performance, we report several indicators: the CPU time (CPU), the total runtime (Time), the best-found objective value (OBJ), the best-known lower bound (LB), and the optimality gap (Gap), all measured under a 30-minute time limit. Here, CPU time refers to the solver time reported by Gurobi, whereas the total runtime includes preprocessing time, network construction time, and CPU time. The optimality gap is computed as $\text{Gap} = \frac{\text{OBJ} - \text{LB}}{\text{OBJ}} \times 100\%$. As the number of test instances is large, we adopt a 30-minute limit for all procedures. We also include results from \citet{Su2024}, which were obtained under a 2-hour limit for type-a instances. Although the runtime limits differ, the 30-minute results of our formulations already provide sufficiently high-quality solutions to demonstrate their computational advantages. For BTSFF and TSFFCS, we additionally report the number of fragments ($|F|$) and the network construction time (Net). All time values reported in the following tables are in seconds.

		All our algorithms are implemented using the Python programming language. All computational experiments are run on a system with an Intel Core i7-7820HQ processor with 2.90 GHz CPU speed and 32 GB of RAM under a Windows 10 64-bit OS\@. All linear formulations are solved with the commercial solver Gurobi 9.0.3 with a single thread; all other Gurobi parameters are set to their default values. 

		\subsection{Computational performance of E-ADARP}\label{numEADARP}
		
		\begin{table}[h] \caption{Average computational performance of different procedures on E-ADARP} \footnotesize \centering \setlength{\tabcolsep}{12pt} \renewcommand{\arraystretch}{1} \begin{threeparttable} \begin{tabular}{@{}ccccccccc@{}} \toprule & & & \multicolumn{3}{c}{IEBF} & \multicolumn{3}{c}{REBF} \\ \midrule Type & $N_{\max}$ & $\gamma$ & CPU & Time & Gap & CPU & Time & Gap \\ \midrule \multirow{6}{*}{a} & \multirow{3}{*}{$\infty$} & 70\% & 201.5 & 201.9 & 0.00\% & 774.1 & 777.1 & 0.44\% \\ & & 40\% & 9.8 & 10.5 & 0.00\% & 107.1 & 109.5 & 0.00\% \\ & & 10\% & 9.4 & 10.2 & 0.00\% & 152.3 & 154.5 & 0.00\% \\ \cmidrule(l){2-9} & \multirow{3}{*}{1} & 70\% & 591.3 & 592.8 & 0.19\% & 658.1 & 661.2 & 0.18\% \\ & & 40\% & 18.8 & 19.9 & 0.00\% & 17.6 & 19.8 & 0.00\% \\ & & 10\% & 8.2 & 9.1 & 0.00\% & 16.2 & 18.1 & 0.00\% \\ \cmidrule(r){1-9} \multirow{6}{*}{u} & \multirow{3}{*}{$\infty$} & 70\% & 273.3 & 277.3 & 0.42\% & 819.6 & 820.2 & 0.56\% \\ & & 40\% & 244.1 & 245.4 & 0.18\% & 305.4 & 308.8 & 0.43\% \\ & & 10\% & 251.0 & 254.8 & 0.23\% & 266.2 & 270.5 & 0.25\% \\ \cmidrule(lr){2-9} & \multirow{3}{*}{1} & 70\% & 415.0 & 416.6 & 0.58\% & 477.5 & 482.4 & NA \\ & & 40\% & 257.6 & 258.9 & 0.31\% & 193.7 & 195.3 & 0.00\% \\ & & 10\% & 235.1 & 236.2 & 0.25\% & 138.7 & 141.1 & 0.00\% \\ \bottomrule \end{tabular} \begin{tablenotes} \scriptsize \item[1] Type: instance type. Gap: optimality gap. \item[2] ``NA'' denotes that no feasible solution is obtained within the time limit for a given instance. For the calculation of averages: (i) instances for which all formulations report ``NA'' are excluded; (ii) a formulation's average is recorded as ``NA'' if it uniquely reports ``NA'' for at least one instance. \end{tablenotes} \end{threeparttable} \label{table:avgE} \end{table}

		Table~\ref{table:avgE} provides the average computational performance of IEBF and REBF under the settings of $N_{\max}=\infty$ and $N_{\max}=1$ for E-ADARP. The detailed results for the two formulations and the benchmark results reported by \citet{Stallhofer2025} and \citet{Su2024} for type-a and type-u instances are provided in Appendix~A.
		These tables further show that both IEBF and REBF outperform the benchmark results of \citet{Su2024} in most cases, while REBF also improves upon the results reported by \citet{Stallhofer2025}, confirming the reliability of our replication.

		Table~\ref{table:avgE} shows that IEBF significantly outperforms REBF when $N_{\max}=\infty$, while under $N_{\max}=1$, IEBF is advantageous only in scenarios with a high recharging probability ($\gamma = 70\%$). These findings highlight the effectiveness of the two key techniques implemented in IEBF (see Section~\ref{IEBF}) when recharging activities are frequent. This pattern is consistent with the intuition that, in low-recharge settings (e.g., $\gamma=10\%$) where vehicles can often complete routes without recharging, the increased number of SoC arc-flow variables (versus node-based ones) would become a computational burden rather than a benefit. 
		
		An additional finding is that instances under the $N_{\max}=\infty$ setting are solved much more easily than those under $N_{\max}=1$. This holds even for REBF, despite the fact that replicating charging stations (an adjustment necessary for $N_{\max}=\infty$) is typically assumed to increase model size and thus computational difficulty.

		\subsection{Computational performance of D-E-ADARP}\label{numDEADARP}

		\begin{table}[h]
			\caption{Average computational performance of different formulations on D-E-ADARP with $N_{\max}=\infty$}
			\footnotesize  
			\centering
			\setlength{\tabcolsep}{6pt}
			\renewcommand{\arraystretch}{1}  
			\begin{threeparttable}
				\begin{tabular}{@{}cccccccccccccc@{}}
					\toprule
					&                    &          & \multicolumn{5}{c}{TSFFCS}                              & \multicolumn{3}{c}{IEBF} & \multicolumn{3}{c}{BTSFF}                                          \\ \midrule
					Type               & Unit               & $\gamma$ & $|F|$                   & Net  & CPU   & Time  & Gap    & CPU    & Time   & Gap    & CPU                  & Time                 & Gap                  \\ \midrule
					\multirow{9}{*}{a} & \multirow{3}{*}{5} & 70\%     & \multirow{3}{*}{3542.3} & 5.1  & 176.4 & 260.2 & 0.17\% & 247.9  & 249.4  & 0.23\% & 228.8                & 444.6                & NA                   \\
					&                    & 40\%     &                         & 5.5  & 8.3   & 99.8  & 0.00\% & 35.9   & 37.5   & 0.00\% & 136.8                & 377.2                & NA                   \\
					&                    & 10\%     &                         & 4.7  & 5.4   & 96.3  & 0.00\% & 29.3   & 30.6   & 0.00\% & 130.9                & 357.0                & NA                   \\ \cmidrule(l){2-14} 
					& \multirow{3}{*}{2} & 70\%     & \multirow{3}{*}{379.8}  & 16.0 & 227.7 & 233.9 & 0.20\% & 235.5  & 236.2  & 0.11\% & \multicolumn{1}{l}{} & \multicolumn{1}{l}{} & \multicolumn{1}{l}{} \\
					&                    & 40\%     &                         & 15.5 & 10.9  & 29.7  & 0.00\% & 29.4   & 30.6   & 0.00\% & \multicolumn{1}{l}{} & \multicolumn{1}{l}{} & \multicolumn{1}{l}{} \\
					&                    & 10\%     &                         & 15.6 & 8.4   & 27.3  & 0.00\% & 13.3   & 14.7   & 0.00\% & \multicolumn{1}{l}{} & \multicolumn{1}{l}{} & \multicolumn{1}{l}{} \\ \cmidrule(l){2-14} 
					& \multirow{3}{*}{1} & 70\%     & \multirow{3}{*}{163.5}  & 60.3 & 486.8 & 530.0 & 0.23\% & 76.7   & 77.4   & 0.00\% & \multicolumn{1}{l}{} & \multicolumn{1}{l}{} & \multicolumn{1}{l}{} \\
					&                    & 40\%     &                         & 58.3 & 41.9  & 101.7 & 0.00\% & 10.2   & 11.0   & 0.00\% &                      & \multicolumn{1}{l}{} & \multicolumn{1}{l}{} \\
					&                    & 10\%     &                         & 58.8 & 34.4  & 94.7  & 0.00\% & 8.1    & 9.1    & 0.00\% &                      &                      &                      \\ \midrule
					\multirow{6}{*}{u} & \multirow{3}{*}{2} & 70\%     & \multirow{3}{*}{313.7}  & 5.2  & 568.4 & 587.8 & 0.23\% & 585.5  & 587.4  & 0.91\% & 340.3                & 399.5                & 0.10\%               \\
					&                    & 40\%     &                         & 4.9  & 102.8 & 123.0 & 0.00\% & 398.4  & 401.2  & 0.53\% & 441.7                & 496.4                & 6.37\%               \\
					&                    & 10\%     &                         & 5.4  & 36.5  & 59.5  & 0.00\% & 309.9  & 313.5  & 0.45\% &                      &                      &                      \\ \cmidrule(l){2-14} 
					& \multirow{3}{*}{1} & 70\%     & \multirow{3}{*}{521.8}  & 18.1 & 282.1 & 340.6 & 0.07\% & 383.1  & 385.7  & 0.80\% & 413.3                & 512.6                & NA                   \\
					&                    & 40\%     &                         & 20.2 & 53.7  & 120.9 & 0.00\% & 291.2  & 293.4  & 0.64\% & \multicolumn{1}{l}{} & \multicolumn{1}{l}{} & \multicolumn{1}{l}{} \\
					&                    & 10\%     &                         & 15.4 & 40.3  & 97.5  & 0.00\% & 268.0  & 270.6  & 0.64\% & \multicolumn{1}{l}{} & \multicolumn{1}{l}{} & \multicolumn{1}{l}{} \\ \bottomrule
				\end{tabular}
				\begin{tablenotes}
					\scriptsize
					\item[1] Type: instance type. Net: network construction time. Gap: optimality gap.
					\item[2] ``NA'' denotes that no feasible solution is obtained within the time limit for a given instance. For the calculation of averages: (i) instances for which all formulations report ``NA'' are excluded; (ii) a formulation's average is recorded as ``NA'' if it uniquely reports ``NA'' for at least one instance.
				\end{tablenotes}
			\end{threeparttable}
			\label{table:avgDE}
		\end{table}

		Table~\ref{table:avgDE} summarizes the average computational performance of three formulations on type-a and type-u instances of D-E-ADARP under different time discretization granularities, assuming $N_{\max}=\infty$ (complete results are provided in Appendix~B).
		There are three main findings.

		First, BTSFF rarely outperforms TSFFCS and IEBF, except in settings where the number of reachable SoC levels is small. To avoid excessive infeasible tests, we therefore focus on settings where SoC discretization is most likely to be advantageous, namely under relatively frequent recharging and coarse discretization. Even under these favorable conditions, BTSFF generally performs worse than the other formulations. Specifically, BTSFF shows an advantage only for type-u instances with a 2-minute time unit and $\gamma=70\%$, whereas TSFFCS and IEBF outperform BTSFF in most other cases, such as when the battery size is larger (type-a), time discretization is finer, or recharging is less frequent. Although BTSFF's compact structure may benefit frequent recharging, explicit SoC discretization enlarges the state space and increases model size, so that this advantage is realized only when the number of reachable SoC levels is limited. To further examine the role of limited SoC levels, we test type-a instances with 25\% battery capacity, 2-minute discretization, $\gamma=70\%$, and $N_{\max}=\infty$. As shown in Appendix~C, BTSFF outperforms both IEBF and TSFFCS in this setting. Overall, BTSFF becomes competitive primarily when both the reachable SoC state space is small and recharging is frequent.
		
		Second, time discretization provides computational gains mainly under relatively coarse granularities and frequent recharging conditions. TSFFCS outperforms IEBF on type-a instances with 5-minute granularity and a high recharging probability ($\gamma = 70\%$), whereas IEBF is generally superior under finer granularities or lower recharging frequencies ($\gamma = 40\%$ or $10\%$). This behavior may be explained by the fact that the discrete-time representation remains compact enough to exploit the fragment-based structure while avoiding the state-space expansion caused by explicit SoC discretization. 
		TSFFCS also shows a stronger advantage on real-world-derived type-u instances than on type-a instances. For instances with the same number of customers, type-u instances have a shorter operating horizon (around 3.5 hours in type-u versus 10 hours in type-a), implying a higher customer intensity. As shown in the DARP study \citet{Zhao2026DialARide}, the performance of event-based formulations deteriorates more rapidly under such conditions than that of fragment-based formulations.

		Overall, the preferred formulation is regime-dependent. TSFFCS performs best under frequent recharging and relatively coarse time discretization, whereas IEBF is generally preferable otherwise. BTSFF becomes competitive only when the number of reachable SoC levels is small. These patterns can be understood as a trade-off between state-space expansion and structural compactness, where different discretization choices shift the balance between these two effects.
		This state-space size is reflected in the extent of SoC, time, and spatial expansion, with the spatial dimension corresponding to the number of events and fragments (see Appendix~D).
		Accordingly, time discretization is beneficial across a wider range of practical regimes, whereas explicit SoC discretization is advantageous only in relatively narrow regimes. Within the discretized instances considered here, no loss of solution quality arises because all formulations are solved on the same rounded data and are not intended to approximate the original continuous problem.

		Several secondary observations are worth noting. First, the runtime of IEBF tends to decrease as time discretization becomes finer. This behavior is likely related to the event-based structure, where coarser parameter rounding can generate more equivalent-cost solutions, thereby increasing branching effort. Second, somewhat counterintuitively, the runtime of TSFFCS does not increase monotonically as time discretization becomes finer. For example, reducing the time step from 5 to 2 minutes for type-a instances, or from 2 to 1 minute for type-u instances, can yield comparable or even faster solution times. This effect is mainly due to the instance generation procedure and the rounding of time-related parameters. In particular, coarse rounding may relax time-window and maximum ride-time constraints, admitting fragments that would be infeasible under finer or continuous parameters. This mechanism is reflected in the fragment counts, which decrease from 5-minute to 2-minute to 1-minute granularity in type-a instances. However, coarse rounding does not always increase the number of fragments and may even reduce it. For instance, in type-u instances the fragment count is higher under 1-minute than under 2-minute granularity because coarse rounding can tighten ride-time constraints (e.g., an 8.5-minute ride time rounds to 8 minutes at 2-minute granularity but to 9 minutes at 1-minute granularity).

		\section{Conclusions}	\label{Conclusions}
		
		This paper studies how discretization reshapes formulation effectiveness for the electric autonomous dial-a-ride problem. We first introduce IEBF, a strengthened event-based formulation for the original continuous-parameter E-ADARP, which improves the existing event-based formulation by using arc-flow SoC variables and unified node-station-node charging arcs. We then develop two fragment-based formulations for the discrete-parameter problem. TSFFCS discretizes time while maintaining continuous SoC variables, whereas BTSFF discretizes both time and SoC. Together, these three formulations provide a controlled framework for isolating the computational effects of time discretization and SoC discretization.
		Computational results show that IEBF outperforms the existing event-based formulation for the continuous-parameter E-ADARP, especially in settings with frequent recharging or higher allowable visits to charging stations. Under discretized settings, the comparison among formulations provides guidance on formulation selection and clarifies the computational impact of time and SoC discretization. TSFFCS tends to outperform IEBF, particularly when recharging is frequent and the time discretization is relatively coarse, indicating that time discretization can improve performance across a wide range of settings.
		In contrast, BTSFF rarely outperforms TSFFCS unless the number of reachable SoC levels is limited, suggesting that explicit SoC discretization is beneficial only in relatively restricted settings. Outside these settings, IEBF remains a robust formulation choice.


		\bibliographystyle{model5names}
		\bibliography{library2}
		\clearpage

			\appendix
		
		\counterwithin{table}{section}
		\counterwithin{figure}{section}
		\renewcommand{\thetable}{\thesection.\arabic{table}}
		\renewcommand{\thefigure}{\thesection.\arabic{figure}}

		\section{Computational results of E-ADARP with $N_{\max}=1$ and $N_{\max}=\infty$ }\label{origEADARP}

		Tables~\ref{table:detailEADARPa1}--\ref{table:detailEADARPuinfty} present the computational performance of IEBF (our improved event-based formulation) and REBF (our replication of the existing event-based formulation of \citet{Stallhofer2025}) as well as the reported results of \citet{Stallhofer2025} and \citet{Su2024} on E-ADARP with $N_{\max}=1$ and $N_{\max}=\infty$  for type-a and type-u instances, respectively. Since \citet{Stallhofer2025} does not test on type-u instances and for scenarios with $N_{\max}=\infty$, we leave these parts empty in these tables. We also add the computational results of TSFFCS in the type-a instances with $N_{\max}=\infty$; because TSFFCS uses rounded-down travel time and discretized recharging time, these solutions should be interpreted as approximate solutions instead of valid lower or upper bounds. They mainly serve to indicate the computational effort of TSFFCS and to illustrate the difference in fragment generation between continuous-parameter E-ADARP and discretized D-E-ADARP. 
		The time limit for IEBF, REBF, and TSFFCS is set to 30 minutes, while the results of \citet{Stallhofer2025} and \citet{Su2024} are taken directly from their studies. The results show that REBF generally outperforms the results of the event-based method reported by \citet{Stallhofer2025}, thereby underscoring the effectiveness of our replicated event-based method (REBF). Additionally, REBF outperforms the B\&P results of \citet{Su2024} in most cases except for the type-u instances with $N_{\max}=\infty$ and $\gamma\in \{40\%, 10\%\}$. The discussion on IEBF and TSFFCS is shown in the manuscript.

		\begin{sidewaystable}[htbp]
			\caption{Computational performance of different procedures on E-ADARP for type-a instances with $N_{\max}=1$}
			\scriptsize  
			\centering
			\setlength{\tabcolsep}{1.0 pt}
			\renewcommand{\arraystretch}{0.5}  
			\begin{threeparttable}
				\begin{tabular}{@{}cccccccccccccccccccccc@{}}
					\toprule
					&       & \multicolumn{6}{c}{IEBF}                            & \multicolumn{6}{c}{REBF}                            & \multicolumn{4}{c}{Results of   \citeauthor{Stallhofer2025}} & \multicolumn{4}{c}{Results of   \citeauthor{Su2024}} \\ \midrule
					$\gamma$              & Name  & E-time & CPU    & Time   & OBJ    & LB     & Gap    & E-time & CPU    & Time   & OBJ    & LB     & Gap    & Time              & OBJ               & LB                & Gap               & Time            & OBJ             & LB              & Gap             \\
					\multirow{14}{*}{70\%} & a2-16 & 0.3    & 2.6    & 3.4    & 240.66 & 240.66 & 0.00\% & 0.1    & 1.4    & 1.5    & 240.66 & 240.66 & 0.00\% & 1.7               & 240.66            & 240.66            & 0.00\%            & 20.8            & 240.66          & 240.66          & 0.00\%          \\
					& a2-20 & 0.5    & 29.8   & 30.4   & 293.27 & 293.27 & 0.00\% & 0.1    & 128.9  & 129.1  & 293.27 & 293.27 & 0.00\% & 445.7             & 293.27            & 293.24            & 0.00\%            & 1758.6          & 293.27          & 293.27          & 0.00\%          \\
					& a2-24 & 0.4    & 3.8    & 4.3    & 353.18 & 353.18 & 0.00\% & 0.2    & 66.2   & 66.4   & 353.18 & 353.18 & 0.00\% & 213.5             & 353.18            & 353.15            & 0.00\%            & 1241.9          & 353.18          & 353.18          & 0.00\%          \\
					& a3-18 & 0.2    & 4.5    & 4.8    & 240.58 & 240.58 & 0.00\% & 0.1    & 9.9    & 10.0   & 240.58 & 240.58 & 0.00\% & 6.3               & 240.58            & 240.56            & 0.00\%            & 14.1            & 240.58          & 240.58          & 0.00\%          \\
					& a3-24 & 0.7    & 4.1    & 4.9    & 275.97 & 275.97 & 0.00\% & 0.3    & 9.9    & 10.3   & 275.97 & 275.97 & 0.00\% & 18.1              & 275.97            & 275.94            & 0.00\%            & 100.8           & 275.97          & 275.97          & 0.00\%          \\
					& a3-30 & 0.7    & 115.6  & 116.4  & 424.93 & 424.93 & 0.00\% & 0.6    & 251.4  & 252.1  & 424.93 & 424.93 & 0.00\% & 6822.9            & 424.93            & 424.89            & 0.00\%            & 1893.3          & 424.93          & 424.93          & 0.00\%          \\
					& a3-36 & 0.8    & 33.0   & 33.9   & 494.04 & 494.04 & 0.00\% & 1.0    & 1380.0 & 1381.1 & 494.04 & 494.04 & 0.00\% & 7200.0            & 494.04            & 487.58            & 1.30\%            & 7200.0          & 494.04          & 494.01          & 0.00\%          \\
					& a4-16 & 0.2    & 1.7    & 1.9    & 223.13 & 223.13 & 0.00\% & 0.2    & 1.7    & 1.9    & 223.13 & 223.13 & 0.00\% & 2.8               & 223.13            & 223.13            & 0.00\%            & 8.4             & 223.13          & 223.13          & 0.00\%          \\
					& a4-24 & 0.7    & 17.3   & 18.1   & 316.65 & 316.65 & 0.00\% & 0.5    & 27.2   & 27.8   & 316.65 & 316.65 & 0.00\% & 182.5             & 316.65            & 316.62            & 0.00\%            & 164.4           & 316.65          & 316.65          & 0.00\%          \\
					& a4-32 & 0.8    & 63.8   & 64.8   & 397.87 & 397.87 & 0.00\% & 1.0    & 132.7  & 133.8  & 397.87 & 397.86 & 0.00\% & 789.2             & 397.87            & 397.83            & 0.00\%            & 1638.6          & 397.87          & 397.87          & 0.00\%          \\
					& a4-40 & 1.8    & 1798.0 & 1800.0 & 467.72 & 461.87 & 1.25\% & 2.0    & 1797.7 & 1800.0 & 467.72 & 462.14 & 1.19\% & 7200.0            & NA                & 456.08            & NA                & 3987.4          & 467.72          & 467.72          & 0.00\%          \\
					& a4-48 & 3.2    & 1795.9 & 1800.0 & NA     & 557.59 & NA     & 4.9    & 1794.6 & 1800.0 & NA     & 556.62 & NA     & 7200.0            & NA                & 555.15            & NA                & 7200.0          & NA              & 476.54          & NA              \\
					& a5-40 & 3.1    & 147.7  & 151.1  & 418.75 & 418.75 & 0.00\% & 3.5    & 115.6  & 119.5  & 418.75 & 418.75 & 0.00\% & 4519.4            & 418.75            & 418.71            & 0.00\%            & 1572.0          & 418.75          & 418.75          & 0.00\%          \\
					& a5-50 & 10.6   & 1789.2 & 1800.0 & NA     & 557.50 & NA     & 24.4   & 1774.8 & 1800.0 & NA     & 560.72 & NA     & 7200.0            & NA                & 559.50            & NA                & 7200.0          & 690.79          & 327.84          & NA              \\
					& Avg   &        & 591.3  & 592.8  &        &        & 0.19\% &        & 658.1  & 661.2  &        &        & 0.18\% & 1836.6            &                   &                   & NA                & 1419.4          &                 &                 & 0.00\%          \\
					\multirow{14}{*}{40\%} & a2-16 & 0.1    & 0.3    & 0.4    & 237.38 & 237.38 & 0.00\% & 0.1    & 0.6    & 0.7    & 237.38 & 237.38 & 0.00\% & 0.4               & 237.38            & 237.38            & 0.00\%            & 12.7            & 237.38          & 237.38          & 0.00\%          \\
					& a2-20 & 0.2    & 1.3    & 1.6    & 280.70 & 280.70 & 0.00\% & 0.1    & 7.2    & 7.3    & 280.70 & 280.70 & 0.00\% & 2.6               & 280.70            & 280.70            & 0.00\%            & 93.8            & 280.70          & 280.70          & 0.00\%          \\
					& a2-24 & 0.6    & 1.9    & 2.5    & 347.04 & 347.03 & 0.00\% & 0.2    & 14.6   & 14.8   & 347.04 & 347.04 & 0.00\% & 23.0              & 347.04            & 347.04            & 0.00\%            & 267.8           & 347.04          & 347.04          & 0.00\%          \\
					& a3-18 & 0.3    & 1.0    & 1.3    & 236.81 & 236.81 & 0.00\% & 0.2    & 1.8    & 2.0    & 236.81 & 236.81 & 0.00\% & 1.5               & 236.82            & 236.80            & 0.00\%            & 5.3             & 236.82          & 236.82          & 0.00\%          \\
					& a3-24 & 0.6    & 1.1    & 1.8    & 274.80 & 274.80 & 0.00\% & 0.4    & 12.8   & 13.5   & 274.80 & 274.80 & 0.00\% & 2.8               & 274.80            & 274.80            & 0.00\%            & 69.7            & 274.80          & 274.80          & 0.00\%          \\
					& a3-30 & 0.9    & 2.6    & 3.6    & 413.34 & 413.34 & 0.00\% & 0.7    & 16.2   & 17.0   & 413.34 & 413.34 & 0.00\% & 56.0              & 413.34            & 413.34            & 0.00\%            & 306.9           & 413.34          & 413.34          & 0.00\%          \\
					& a3-36 & 0.9    & 15.6   & 16.7   & 482.75 & 482.75 & 0.00\% & 0.9    & 46.9   & 47.9   & 482.75 & 482.75 & 0.00\% & 376.2             & 482.75            & 482.70            & 0.00\%            & 5154.1          & 482.75          & 482.75          & 0.00\%          \\
					& a4-16 & 0.4    & 1.1    & 1.5    & 222.49 & 222.49 & 0.00\% & 0.2    & 0.9    & 1.1    & 222.49 & 222.49 & 0.00\% & 1.3               & 222.49            & 222.49            & 0.00\%            & 3.3             & 222.49          & 222.49          & 0.00\%          \\
					& a4-24 & 0.8    & 2.3    & 3.1    & 311.03 & 311.03 & 0.00\% & 0.6    & 5.5    & 6.2    & 311.03 & 311.03 & 0.00\% & 7.1               & 311.03            & 311.01            & 0.00\%            & 18.4            & 311.03          & 311.03          & 0.00\%          \\
					& a4-32 & 0.9    & 7.7    & 8.7    & 394.26 & 394.23 & 0.01\% & 1.1    & 18.2   & 19.5   & 394.26 & 394.26 & 0.00\% & 31.9              & 394.26            & 394.22            & 0.00\%            & 199.3           & 394.26          & 394.26          & 0.00\%          \\
					& a4-40 & 1.6    & 8.9    & 10.7   & 453.84 & 453.84 & 0.00\% & 2.6    & 26.3   & 29.2   & 453.84 & 453.84 & 0.00\% & 88.5              & 453.84            & 453.80            & 0.00\%            & 792.0           & 453.84          & 453.84          & 0.00\%          \\
					& a4-48 & 5.0    & 22.1   & 27.7   & 554.60 & 554.54 & 0.00\% & 8.6    & 56.1   & 65.4   & 554.60 & 554.60 & 0.00\% & 253.7             & 554.60            & 554.57            & 0.00\%            & 2292.0          & 554.60          & 554.60          & 0.00\%          \\
					& a5-40 & 1.5    & 11.0   & 12.8   & 414.50 & 414.50 & 0.00\% & 3.7    & 17.5   & 21.5   & 414.50 & 414.50 & 0.00\% & 65.9              & 414.51            & 414.47            & 0.00\%            & 323.6           & 414.51          & 414.51          & 0.00\%          \\
					& a5-50 & 4.3    & 140.7  & 145.7  & 559.51 & 559.48 & 0.00\% & 11.0   & 81.5   & 93.2   & 559.51 & 559.51 & 0.00\% & 321.1             & 559.51            & 559.45            & 0.00\%            & 1957.1          & 559.51          & 559.51          & 0.00\%          \\
					& Avg   &        & 18.8   & 19.9   &        &        & 0.00\% &        & 17.6   & 19.8   &        &        & 0.00\% & 88.0              &                   &                   & 0.00\%            & 821.1           &                 &                 & 0.00\%          \\
					\multirow{14}{*}{10\%} & a2-16 & 0.1    & 0.2    & 0.4    & 237.38 & 237.38 & 0.00\% & 0.1    & 0.9    & 0.9    & 237.38 & 237.38 & 0.00\% & 0.4               & 237.38            & 237.38            & 0.00\%            & 11.1            & 237.38          & 237.38          & 0.00\%          \\
					& a2-20 & 0.3    & 0.6    & 0.9    & 279.08 & 279.08 & 0.00\% & 0.3    & 6.3    & 6.6    & 279.08 & 279.08 & 0.00\% & 2.0               & 279.08            & 279.08            & 0.00\%            & 70.9            & 279.08          & 279.08          & 0.00\%          \\
					& a2-24 & 0.5    & 0.8    & 1.3    & 346.21 & 346.21 & 0.00\% & 0.2    & 7.7    & 7.9    & 346.22 & 346.21 & 0.00\% & 17.2              & 346.21            & 346.21            & 0.00\%            & 243.3           & 346.21          & 346.21          & 0.00\%          \\
					& a3-18 & 0.2    & 0.6    & 0.9    & 236.81 & 236.81 & 0.00\% & 0.1    & 1.6    & 1.9    & 236.81 & 236.81 & 0.00\% & 0.9               & 236.82            & 236.81            & 0.00\%            & 5.0             & 236.82          & 236.82          & 0.00\%          \\
					& a3-24 & 0.5    & 0.8    & 1.4    & 274.80 & 274.79 & 0.00\% & 0.3    & 3.8    & 4.2    & 274.80 & 274.80 & 0.00\% & 2.4               & 274.80            & 274.80            & 0.00\%            & 81.2            & 274.80          & 274.80          & 0.00\%          \\
					& a3-30 & 0.6    & 2.0    & 2.7    & 413.27 & 413.27 & 0.00\% & 0.5    & 10.4   & 10.9   & 413.27 & 413.27 & 0.00\% & 72.3              & 413.27            & 413.23            & 0.00\%            & 221.7           & 413.27          & 413.27          & 0.00\%          \\
					& a3-36 & 0.9    & 3.1    & 4.1    & 481.17 & 481.17 & 0.00\% & 1.1    & 28.6   & 29.8   & 481.17 & 481.17 & 0.00\% & 101.8             & 481.17            & 481.13            & 0.00\%            & 730.8           & 481.17          & 481.17          & 0.00\%          \\
					& a4-16 & 0.3    & 0.7    & 1.0    & 222.49 & 222.49 & 0.00\% & 0.2    & 0.9    & 1.1    & 222.49 & 222.49 & 0.00\% & 0.8               & 222.49            & 222.49            & 0.00\%            & 5.5             & 222.49          & 222.49          & 0.00\%          \\
					& a4-24 & 0.6    & 1.6    & 2.2    & 310.84 & 310.84 & 0.00\% & 0.4    & 2.8    & 3.2    & 310.84 & 310.84 & 0.00\% & 4.5               & 310.84            & 310.81            & 0.00\%            & 25.0            & 310.84          & 310.84          & 0.00\%          \\
					& a4-32 & 1.4    & 4.9    & 6.5    & 393.95 & 393.94 & 0.00\% & 0.9    & 25.7   & 26.7   & 393.95 & 393.95 & 0.00\% & 30.4              & 393.96            & 393.92            & 0.00\%            & 143.2           & 393.96          & 393.96          & 0.00\%          \\
					& a4-40 & 2.0    & 8.6    & 11.0   & 453.84 & 453.84 & 0.00\% & 1.9    & 32.9   & 35.1   & 453.84 & 453.84 & 0.00\% & 93.0              & 453.84            & 453.80            & 0.00\%            & 653.0           & 453.84          & 453.84          & 0.00\%          \\
					& a4-48 & 2.1    & 13.3   & 16.2   & 554.54 & 554.54 & 0.00\% & 5.3    & 63.4   & 69.3   & 554.54 & 554.54 & 0.00\% & 178.2             & 554.54            & 554.49            & 0.00\%            & 1334.4          & 554.54          & 554.54          & 0.00\%          \\
					& a5-40 & 2.1    & 17.4   & 19.8   & 414.50 & 414.50 & 0.00\% & 2.4    & 37.9   & 40.6   & 414.50 & 414.50 & 0.00\% & 91.0              & 414.51            & 414.47            & 0.00\%            & 448.3           & 414.51          & 414.51          & 0.00\%          \\
					& a5-50 & 2.7    & 47.1   & 50.5   & 559.17 & 559.17 & 0.00\% & 11.6   & 35.9   & 48.2   & 559.17 & 559.17 & 0.00\% & 170.1             & 559.17            & 559.11            & 0.00\%            & 496.2           & 559.17          & 559.17          & 0.00\%          \\
					& Avg   &        & 8.2    & 9.1    &        &        & 0.00\% &        & 16.2   & 18.1   &        &        & 0.00\% & 54.6              &                   &                   & 0.00\%            & 319.3           &                 &                 & 0.00\%          \\ \bottomrule
				\end{tabular}
				\end{threeparttable}
				\label{table:detailEADARPa1}
			\end{sidewaystable}

			\begin{sidewaystable}[htbp]
				\caption{Computational performance of different procedures on E-ADARP for type-u instances with $N_{\max}=1$}
				\scriptsize  
				\centering
				\setlength{\tabcolsep}{1.0 pt}
				\renewcommand{\arraystretch}{0.5}  
				\begin{threeparttable}
					\begin{tabular}{@{}cccccccccccccccccc@{}}
						\toprule
						&       & \multicolumn{6}{c}{IEBF}                              & \multicolumn{6}{c}{REBF}                            & \multicolumn{4}{c}{Results of   \citeauthor{Su2024}} \\ \midrule
						$\gamma$              & Name  & E-time & CPU    & Time   & OBJ    & LB     & Gap      & E-time & CPU    & Time   & OBJ    & LB     & Gap    & Time            & OBJ             & LB              & Gap             \\
						\multirow{14}{*}{70\%} & u2-16 & 0.5    & 19.9   & 20.4   & 59.19  & 59.19  & 0.00\%   & 1.0    & 25.8   & 26.8   & 59.19  & 59.19  & 0.00\% & 679.8           & 59.19           & 59.19           & 0.00\%          \\
						& u2-20 & 1.0    & 12.6   & 13.7   & 56.86  & 56.86  & 0.00\%   & 0.6    & 27.5   & 28.2   & 56.86  & 56.86  & 0.00\% & 337.0           & 56.86           & 56.86           & 0.00\%          \\
						& u2-24 & 1.1    & 19.8   & 21.0   & 92.01  & 92.01  & 0.00\%   & 0.4    & 189.0  & 189.4  & 92.01  & 92.01  & 0.00\% & 7200.0          & 92.01           & 91.60           & 0.52\%          \\
						& u3-18 & 0.7    & 15.8   & 16.5   & 50.99  & 50.99  & 0.00\%   & 0.3    & 3.4    & 3.7    & 50.99  & 50.99  & 0.00\% & 45.2            & 50.99           & 50.99           & 0.00\%          \\
						& u3-24 & 6.3    & 35.6   & 42.4   & 68.34  & 68.34  & 0.00\%   & 0.8    & 37.4   & 38.3   & 68.34  & 68.34  & 0.00\% & 541.9           & 68.34           & 68.34           & 0.00\%          \\
						& u3-30 & 2.1    & 50.9   & 53.3   & 77.41  & 77.41  & 0.00\%   & 2.0    & 53.1   & 55.2   & 77.41  & 77.41  & 0.00\% & 1725.5          & 77.41           & 77.41           & 0.00\%          \\
						& u3-36 & 3.3    & 390.4  & 394.0  & 105.79 & 105.79 & 0.00\%   & 3.1    & 648.3  & 651.6  & 105.79 & 105.78 & 0.00\% & 7200.0          & 105.79          & 105.78          & 0.01\%          \\
						& u4-16 & 0.6    & 26.6   & 27.2   & 53.87  & 53.87  & 0.00\%   & 0.3    & 4.3    & 4.6    & 53.87  & 53.87  & 0.00\% & 4.0             & 53.87           & 53.87           & 0.00\%          \\
						& u4-24 & 1.6    & 5.5    & 7.2    & 89.96  & 89.96  & 0.00\%   & 0.5    & 2.4    & 3.0    & 89.96  & 89.96  & 0.00\% & 82.3            & 89.96           & 89.96           & 0.00\%          \\
						& u4-32 & 2.5    & 16.7   & 19.5   & 99.50  & 99.50  & 0.00\%   & 2.3    & 46.7   & 49.5   & 99.50  & 99.50  & 0.00\% & 1028.6          & 99.50           & 99.50           & 0.00\%          \\
						& u4-40 & 4.0    & 211.1  & 215.4  & 134.81 & 134.80 & 0.00\%   & 1.5    & 746.6  & 748.2  & 134.81 & 134.81 & 0.00\% & 7200.0          & 135.76          & 134.96          & 0.59\%          \\
						& u4-48 & 6.4    & 1792.7 & 1800.1 & 149.33 & 142.86 & 4.34\%   & 28.7   & 1770.5 & 1800.0 & NA     & 144.12 & NA     & 7200.0          & 185.16          & NA              & NA              \\
						& u5-40 & 4.2    & 345.4  & 350.1  & 122.93 & 122.93 & 0.00\%   & 5.5    & 761.0  & 766.9  & 122.93 & 122.93 & 0.00\% & 6513.9          & 122.93          & 122.93          & 0.00\%          \\
						& u5-50 & 8.5    & 1703.8 & 1713.4 & 143.35 & 143.35 & 0.00\%   & 17.1   & 1489.5 & 1507.7 & 143.35 & 143.35 & 0.00\% & 7200.0          & 195.72          & 132.79          & NA              \\
						& Avg   &        & 415.0  & 416.6  &        &        & 0.58\%   &        & 477.5  & 482.4  &        &        & 0.00\% & 3198.4          &                 &                 & NA              \\
						\multirow{14}{*}{40\%} & u2-16 & 0.5    & 4.5    & 5.0    & 57.65  & 57.65  & 0.00\%   & 0.2    & 4.0    & 4.2    & 57.65  & 57.65  & 0.00\% & 53.2            & 57.65           & 57.65           & 0.00\%          \\
						& u2-20 & 0.9    & 6.5    & 7.6    & 56.34  & 56.34  & 0.00\%   & 0.9    & 8.6    & 9.6    & 56.34  & 56.34  & 0.00\% & 281.3           & 56.34           & 56.34           & 0.00\%          \\
						& u2-24 & 1.0    & 11.4   & 12.5   & 91.06  & 91.06  & 0.00\%   & 0.4    & 24.0   & 24.4   & 91.06  & 91.06  & 0.00\% & 6917.0          & 91.06           & 91.06           & 0.00\%          \\
						& u3-18 & 0.6    & 4.7    & 5.3    & 50.74  & 50.74  & 0.00\%   & 0.3    & 2.0    & 2.3    & 50.74  & 50.74  & 0.00\% & 15.0            & 50.74           & 50.74           & 0.00\%          \\
						& u3-24 & 1.5    & 12.1   & 13.8   & 67.56  & 67.56  & 0.00\%   & 0.7    & 17.8   & 18.6   & 67.56  & 67.56  & 0.00\% & 58.5            & 67.56           & 67.56           & 0.00\%          \\
						& u3-30 & 1.6    & 28.8   & 30.5   & 76.75  & 76.75  & 0.00\%   & 1.5    & 38.9   & 40.7   & 76.75  & 76.75  & 0.00\% & 500.2           & 76.75           & 76.75           & 0.00\%          \\
						& u3-36 & 2.8    & 36.2   & 39.2   & 104.06 & 104.06 & 0.00\%   & 2.0    & 42.5   & 44.7   & 104.06 & 104.06 & 0.00\% & 1769.0          & 104.06          & 104.06          & 0.00\%          \\
						& u4-16 & 0.6    & 5.6    & 6.2    & 53.58  & 53.58  & 0.00\%   & 0.3    & 2.3    & 2.6    & 53.58  & 53.58  & 0.00\% & 2.2             & 53.58           & 53.58           & 0.00\%          \\
						& u4-24 & 1.3    & 2.4    & 3.7    & 89.83  & 89.83  & 0.00\%   & 0.3    & 2.0    & 2.3    & 89.83  & 89.83  & 0.00\% & 30.8            & 89.83           & 89.83           & 0.00\%          \\
						& u4-32 & 2.3    & 11.1   & 13.7   & 99.29  & 99.29  & 0.00\%   & 2.2    & 17.3   & 19.7   & 99.29  & 99.29  & 0.00\% & 393.4           & 99.29           & 99.29           & 0.00\%          \\
						& u4-40 & 2.5    & 26.2   & 28.9   & 133.78 & 133.78 & 0.00\%   & 1.7    & 47.5   & 49.3   & 133.78 & 133.78 & 0.00\% & 7200.0          & 133.78          & 133.70          & 0.06\%          \\
						& u4-48 & 5.1    & 1794.2 & 1800.0 & 148.63 & 145.15 & 2.34\%   & 8.3    & 1491.6 & 1500.7 & 147.40 & 147.39 & 0.00\% & 7200.0          & 147.63          & 146.37          & 0.85\%          \\
						& u5-40 & 4.5    & 60.0   & 65.0   & 121.96 & 121.96 & 0.00\%   & 3.9    & 66.3   & 70.6   & 121.96 & 121.96 & 0.00\% & 723.7           & 121.96          & 121.96          & 0.00\%          \\
						& u5-50 & 7.1    & 344.6  & 352.7  & 142.83 & 142.83 & 0.00\%   & 10.3   & 131.0  & 142.2  & 142.83 & 142.83 & 0.00\% & 7200.0          & 142.84          & 142.75          & 0.06\%          \\
						& Avg   &        & 257.6  & 258.9  &        &        & 0.31\%   &        & 193.7  & 195.3  &        &        & 0.00\% & 2753.7          &                 &                 & 0.11\%          \\
						\multirow{14}{*}{10\%} & u2-16 & 0.5    & 2.2    & 2.8    & 57.61  & 57.61  & 0.00\%   & 0.2    & 1.6    & 1.8    & 57.61  & 57.61  & 0.00\% & 22.2            & 57.61           & 57.61           & 0.00\%          \\
						& u2-20 & 0.9    & 9.0    & 10.0   & 55.59  & 55.59  & 0.00\%   & 0.7    & 7.3    & 8.2    & 55.59  & 55.59  & 0.00\% & 69.3            & 55.59           & 55.59           & 0.00\%          \\
						& u2-24 & 1.0    & 10.2   & 11.2   & 90.73  & 90.73  & 0.00\%   & 0.4    & 18.7   & 19.3   & 90.73  & 90.73  & 0.00\% & 1649.0          & 90.66           & 90.66           & 0.00\%          \\
						& u3-18 & 0.6    & 2.9    & 3.5    & 50.74  & 50.74  & 0.00\%   & 0.3    & 3.7    & 4.0    & 50.74  & 50.74  & 0.00\% & 14.0            & 50.74           & 50.74           & 0.00\%          \\
						& u3-24 & 1.1    & 16.9   & 18.1   & 67.56  & 67.56  & 0.00\%   & 1.0    & 21.3   & 22.4   & 67.56  & 67.56  & 0.00\% & 40.3            & 67.56           & 67.56           & 0.00\%          \\
						& u3-30 & 1.6    & 23.5   & 25.3   & 76.75  & 76.75  & 0.00\%   & 1.2    & 32.3   & 33.6   & 76.75  & 76.75  & 0.00\% & 570.8           & 76.75           & 76.75           & 0.00\%          \\
						& u3-36 & 2.3    & 25.8   & 28.4   & 104.04 & 104.04 & 8.01E-12 & 2.5    & 49.3   & 52.3   & 104.04 & 104.04 & 0.00\% & 4204.6          & 104.04          & 104.04          & 0.00\%          \\
						& u4-16 & 0.7    & 3.4    & 4.2    & 53.58  & 53.58  & 0.00\%   & 0.3    & 2.2    & 2.5    & 53.58  & 53.58  & 0.00\% & 1.9             & 53.58           & 53.58           & 0.00\%          \\
						& u4-24 & 1.0    & 1.7    & 2.8    & 89.83  & 89.83  & 0.00\%   & 0.3    & 1.6    & 1.9    & 89.83  & 89.83  & 0.00\% & 25.5            & 89.83           & 89.83           & 0.00\%          \\
						& u4-32 & 1.9    & 7.7    & 9.9    & 99.29  & 99.29  & 0.00\%   & 1.8    & 18.1   & 20.2   & 99.29  & 99.29  & 0.00\% & 366.2           & 99.29           & 99.29           & 0.00\%          \\
						& u4-40 & 2.7    & 14.4   & 17.3   & 133.11 & 133.11 & 0.00\%   & 1.5    & 27.1   & 28.8   & 133.11 & 133.11 & 0.00\% & 1376.6          & 133.11          & 133.11          & 0.00\%          \\
						& u4-48 & 4.6    & 1794.8 & 1800.0 & 147.95 & 145.10 & 1.92\%   & 9.4    & 1029.7 & 1040.0 & 147.26 & 147.26 & 0.00\% & 7200.0          & 148.08          & 147.02          & 0.72\%          \\
						& u5-40 & 4.2    & 45.3   & 49.9   & 121.86 & 121.86 & 0.00\%   & 4.6    & 68.2   & 73.3   & 121.86 & 121.86 & 0.00\% & 496.9           & 121.86          & 121.86          & 0.00\%          \\
						& u5-50 & 6.6    & 152.4  & 159.9  & 142.83 & 142.83 & 0.00\%   & 11.0   & 144.7  & 156.7  & 142.83 & 142.83 & 0.00\% & 7200.0          & 142.82          & 142.77          & 0.04\%          \\
						& Avg   &        & 235.1  & 236.2  &        &        & 0.25\%   &        & 138.7  & 141.1  &        &        & 0.00\% & 1946.5          &                 &                 & 0.10\%          \\ \bottomrule
					\end{tabular}
					\end{threeparttable}
					\label{table:detailEADARPu1}
				\end{sidewaystable}

				\begin{sidewaystable}[htbp]
					\caption{Computational performance of different procedures on E-ADARP for type-a instances with $N_{\max}=\infty$}
					\scriptsize  
					\centering
					\setlength{\tabcolsep}{1.0 pt}
					\renewcommand{\arraystretch}{0.5}  
					\begin{threeparttable}
						\begin{tabular}{@{}cccccccccccccccccccccccccc@{}}
							\toprule
							&       & \multicolumn{8}{c}{TSFFCS (Approx)}                                         & \multicolumn{6}{c}{IEBF}                            & \multicolumn{6}{c}{REBF}                            & \multicolumn{4}{c}{Results of   \citeauthor{Su2024}} \\ \midrule
							$\gamma$              & Name  & $|F|$ & F-time & Net  & CPU    & Time   & OBJ    & LB     & Gap    & E-time & CPU    & Time   & OBJ    & LB     & Gap    & E-time & CPU    & Time   & OBJ    & LB     & Gap    & Time            & OBJ             & LB              & Gap             \\
							\multirow{14}{*}{70\%} & a2-16 & 32    & 0.1    & 1.9  & 4.5    & 6.4    & 240.66 & 240.66 & 0.00\% & 0.3    & 0.5    & 0.8    & 240.66 & 240.66 & 0.00\% & 0.3    & 270.4  & 270.7  & 240.66 & 240.65 & 0.00\% & 225.4           & 240.66          & 240.66          & 0.00\%          \\
							& a2-20 & 50    & 0.4    & 8.4  & 15.8   & 24.8   & 285.15 & 285.15 & 0.00\% & 0.2    & 2.5    & 2.8    & 285.86 & 285.86 & 0.00\% & 0.3    & 1799.7 & 1800.0 & 285.86 & 282.64 & 1.13\% & 469.6           & 285.86          & 285.86          & 0.00\%          \\
							& a2-24 & 64    & 0.2    & 2.5  & 2.6    & 5.3    & 346.44 & 346.44 & 0.00\% & 0.3    & 1.9    & 2.3    & 350.32 & 350.32 & 0.00\% & 0.4    & 1799.6 & 1800.0 & 350.32 & 346.88 & 0.98\% & 3513.6          & 350.32          & 350.32          & 0.00\%          \\
							& a3-18 & 65    & 0.2    & 1.2  & 1.3    & 2.6    & 238.35 & 238.35 & 0.00\% & 0.2    & 3.2    & 3.4    & 238.82 & 238.82 & 0.00\% & 1.7    & 46.5   & 48.3   & 238.82 & 238.82 & 0.00\% & 62.1            & 238.82          & 238.82          & 0.00\%          \\
							& a3-24 & 106   & 0.3    & 2.9  & 1.7    & 5.0    & 275.20 & 275.20 & 0.00\% & 1.3    & 1.8    & 3.2    & 275.20 & 275.20 & 0.00\% & 0.6    & 37.2   & 37.9   & 275.20 & 275.20 & 0.00\% & 244.7           & 275.20          & 275.20          & 0.00\%          \\
							& a3-30 & 89    & 0.2    & 3.7  & 2.7    & 6.6    & 413.45 & 413.42 & 0.00\% & 0.7    & 3.0    & 3.8    & 413.45 & 413.41 & 0.00\% & 0.9    & 188.5  & 189.5  & 413.45 & 413.41 & 0.00\% & 1556.4          & 413.35          & 413.35          & 0.00\%          \\
							& a3-36 & 116   & 0.3    & 15.1 & 10.0   & 25.4   & 481.99 & 481.99 & 0.00\% & 1.1    & 8.8    & 10.0   & 483.08 & 483.08 & 0.00\% & 1.5    & 1798.5 & 1800.0 & 483.08 & 482.79 & 0.06\% & 1254.4          & 483.08          & 483.08          & 0.00\%          \\
							& a4-16 & 82    & 0.2    & 0.8  & 0.4    & 1.4    & 221.48 & 221.48 & 0.00\% & 0.3    & 0.8    & 1.1    & 222.49 & 222.49 & 0.00\% & 0.5    & 8.5    & 9.0    & 222.49 & 222.47 & 0.00\% & 11.0            & 222.49          & 222.49          & 0.00\%          \\
							& a4-24 & 95    & 0.3    & 2.5  & 9.0    & 11.8   & 315.25 & 315.25 & 0.00\% & 0.8    & 6.4    & 7.3    & 315.40 & 315.40 & 0.00\% & 0.8    & 1799.1 & 1800.0 & 315.40 & 313.40 & 0.63\% & 193.6           & 315.40          & 315.40          & 0.00\%          \\
							& a4-32 & 239   & 0.8    & 3.4  & 4.0    & 8.3    & 394.94 & 394.94 & 0.00\% & 0.7    & 28.3   & 29.2   & 394.94 & 394.94 & 0.00\% & 1.8    & 723.1  & 725.0  & 394.94 & 394.94 & 0.00\% & 427.7           & 394.94          & 394.94          & 0.00\%          \\
							& a4-40 & 239   & 1.0    & 9.8  & 16.8   & 27.7   & 456.08 & 456.08 & 0.00\% & 1.6    & 25.4   & 27.3   & 457.76 & 457.76 & 0.00\% & 4.1    & 1795.5 & 1800.0 & 457.76 & 454.47 & 0.72\% & 1604.2          & 457.76          & 457.76          & 0.00\%          \\
							& a4-48 & 346   & 1.4    & 15.6 & 108.0  & 125.1  & 558.68 & 558.68 & 0.00\% & 2.1    & 46.7   & 49.3   & 557.06 & 557.06 & 0.00\% & 29.7   & 1766.5 & 1800.0 & 558.24 & 554.54 & 0.66\% & 7200.0          & 570.31          & 556.99          & 1.95\%          \\
							& a5-40 & 319   & 3.1    & 23.8 & 10.1   & 37.1   & 415.88 & 415.88 & 0.00\% & 1.4    & 33.4   & 35.2   & 415.88 & 415.88 & 0.00\% & 4.9    & 1794.8 & 1800.0 & 415.88 & 414.60 & 0.31\% & 1171.2          & 415.88          & 415.88          & 0.00\%          \\
							& a5-50 & 692   & 4.5    & 12.5 & 1783.6 & 1800.0 & 566.75 & 563.60 & 0.56\% & 3.1    & 1531.5 & 1535.5 & 567.61 & 567.61 & 0.00\% & 10.7   & 1788.5 & 1800.0 & 567.61 & 559.17 & 1.49\% & 7200.0          & 580.00          & 565.89          & 2.30\%          \\
							& Avg   & 132.9 & 0.9    & 5.8  & 234.7  & 235.8  &        &        & 0.07\% &        & 201.5  & 201.9  &        &        & 0.00\% &        & 774.1  & 777.1  &        &        & 0.44\% & 1789.7          &                 &                 & 0.52\%          \\
							\multirow{14}{*}{40\%} & a2-16 & 32    & 0.1    & 1.4  & 0.6    & 2.1    & 237.38 & 237.38 & 0.00\% & 0.1    & 0.3    & 0.4    & 237.38 & 237.38 & 0.00\% & 0.3    & 9.7    & 10.0   & 237.38 & 237.38 & 0.00\% & 34.6            & 237.38          & 237.38          & 0.00\%          \\
							& a2-20 & 50    & 0.2    & 2.6  & 2.3    & 5.1    & 280.53 & 280.53 & 0.00\% & 0.2    & 1.1    & 1.4    & 280.58 & 280.57 & 0.00\% & 0.3    & 42.0   & 42.3   & 280.58 & 280.58 & 0.00\% & 255.4           & 280.70          & 280.70          & 0.00\%          \\
							& a2-24 & 64    & 0.3    & 3.7  & 1.5    & 5.6    & 344.03 & 344.03 & 0.00\% & 0.4    & 1.3    & 1.8    & 346.28 & 346.28 & 0.00\% & 0.4    & 36.0   & 36.4   & 346.28 & 346.28 & 0.00\% & 964.6           & 346.28          & 346.28          & 0.00\%          \\
							& a3-18 & 65    & 0.3    & 1.5  & 0.6    & 2.4    & 236.33 & 236.33 & 0.00\% & 0.2    & 0.6    & 0.9    & 236.81 & 236.81 & 0.00\% & 0.3    & 11.4   & 11.7   & 236.81 & 236.81 & 0.00\% & 30.1            & 236.82          & 236.82          & 0.00\%          \\
							& a3-24 & 106   & 0.4    & 4.4  & 1.2    & 6.0    & 274.80 & 274.80 & 0.00\% & 0.4    & 1.3    & 1.7    & 274.80 & 274.80 & 0.00\% & 0.6    & 23.7   & 24.3   & 274.80 & 274.80 & 0.00\% & 86.7            & 274.80          & 274.80          & 0.00\%          \\
							& a3-30 & 89    & 0.2    & 3.8  & 1.7    & 5.8    & 413.27 & 413.27 & 0.00\% & 1.6    & 3.5    & 5.3    & 413.28 & 413.26 & 0.00\% & 0.9    & 55.3   & 56.3   & 413.28 & 413.27 & 0.00\% & 742.9           & 413.28          & 413.28          & 0.00\%          \\
							& a3-36 & 116   & 0.4    & 5.5  & 2.0    & 8.0    & 480.37 & 480.37 & 0.00\% & 2.5    & 5.3    & 8.0    & 481.17 & 481.17 & 0.00\% & 1.9    & 636.8  & 638.8  & 481.17 & 481.17 & 0.00\% & 1522.8          & 481.17          & 481.17          & 0.00\%          \\
							& a4-16 & 82    & 0.2    & 0.8  & 0.4    & 1.4    & 221.48 & 221.48 & 0.00\% & 0.4    & 0.9    & 1.4    & 222.49 & 222.49 & 0.00\% & 0.5    & 9.2    & 9.7    & 222.49 & 222.49 & 0.00\% & 9.6             & 222.49          & 222.49          & 0.00\%          \\
							& a4-24 & 95    & 0.3    & 3.7  & 6.7    & 10.7   & 311.03 & 311.03 & 0.00\% & 0.6    & 2.4    & 3.1    & 311.03 & 311.03 & 0.00\% & 1.1    & 19.3   & 20.5   & 311.03 & 311.03 & 0.00\% & 34.5            & 311.03          & 311.03          & 0.00\%          \\
							& a4-32 & 239   & 2.4    & 5.1  & 3.0    & 10.7   & 394.26 & 394.26 & 0.00\% & 1.0    & 6.7    & 7.8    & 394.26 & 394.26 & 0.00\% & 1.4    & 165.9  & 167.4  & 394.26 & 394.26 & 0.00\% & 205.0           & 394.26          & 394.26          & 0.00\%          \\
							& a4-40 & 239   & 1.3    & 7.4  & 2.7    & 11.5   & 452.16 & 452.16 & 0.00\% & 1.3    & 7.8    & 9.5    & 453.84 & 453.84 & 0.00\% & 4.3    & 149.1  & 153.7  & 453.84 & 453.84 & 0.00\% & 1027.1          & 453.84          & 453.84          & 0.00\%          \\
							& a4-48 & 346   & 1.5    & 10.1 & 5.4    & 17.0   & 554.54 & 554.54 & 0.00\% & 3.0    & 18.6   & 22.4   & 554.54 & 554.54 & 0.00\% & 8.1    & 142.5  & 151.4  & 554.54 & 554.54 & 0.00\% & 4630.9          & 554.54          & 554.54          & 0.00\%          \\
							& a5-40 & 319   & 1.4    & 13.0 & 8.6    & 23.0   & 414.50 & 414.50 & 0.00\% & 1.5    & 10.1   & 11.8   & 414.50 & 414.50 & 0.00\% & 3.3    & 66.8   & 70.4   & 414.50 & 414.50 & 0.00\% & 539.0           & 414.51          & 414.51          & 0.00\%          \\
							& a5-50 & 692   & 5.2    & 10.3 & 8.7    & 24.3   & 559.20 & 559.17 & 0.00\% & 2.5    & 67.0   & 70.3   & 559.48 & 559.48 & 0.00\% & 12.9   & 224.2  & 237.7  & 559.48 & 559.48 & 0.00\% & 2661.4          & 559.48          & 559.48          & 0.00\%          \\
							& Avg   & 132.9 & 1.0    & 2.9  & 2.3    & 5.7    &        &        & 0.00\% &        & 9.8    & 10.5   &        &        & 0.00\% &        & 107.1  & 109.5  &        &        & 0.00\% & 893.6           &                 &                 & 0.00\%          \\
							\multirow{14}{*}{10\%} & a2-16 & 32    & 0.1    & 1.0  & 1.0    & 2.1    & 237.38 & 237.38 & 0.00\% & 0.1    & 0.2    & 0.4    & 237.38 & 237.38 & 0.00\% & 0.2    & 7.4    & 7.6    & 237.38 & 237.38 & 0.00\% &                 &                 &                 &                 \\
							& a2-20 & 50    & 0.4    & 7.2  & 2.3    & 10.0   & 279.08 & 279.08 & 0.00\% & 0.2    & 0.5    & 0.7    & 279.08 & 279.08 & 0.00\% & 0.4    & 23.9   & 24.3   & 279.08 & 279.08 & 0.00\% &                 &                 &                 &                 \\
							& a2-24 & 64    & 0.6    & 10.1 & 2.8    & 13.7   & 343.95 & 343.95 & 0.00\% & 0.4    & 0.8    & 1.2    & 346.21 & 346.21 & 0.00\% & 0.4    & 37.0   & 37.5   & 346.21 & 346.21 & 0.00\% &                 &                 &                 &                 \\
							& a3-18 & 65    & 0.2    & 1.4  & 0.5    & 2.2    & 236.33 & 236.33 & 0.00\% & 0.2    & 0.7    & 0.9    & 236.81 & 236.81 & 0.00\% & 0.3    & 11.9   & 12.2   & 236.81 & 236.81 & 0.00\% &                 &                 &                 &                 \\
							& a3-24 & 106   & 0.4    & 3.1  & 1.0    & 4.5    & 274.80 & 274.80 & 0.00\% & 0.5    & 0.9    & 1.4    & 274.80 & 274.80 & 0.00\% & 0.7    & 28.3   & 29.0   & 274.80 & 274.80 & 0.00\% &                 &                 &                 &                 \\
							& a3-30 & 89    & 0.2    & 4.2  & 1.2    & 5.7    & 413.27 & 413.27 & 0.00\% & 0.6    & 1.6    & 2.3    & 413.27 & 413.27 & 0.00\% & 0.8    & 81.7   & 82.6   & 413.27 & 413.27 & 0.00\% &                 &                 &                 &                 \\
							& a3-36 & 116   & 0.3    & 7.0  & 4.9    & 12.3   & 480.37 & 480.37 & 0.00\% & 0.8    & 5.0    & 5.9    & 481.17 & 481.17 & 0.00\% & 1.3    & 710.9  & 712.4  & 481.17 & 481.17 & 0.00\% &                 &                 &                 &                 \\
							& a4-16 & 82    & 0.4    & 1.2  & 0.5    & 2.1    & 221.48 & 221.48 & 0.00\% & 0.4    & 1.4    & 1.9    & 222.49 & 222.49 & 0.00\% & 0.4    & 3.7    & 4.1    & 222.49 & 222.49 & 0.00\% &                 &                 &                 &                 \\
							& a4-24 & 95    & 0.4    & 3.1  & 1.0    & 4.6    & 310.69 & 310.69 & 0.00\% & 3.1    & 3.2    & 6.5    & 310.84 & 310.84 & 0.00\% & 0.6    & 22.6   & 23.5   & 310.84 & 310.84 & 0.00\% &                 &                 &                 &                 \\
							& a4-32 & 239   & 1.0    & 4.3  & 1.7    & 7.2    & 393.95 & 393.95 & 0.00\% & 2.8    & 11.0   & 14.1   & 393.95 & 393.95 & 0.00\% & 1.5    & 60.6   & 62.3   & 393.95 & 393.95 & 0.00\% &                 &                 &                 &                 \\
							& a4-40 & 239   & 1.2    & 6.4  & 2.9    & 10.6   & 452.16 & 452.16 & 0.00\% & 1.6    & 8.1    & 10.1   & 453.84 & 453.84 & 0.00\% & 3.0    & 284.2  & 287.6  & 453.84 & 453.84 & 0.00\% &                 &                 &                 &                 \\
							& a4-48 & 346   & 1.3    & 9.9  & 9.2    & 20.5   & 554.54 & 554.54 & 0.00\% & 2.6    & 15.1   & 18.2   & 554.54 & 554.54 & 0.00\% & 10.3   & 477.9  & 488.7  & 554.54 & 554.54 & 0.00\% &                 &                 &                 &                 \\
							& a5-40 & 319   & 3.9    & 7.9  & 4.2    & 16.4   & 414.50 & 414.50 & 0.00\% & 1.5    & 11.6   & 13.3   & 414.50 & 414.50 & 0.00\% & 7.0    & 92.4   & 99.7   & 414.50 & 414.50 & 0.00\% &                 &                 &                 &                 \\
							& a5-50 & 692   & 3.9    & 10.2 & 6.2    & 20.5   & 559.17 & 559.17 & 0.00\% & 3.0    & 63.2   & 66.7   & 559.17 & 559.17 & 0.00\% & 10.1   & 167.1  & 178.3  & 559.17 & 559.17 & 0.00\% &                 &                 &                 &                 \\
							& Avg   & 132.9 & 0.9    & 2.9  & 1.9    & 5.4    &        &        & 0.00\% &        & 9.4    & 10.2   &        &        & 0.00\% &        & 152.3  & 154.5  &        &        & 0.00\% &                 &                 &                 &                 \\ \bottomrule
						\end{tabular}%
						\begin{tablenotes}
							\item[1] Approx: the optimal solution of TSFFCS is an approximate solution for E-ADARP. 
						\end{tablenotes}
					\end{threeparttable}
					\label{table:detailEADARPainfty}
				\end{sidewaystable}

				\begin{sidewaystable}[htbp]
					\caption{Computational performance of different procedures on E-ADARP for type-u instances with $N_{\max}=\infty$}
					\scriptsize  
					\centering
					\setlength{\tabcolsep}{1.0 pt}
					\renewcommand{\arraystretch}{0.5}  
					\begin{threeparttable}
						\begin{tabular}{@{}cccccccccccccccccc@{}}
							\toprule
							&       & \multicolumn{6}{c}{IEBF}                            & \multicolumn{6}{c}{REBF}                            & \multicolumn{4}{c}{Results of   \citeauthor{Su2024}} \\ \midrule
							$\gamma$              & Name  & E-time & CPU    & Time   & OBJ    & LB     & Gap    & E-time & CPU    & Time   & OBJ    & LB     & Gap    & Time            & OBJ             & LB              & Gap             \\
							\multirow{14}{*}{70\%} & u2-16 & 0.6    & 3.9    & 4.5    & 58.17  & 58.17  & 0.00\% & 0.6    & 77.0   & 77.7   & 58.17  & 58.17  & 0.00\% & 172.2           & 58.17           & 58.17           & 0.00\%          \\
							& u2-20 & 1.2    & 15.3   & 16.6   & 56.86  & 56.86  & 0.00\% & 1.2    & 68.8   & 70.1   & 56.86  & 56.86  & 0.00\% & 397.0           & 56.86           & 56.86           & 0.00\%          \\
							& u2-24 & 0.8    & 8.0    & 8.9    & 91.33  & 91.33  & 0.00\% & 1.1    & 1798.9 & 1800.1 & 91.33  & 91.14  & 0.21\% & 5250.1          & 91.33           & 91.33           & 0.00\%          \\
							& u3-18 & 0.6    & 3.2    & 3.9    & 50.99  & 50.99  & 0.00\% & 0.6    & 40.8   & 41.4   & 50.99  & 50.99  & 0.00\% & 47.4            & 50.99           & 50.99           & 0.00\%          \\
							& u3-24 & 1.0    & 26.2   & 27.3   & 68.06  & 68.06  & 0.00\% & 1.7    & 141.9  & 143.7  & 68.06  & 68.06  & 0.00\% & 558.9           & 68.06           & 68.06           & 0.00\%          \\
							& u3-30 & 1.6    & 42.2   & 44.0   & 77.29  & 77.29  & 0.00\% & 3.8    & 1796.3 & 1800.3 & 77.29  & 77.16  & 0.17\% & 807.9           & 77.29           & 77.29           & 0.00\%          \\
							& u3-36 & 2.3    & 163.0  & 165.5  & 104.97 & 104.97 & 0.00\% & 4.5    & 1795.2 & 1800.0 & 104.97 & 104.75 & 0.21\% & 7200.0          & 106.72          & 104.85          & 1.07\%          \\
							& u4-16 & 0.9    & 6.5    & 7.4    & 53.87  & 53.87  & 0.00\% & 0.6    & 10.0   & 10.6   & 53.87  & 53.87  & 0.00\% & 16.7            & 53.87           & 53.87           & 0.00\%          \\
							& u4-24 & 1.1    & 2.5    & 3.7    & 89.83  & 89.83  & 0.00\% & 1.0    & 7.6    & 8.7    & 89.83  & 89.83  & 0.00\% & 74.6            & 89.83           & 89.83           & 0.00\%          \\
							& u4-32 & 2.1    & 28.2   & 30.5   & 99.50  & 99.50  & 0.00\% & 10.2   & 127.3  & 137.9  & 99.50  & 99.50  & 0.00\% & 1600.3          & 99.50           & 99.50           & 0.00\%          \\
							& u4-40 & 3.1    & 67.9   & 71.4   & 134.38 & 134.38 & 0.00\% & 3.7    & 1796.2 & 1800.1 & 134.38 & 133.49 & 0.66\% & 7200.0          & 134.38          & 134.16          & 0.16\%          \\
							& u4-48 & 6.8    & 1792.4 & 1800.1 & 149.02 & 144.33 & 3.15\% & 23.1   & 1776.3 & 1800.4 & 150.39 & 144.49 & 3.92\% & 7200.0          & 152.72          & 145.99          & 2.35\%          \\
							& u5-40 & 5.0    & 330.0  & 335.3  & 122.93 & 122.93 & 0.00\% & 10.0   & 1789.7 & 1800.1 & 122.93 & 122.25 & 0.55\% & 7200.0          & 123.00          & 122.72          & 0.23\%          \\
							& u5-50 & 25.5   & 412.7  & 439.4  & 142.89 & 142.89 & 0.00\% & 36.3   & 477.1  & 514.6  & 142.89 & 142.89 & 0.00\% & 5628.9          & 142.89          & 142.89          & 0.00\%          \\
							& Avg   &        & 273.3  & 277.3  &        &        & 0.42\% &        & 819.6  & 820.2  &        &        & 0.56\% & 3014.1          &                 &                 & 0.41\%          \\
							\multirow{14}{*}{40\%} & u2-16 & 0.4    & 3.0    & 3.5    & 57.65  & 57.65  & 0.00\% & 0.5    & 24.1   & 24.7   & 57.65  & 57.65  & 0.00\% & 35.7            & 57.65           & 57.65           & 0.00\%          \\
							& u2-20 & 1.8    & 10.5   & 12.4   & 56.34  & 56.34  & 0.00\% & 1.4    & 50.7   & 52.3   & 56.34  & 56.34  & 0.00\% & 329.9           & 56.34           & 56.34           & 0.00\%          \\
							& u2-24 & 1.3    & 8.1    & 9.5    & 90.84  & 90.84  & 0.00\% & 0.9    & 451.2  & 452.1  & 90.84  & 90.84  & 0.00\% & 2408.5          & 90.84           & 90.84           & 0.00\%          \\
							& u3-18 & 0.7    & 2.5    & 3.2    & 50.74  & 50.74  & 0.00\% & 0.9    & 31.2   & 32.1   & 50.74  & 50.74  & 0.00\% & 62.6            & 50.74           & 50.74           & 0.00\%          \\
							& u3-24 & 1.2    & 22.0   & 23.3   & 67.56  & 67.56  & 0.00\% & 1.7    & 65.2   & 66.9   & 67.56  & 67.56  & 0.00\% & 141.0           & 67.56           & 67.56           & 0.00\%          \\
							& u3-30 & 1.8    & 25.5   & 27.4   & 76.75  & 76.75  & 0.00\% & 2.3    & 204.3  & 206.8  & 76.75  & 76.75  & 0.00\% & 1457.8          & 76.75           & 76.75           & 0.00\%          \\
							& u3-36 & 2.7    & 27.3   & 30.2   & 104.06 & 104.06 & 0.00\% & 6.1    & 237.1  & 243.4  & 104.06 & 104.06 & 0.00\% & 3729.6          & 104.06          & 104.06          & 0.00\%          \\
							& u4-16 & 0.6    & 4.4    & 5.1    & 53.58  & 53.58  & 0.00\% & 1.1    & 7.6    & 8.8    & 53.58  & 53.58  & 0.00\% & 7.1             & 53.58           & 53.58           & 0.00\%          \\
							& u4-24 & 1.6    & 2.4    & 4.1    & 89.83  & 89.83  & 0.00\% & 1.0    & 5.4    & 6.6    & 89.83  & 89.83  & 0.00\% & 39.0            & 89.83           & 89.83           & 0.00\%          \\
							& u4-32 & 2.7    & 16.0   & 19.0   & 99.29  & 99.29  & 0.00\% & 5.6    & 132.7  & 138.5  & 99.29  & 99.29  & 0.00\% & 438.3           & 99.29           & 99.29           & 0.00\%          \\
							& u4-40 & 3.1    & 20.8   & 24.1   & 133.36 & 133.36 & 0.00\% & 2.9    & 793.9  & 797.0  & 133.36 & 133.36 & 0.00\% & 1581.5          & 133.36          & 133.36          & 0.00\%          \\
							& u4-48 & 5.7    & 1792.8 & 1800.1 & 147.32 & 145.35 & 1.34\% & 22.8   & 1776.4 & 1800.3 & 148.40 & 143.61 & 3.22\% & 7200.0          & 147.56          & 146.97          & 0.40\%          \\
							& u5-40 & 3.8    & 55.7   & 59.9   & 121.96 & 121.96 & 0.00\% & 9.0    & 149.0  & 158.4  & 121.96 & 121.96 & 0.00\% & 1073.8          & 121.96          & 121.96          & 0.00\%          \\
							& u5-50 & 6.0    & 234.0  & 241.0  & 142.83 & 142.83 & 0.00\% & 16.8   & 334.0  & 351.8  & 142.83 & 142.83 & 0.00\% & 6587.9          & 142.83          & 142.83          & 0.00\%          \\
							& Avg   &        & 244.1  & 245.4  &        &        & 0.18\% &        & 305.4  & 308.8  &        &        & 0.43\% & 1822.4          &                 &                 & 0.05\%          \\
							\multirow{14}{*}{10\%} & u2-16 & 0.5    & 2.3    & 2.8    & 57.61  & 57.61  & 0.00\% & 0.6    & 16.6   & 17.2   & 57.61  & 57.61  & 0.00\% & 69.5            & 57.61           & 57.61           & 0.00\%          \\
							& u2-20 & 1.5    & 6.7    & 8.4    & 55.59  & 55.59  & 0.00\% & 1.3    & 34.6   & 36.0   & 55.59  & 55.59  & 0.00\% & 215.3           & 55.59           & 55.59           & 0.00\%          \\
							& u2-24 & 1.1    & 7.9    & 9.0    & 90.66  & 90.66  & 0.00\% & 1.1    & 139.3  & 140.4  & 90.66  & 90.66  & 0.00\% & 1649.0          & 90.66           & 90.66           & 0.00\%          \\
							& u3-18 & 0.7    & 2.4    & 3.1    & 50.74  & 50.74  & 0.00\% & 0.7    & 40.3   & 41.0   & 50.74  & 50.74  & 0.00\% & 46.0            & 50.74           & 50.74           & 0.00\%          \\
							& u3-24 & 1.1    & 20.9   & 22.1   & 67.56  & 67.56  & 0.00\% & 1.5    & 66.7   & 68.3   & 67.56  & 67.56  & 0.00\% & 109.6           & 67.56           & 67.56           & 0.00\%          \\
							& u3-30 & 1.8    & 21.0   & 23.0   & 76.75  & 76.75  & 0.00\% & 2.1    & 410.2  & 412.5  & 76.75  & 76.75  & 0.00\% & 750.8           & 76.75           & 76.75           & 0.00\%          \\
							& u3-36 & 2.3    & 23.4   & 25.9   & 103.93 & 103.93 & 0.00\% & 6.0    & 181.7  & 188.0  & 103.93 & 103.93 & 0.00\% & 4326.5          & 103.93          & 103.93          & 0.00\%          \\
							& u4-16 & 0.6    & 5.0    & 5.6    & 53.58  & 53.58  & 0.00\% & 0.6    & 13.0   & 13.7   & 53.58  & 53.58  & 0.00\% & 6.5             & 53.58           & 53.58           & 0.00\%          \\
							& u4-24 & 1.3    & 1.6    & 3.0    & 89.83  & 89.83  & 0.00\% & 1.8    & 9.2    & 11.5   & 89.83  & 89.83  & 0.00\% & 63.2            & 89.83           & 89.83           & 0.00\%          \\
							& u4-32 & 2.3    & 6.8    & 9.4    & 99.29  & 99.29  & 0.00\% & 3.7    & 85.2   & 89.2   & 99.29  & 99.29  & 0.00\% & 416.4           & 99.29           & 99.29           & 0.00\%          \\
							& u4-40 & 2.6    & 9.7    & 12.6   & 133.11 & 133.11 & 0.00\% & 3.0    & 68.3   & 71.5   & 133.11 & 133.11 & 0.00\% & 2586.1          & 133.11          & 133.11          & 0.00\%          \\
							& u4-48 & 10.5   & 1788.7 & 1800.3 & 147.95 & 145.38 & 1.74\% & 15.0   & 1784.2 & 1800.1 & 147.30 & 144.49 & 1.91\% & 7200.0          & 147.33          & 146.74          & 0.40\%          \\
							& u5-40 & 4.9    & 45.6   & 50.9   & 121.86 & 121.86 & 0.00\% & 25.1   & 109.7  & 135.2  & 121.86 & 121.86 & 0.00\% & 1591.2          & 121.86          & 121.86          & 0.00\%          \\
							& u5-50 & 22.4   & 286.4  & 310.1  & 142.83 & 142.83 & 0.00\% & 30.4   & 385.4  & 418.9  & 142.83 & 142.83 & 0.00\% & 7200.0          & 142.82          & 142.75          & 0.05\%          \\
							& Avg   &        & 251.0  & 254.8  &        &        & 0.23\% &        & 266.2  & 270.5  &        &        & 0.25\% & 1974.0          &                 &                 & 0.06\%          \\ \bottomrule
						\end{tabular}
						\end{threeparttable}
						\label{table:detailEADARPuinfty}
					\end{sidewaystable}
					
					\clearpage

					\section{Computational results of D-E-ADARP with $N_{\max}=\infty$}\label{DEADARP}
					
					Tables~\ref{table:detailDEADARPa5}--\ref{table:detailDEADARPu1} present the computational performance of TSFFCS and IEBF for solving D-E-ADARP instances of types a and u under $N_{\max}=\infty$. For type-a instances of D-E-ADARP with a time granularity of 5 minutes, the results of BTSFF are included. All experiments are subject to a total runtime limit of 30 minutes.
					
					In the following tables at this section and subsequent section, we also include the indicator ``E-time'', which denotes the time required to generate events, event arcs, and the event-based network for the E-ADARP. Additionally, we report the fragment generation time (F-time). ``Avg'' denotes the average values for corresponding column. ``NA'' is used to indicate infeasible or missing values. To ensure a fair comparison, the following rules are applied when calculating averages: (i) if all formulations report ``NA'' for a given instance, that instance is excluded from the average calculation; (ii) if a formulation uniquely reports ``NA'' for an instance while others report feasible values, the average for that formulation is recorded as ``NA''.
					
					\begin{sidewaystable}[htbp]
						\caption{Computational performance of different formulations on D-E-ADARP for type-a instances with 5-minute granularity}
						\scriptsize  
						\centering
						\setlength{\tabcolsep}{1.0 pt}
						\renewcommand{\arraystretch}{0.5}  
						\begin{threeparttable}
							\begin{tabular}{@{}cccccccccccccccccccccc@{}}
								\toprule
								&       & \multicolumn{8}{c}{TSFFCS}                                          & \multicolumn{6}{c}{IEBF}                           & \multicolumn{6}{c}{BTSFF}                         \\ \midrule
								$\gamma$              & Name  & $|F|$  & F-time & Net  & CPU    & Time   & OBJ    & LB     & Gap    & E-time & CPU    & Time   & OBJ   & LB     & Gap    & Net   & CPU   & Time   & OBJ    & LB     & Gap    \\
								\multirow{14}{*}{70\%} & a2-16 & 44     & 0.1    & 1.0  & 1.4    & 2.7    & 231.01 & 231.01 & 0.00\% & 0.2    & 0.9    & 1.1    & 231.0 & 231.01 & 0.00\% & 51.5  & 8.6   & 60.5   & 231.01 & 231.01 & 0.00\% \\
								& a2-20 & 67     & 0.2    & 9.2  & 31.6   & 41.2   & 284.24 & 284.24 & 0.00\% & 0.3    & 3.2    & 3.5    & 284.2 & 284.24 & 0.00\% & 88.6  & 44.0  & 132.9  & 284.24 & 284.24 & 0.00\% \\
								& a2-24 & 104    & 0.3    & 2.5  & 3.4    & 6.5    & 344.49 & 344.49 & 0.00\% & 0.4    & 3.4    & 3.7    & 344.5 & 344.49 & 0.00\% & 127.3 & 146.9 & 274.8  & 344.49 & 344.49 & 0.00\% \\
								& a3-18 & 114    & 0.4    & 1.2  & 1.2    & 2.9    & 225.53 & 225.53 & 0.00\% & 0.3    & 2.2    & 2.5    & 225.5 & 225.53 & 0.00\% & 51.0  & 9.1   & 60.9   & 225.53 & 225.53 & 0.00\% \\
								& a3-24 & 224    & 1.5    & 2.5  & 2.7    & 6.9    & 273.49 & 273.49 & 0.00\% & 0.6    & 9.9    & 10.6   & 273.5 & 273.49 & 0.00\% & 133.2 & 24.5  & 159.4  & 273.49 & 273.49 & 0.00\% \\
								& a3-30 & 183    & 0.8    & 3.7  & 7.3    & 12.4   & 408.41 & 408.41 & 0.00\% & 1.1    & 13.1   & 14.4   & 408.4 & 408.41 & 0.00\% & 212.2 & 120.6 & 334.3  & 408.41 & 408.41 & 0.00\% \\
								& a3-36 & 284    & 1.6    & 5.5  & 7.0    & 14.8   & 446.90 & 446.90 & 0.00\% & 1.1    & 7.4    & 8.9    & 446.9 & 446.90 & 0.00\% & 278.3 & 657.6 & 939.7  & 446.90 & 446.90 & 0.00\% \\
								& a4-16 & 348    & 1.2    & 0.9  & 1.0    & 3.2    & 221.26 & 221.26 & 0.00\% & 0.3    & 5.3    & 5.7    & 221.3 & 221.26 & 0.00\% & 32.5  & 5.2   & 39.6   & 221.26 & 221.26 & 0.00\% \\
								& a4-24 & 286    & 1.4    & 2.6  & 3.1    & 7.3    & 310.78 & 310.78 & 0.00\% & 0.5    & 8.8    & 9.4    & 310.8 & 310.78 & 0.00\% & 116.8 & 22.9  & 141.7  & 310.78 & 310.78 & 0.00\% \\
								& a4-32 & 5283   & 75.4   & 7.0  & 15.1   & 98.2   & 389.73 & 389.73 & 0.00\% & 2.5    & 36.2   & 39.1   & 389.7 & 389.73 & 0.00\% & 231.0 & 104.5 & 420.6  & 389.73 & 389.73 & 0.00\% \\
								& a4-40 & 783    & 10.9   & 7.5  & 21.9   & 41.4   & 457.45 & 457.45 & 0.00\% & 1.9    & 63.2   & 66.1   & 457.4 & 457.45 & 0.00\% & 343.5 & 411.3 & 766.6  & 457.45 & 457.45 & 0.00\% \\
								& a4-48 & 1216   & 18.9   & 14.1 & 84.7   & 119.4  & 547.84 & 547.84 & 0.00\% & 3.2    & 267.6  & 272.8  & 547.8 & 547.84 & 0.00\% & 551.5 & 724.2 & 1293.9 & 547.84 & 547.84 & 0.00\% \\
								& a5-40 & 14204  & 327.6  & 15.9 & 40.2   & 384.8  & 414.77 & 414.77 & 0.00\% & 2.1    & 48.6   & 51.5   & 414.8 & 414.77 & 0.00\% & 396.7 & 138.6 & 944.0  & 414.77 & 414.77 & 0.00\% \\
								& a5-50 & 13068  & 427.2  & 23.7 & 1347.0 & 1800.0 & 557.05 & 549.98 & 1.27\% & 5.8    & 1790.2 & 1800.0 & 557.4 & 547.68 & 1.75\% & 603.4 & 698.9 & 1800.0 & NA     & -77.36 & NA     \\
								& Avg   & 3542.3 & 92.0   & 5.1  & 176.4  & 260.2  &        &        & 0.17\% &        & 247.9  & 249.4  &       &        & 0.23\% & 146.5 & 228.8 & 444.6  &        &        & NA \\
								\multirow{14}{*}{40\%} & a2-16 & 44     & 0.1    & 1.3  & 0.4    & 2.0    & 227.72 & 227.72 & 0.00\% & 0.2    & 0.5    & 0.7    & 227.7 & 227.72 & 0.00\% & 81.4  & 7.0   & 88.7   & 227.72 & 227.72 & 0.00\% \\
								& a2-20 & 67     & 0.2    & 2.0  & 1.4    & 3.7    & 278.58 & 278.58 & 0.00\% & 0.3    & 1.9    & 2.3    & 278.6 & 278.58 & 0.00\% & 107.8 & 35.7  & 144.0  & 278.58 & 278.58 & 0.00\% \\
								& a2-24 & 104    & 0.4    & 2.7  & 1.9    & 5.2    & 341.92 & 341.92 & 0.00\% & 0.5    & 2.4    & 3.0    & 341.9 & 341.92 & 0.00\% & 127.0 & 111.1 & 239.4  & 341.92 & 341.92 & 0.00\% \\
								& a3-18 & 114    & 0.4    & 1.2  & 0.5    & 2.3    & 224.68 & 224.68 & 0.00\% & 0.4    & 1.0    & 1.4    & 224.7 & 224.68 & 0.00\% & 45.8  & 9.2   & 55.7   & 224.68 & 224.68 & 0.00\% \\
								& a3-24 & 224    & 1.0    & 2.9  & 1.4    & 5.7    & 273.09 & 273.09 & 0.00\% & 0.5    & 3.9    & 4.5    & 273.1 & 273.09 & 0.00\% & 131.9 & 30.5  & 163.9  & 273.09 & 273.09 & 0.00\% \\
								& a3-30 & 183    & 0.8    & 4.0  & 5.7    & 11.0   & 406.74 & 406.70 & 0.00\% & 0.7    & 7.8    & 8.8    & 406.7 & 406.73 & 0.00\% & 213.2 & 132.2 & 346.8  & 406.74 & 406.74 & 0.00\% \\
								& a3-36 & 284    & 1.6    & 6.5  & 6.1    & 15.0   & 446.02 & 446.02 & 0.00\% & 1.1    & 5.1    & 6.4    & 446.0 & 446.02 & 0.00\% & 261.1 & 260.6 & 526.3  & 446.02 & 446.02 & 0.00\% \\
								& a4-16 & 348    & 1.3    & 1.0  & 1.0    & 3.4    & 221.26 & 221.26 & 0.00\% & 0.3    & 4.2    & 4.5    & 221.3 & 221.26 & 0.00\% & 43.5  & 7.1   & 53.5   & 221.26 & 221.26 & 0.00\% \\
								& a4-24 & 286    & 1.6    & 2.8  & 1.7    & 6.5    & 308.82 & 308.82 & 0.00\% & 0.6    & 3.4    & 4.1    & 308.8 & 308.82 & 0.00\% & 117.3 & 30.9  & 150.2  & 308.82 & 308.82 & 0.00\% \\
								& a4-32 & 5283   & 75.2   & 6.3  & 5.5    & 87.7   & 387.06 & 387.06 & 0.00\% & 3.5    & 18.0   & 22.0   & 387.1 & 387.06 & 0.00\% & 191.4 & 132.3 & 410.8  & 387.06 & 387.06 & 0.00\% \\
								& a4-40 & 783    & 8.7    & 10.4 & 9.9    & 30.1   & 454.77 & 454.77 & 0.00\% & 1.7    & 38.1   & 40.7   & 454.8 & 454.77 & 0.00\% & 366.7 & 332.6 & 711.8  & 454.77 & 454.77 & 0.00\% \\
								& a4-48 & 1216   & 19.0   & 12.6 & 16.4   & 49.9   & 544.67 & 544.65 & 0.00\% & 5.6    & 64.3   & 71.8   & 544.7 & 544.67 & 0.00\% & 555.2 & 380.9 & 955.6  & 544.67 & 544.67 & 0.00\% \\
								& a5-40 & 14204  & 290.5  & 20.8 & 28.5   & 341.0  & 413.37 & 413.37 & 0.00\% & 3.7    & 30.1   & 34.6   & 413.4 & 413.37 & 0.00\% & 415.3 & 212.8 & 1032.2 & 413.37 & 413.37 & 0.00\% \\
								& a5-50 & 13068  & 435.4  & 22.7 & 36.8   & 497.4  & 547.14 & 547.14 & 0.00\% & 3.3    & 238.5  & 243.8  & 547.1 & 547.14 & 0.00\% & 695.1 & 578.5 & 1800.0 & NA     & -59.19 & NA     \\
								& Avg   & 3542.3 & 88.9   & 5.5  & 8.3    & 99.8   &        &        & 0.00\% &        & 35.9   & 37.5   &       &        & 0.00\% & 156.6 & 136.8 & 377.2  &        &        & NA \\
								\multirow{14}{*}{10\%} & a2-16 & 44     & 0.1    & 1.0  & 0.4    & 1.6    & 227.72 & 227.72 & 0.00\% & 0.2    & 0.3    & 0.6    & 227.7 & 227.72 & 0.00\% & 59.0  & 7.3   & 66.7   & 227.72 & 227.72 & 0.00\% \\
								& a2-20 & 67     & 0.2    & 1.8  & 0.8    & 2.9    & 278.58 & 278.58 & 0.00\% & 0.3    & 0.8    & 1.2    & 278.6 & 278.58 & 0.00\% & 72.8  & 72.0  & 145.1  & 278.58 & 278.58 & 0.00\% \\
								& a2-24 & 104    & 0.3    & 2.4  & 1.2    & 4.2    & 341.85 & 341.85 & 0.00\% & 0.3    & 2.4    & 2.8    & 341.9 & 341.85 & 0.00\% & 94.2  & 76.9  & 171.5  & 341.85 & 341.85 & 0.00\% \\
								& a3-18 & 114    & 0.5    & 1.1  & 0.5    & 2.2    & 224.68 & 224.68 & 0.00\% & 0.4    & 1.1    & 1.5    & 224.7 & 224.68 & 0.00\% & 43.8  & 8.1   & 52.4   & 224.68 & 224.68 & 0.00\% \\
								& a3-24 & 224    & 1.0    & 2.5  & 1.1    & 4.9    & 273.09 & 273.09 & 0.00\% & 0.7    & 3.1    & 3.9    & 273.1 & 273.09 & 0.00\% & 107.1 & 30.4  & 138.6  & 273.09 & 273.09 & 0.00\% \\
								& a3-30 & 183    & 0.8    & 3.7  & 3.1    & 8.0    & 406.66 & 406.66 & 0.00\% & 0.9    & 7.2    & 8.3    & 406.7 & 406.66 & 0.00\% & 183.8 & 125.7 & 310.7  & 406.66 & 406.66 & 0.00\% \\
								& a3-36 & 284    & 1.6    & 8.6  & 7.8    & 18.8   & 446.02 & 446.02 & 0.00\% & 1.1    & 3.7    & 5.0    & 446.0 & 446.02 & 0.00\% & 235.6 & 185.0 & 425.3  & 446.02 & 446.02 & 0.00\% \\
								& a4-16 & 348    & 1.8    & 1.1  & 1.2    & 4.2    & 221.26 & 221.26 & 0.00\% & 0.2    & 4.1    & 4.5    & 221.3 & 221.26 & 0.00\% & 53.1  & 7.3   & 62.4   & 221.26 & 221.26 & 0.00\% \\
								& a4-24 & 286    & 1.7    & 2.8  & 1.2    & 6.0    & 308.56 & 308.56 & 0.00\% & 0.9    & 4.2    & 5.3    & 308.6 & 308.56 & 0.00\% & 102.1 & 26.9  & 130.6  & 308.56 & 308.56 & 0.00\% \\
								& a4-32 & 5283   & 67.7   & 6.2  & 5.5    & 80.1   & 387.06 & 387.06 & 0.00\% & 2.6    & 14.8   & 17.9   & 387.1 & 387.06 & 0.00\% & 184.5 & 104.7 & 356.6  & 387.06 & 387.06 & 0.00\% \\
								& a4-40 & 783    & 8.4    & 7.2  & 3.9    & 20.5   & 452.97 & 452.97 & 0.00\% & 1.8    & 28.4   & 31.0   & 453.0 & 452.97 & 0.00\% & 304.7 & 207.8 & 527.5  & 452.97 & 452.97 & 0.00\% \\
								& a4-48 & 1216   & 13.2   & 13.7 & 12.8   & 41.4   & 544.47 & 544.46 & 0.00\% & 2.9    & 54.9   & 59.5   & 544.5 & 544.47 & 0.00\% & 470.6 & 427.5 & 915.4  & 544.47 & 544.47 & 0.00\% \\
								& a5-40 & 14204  & 311.2  & 16.1 & 20.5   & 349.1  & 413.37 & 413.37 & 0.00\% & 1.9    & 17.8   & 20.5   & 413.4 & 413.37 & 0.00\% & 418.9 & 220.3 & 999.4  & 413.37 & 413.37 & 0.00\% \\
								& a5-50 & 13068  & 419.9  & 18.5 & 19.3   & 459.8  & 546.02 & 546.02 & 0.00\% & 4.2    & 193.9  & 201.5  & 546.0 & 546.02 & 0.00\% & 649.1 & 640.2 & 1800.0 & NA     & -59.19 & NA     \\
								& Avg   & 3542.3 & 88.8   & 4.7  & 5.4    & 96.3   &        &        & 0.00\% &        & 29.3   & 30.6   &       &        & 0.00\% & 145.0 & 130.9 & 357.0  &        &        & NA \\ \bottomrule
							\end{tabular}
							\end{threeparttable}
							\label{table:detailDEADARPa5}
						\end{sidewaystable}

						\begin{sidewaystable}[htbp]
							\caption{Computational performance of different formulations on D-E-ADARP for type-a instances with 2-minute granularity}
							\scriptsize  
							\centering
							\setlength{\tabcolsep}{1.0 pt}
							\renewcommand{\arraystretch}{0.5}  
							\begin{threeparttable}
								\begin{tabular}{@{}cccccccccccccccc@{}}
									\toprule
									&       & \multicolumn{8}{c}{TSFFCS}                                         & \multicolumn{6}{c}{IEBF}                           \\ \midrule
									$\gamma$              & Name  & $|F|$ & F-time & Net  & CPU    & Time   & OBJ    & LB     & Gap    & E-time & CPU    & Time   & OBJ   & LB     & Gap    \\
									\multirow{14}{*}{70\%} & a2-16 & 38    & 0.1    & 6.6  & 5.0    & 11.8   & 237.52 & 237.50 & 0.00\% & 0.2    & 0.7    & 1.0    & 237.5 & 237.51 & 0.00\% \\
									& a2-20 & 54    & 0.2    & 10.6 & 196.1  & 207.2  & 286.20 & 286.20 & 0.00\% & 0.4    & 4.8    & 5.2    & 286.2 & 286.20 & 0.00\% \\
									& a2-24 & 69    & 0.2    & 11.3 & 9.6    & 21.4   & 346.01 & 346.01 & 0.00\% & 0.4    & 3.3    & 3.7    & 346.0 & 346.01 & 0.00\% \\
									& a3-18 & 77    & 0.3    & 5.0  & 3.3    & 8.8    & 227.03 & 227.03 & 0.00\% & 0.3    & 1.1    & 1.4    & 227.0 & 227.03 & 0.00\% \\
									& a3-24 & 132   & 0.5    & 13.6 & 8.8    & 23.3   & 275.24 & 275.24 & 0.00\% & 0.5    & 2.3    & 2.9    & 275.2 & 275.24 & 0.00\% \\
									& a3-30 & 101   & 0.3    & 30.1 & 19.2   & 50.1   & 415.10 & 415.10 & 0.00\% & 2.5    & 6.6    & 9.2    & 415.1 & 415.10 & 0.00\% \\
									& a3-36 & 188   & 0.8    & 32.7 & 15.4   & 49.5   & 441.21 & 441.21 & 0.00\% & 1.1    & 4.8    & 6.0    & 441.2 & 441.21 & 0.00\% \\
									& a4-16 & 145   & 0.3    & 4.5  & 1.8    & 6.8    & 222.34 & 222.34 & 0.00\% & 0.3    & 0.9    & 1.2    & 222.3 & 222.34 & 0.00\% \\
									& a4-24 & 124   & 0.5    & 13.8 & 8.6    & 23.3   & 310.21 & 310.21 & 0.00\% & 0.7    & 6.3    & 7.1    & 310.2 & 310.21 & 0.00\% \\
									& a4-32 & 1107  & 8.1    & 21.2 & 30.0   & 59.9   & 391.46 & 391.46 & 0.00\% & 1.3    & 25.3   & 26.9   & 391.5 & 391.46 & 0.00\% \\
									& a4-40 & 426   & 4.5    & 34.5 & 47.5   & 87.8   & 456.22 & 456.22 & 0.00\% & 6.6    & 47.4   & 55.9   & 456.2 & 456.22 & 0.00\% \\
									& a4-48 & 557   & 3.0    & 58.0 & 127.1  & 189.9  & 555.68 & 555.68 & 0.00\% & 3.4    & 87.5   & 92.9   & 555.7 & 555.68 & 0.00\% \\
									& a5-40 & 436   & 3.6    & 39.1 & 30.5   & 74.4   & 417.45 & 417.45 & 0.00\% & 2.4    & 34.8   & 37.9   & 417.5 & 417.45 & 0.00\% \\
									& a5-50 & 2125  & 17.5   & 66.6 & 1714.6 & 1800.0 & 559.31 & 550.69 & 1.54\% & 4.7    & 1793.0 & 1800.0 & 558.8 & 554.20 & 0.83\% \\
									& Avg   & 379.8 & 3.2    & 16.0 & 227.7  & 233.9  &        &        & 0.20\% &        & 235.5  & 236.2  &       &        & 0.11\% \\
									\multirow{14}{*}{40\%} & a2-16 & 38    & 0.1    & 5.7  & 1.7    & 7.7    & 237.44 & 237.44 & 0.00\% & 0.2    & 0.4    & 0.6    & 237.4 & 237.44 & 0.00\% \\
									& a2-20 & 54    & 0.1    & 8.9  & 6.7    & 15.9   & 280.87 & 280.87 & 0.00\% & 0.3    & 1.2    & 1.5    & 280.9 & 280.87 & 0.00\% \\
									& a2-24 & 69    & 0.2    & 11.7 & 7.3    & 19.4   & 343.53 & 343.53 & 0.00\% & 0.3    & 1.2    & 1.6    & 343.5 & 343.53 & 0.00\% \\
									& a3-18 & 77    & 0.2    & 4.7  & 2.1    & 7.2    & 226.18 & 226.18 & 0.00\% & 0.2    & 0.8    & 1.1    & 226.2 & 226.18 & 0.00\% \\
									& a3-24 & 132   & 0.5    & 14.3 & 5.3    & 20.3   & 274.84 & 274.84 & 0.00\% & 0.5    & 1.4    & 1.9    & 274.8 & 274.84 & 0.00\% \\
									& a3-30 & 101   & 0.3    & 24.8 & 8.7    & 34.4   & 414.10 & 414.10 & 0.00\% & 0.6    & 5.2    & 5.9    & 414.1 & 414.10 & 0.00\% \\
									& a3-36 & 188   & 0.8    & 26.6 & 12.0   & 40.3   & 440.94 & 440.94 & 0.00\% & 1.0    & 5.8    & 6.9    & 440.9 & 440.91 & 0.00\% \\
									& a4-16 & 145   & 0.4    & 2.9  & 1.2    & 4.6    & 222.34 & 222.34 & 0.00\% & 0.4    & 1.1    & 1.6    & 222.3 & 222.33 & 0.00\% \\
									& a4-24 & 124   & 0.3    & 12.1 & 7.3    & 20.0   & 308.73 & 308.73 & 0.00\% & 0.6    & 3.6    & 4.4    & 308.7 & 308.73 & 0.00\% \\
									& a4-32 & 1107  & 8.9    & 19.8 & 17.5   & 47.0   & 390.14 & 390.14 & 0.00\% & 1.0    & 10.9   & 12.2   & 390.1 & 390.14 & 0.00\% \\
									& a4-40 & 426   & 5.6    & 46.4 & 16.5   & 69.5   & 452.30 & 452.30 & 0.00\% & 1.6    & 14.1   & 16.2   & 452.3 & 452.30 & 0.00\% \\
									& a4-48 & 557   & 3.1    & 53.9 & 36.3   & 94.8   & 553.57 & 553.57 & 0.00\% & 2.8    & 56.7   & 60.9   & 553.6 & 553.57 & 0.00\% \\
									& a5-40 & 436   & 4.8    & 34.4 & 23.3   & 64.1   & 416.38 & 416.38 & 0.00\% & 4.8    & 19.9   & 25.4   & 416.4 & 416.38 & 0.00\% \\
									& a5-50 & 2125  & 22.9   & 63.4 & 54.7   & 142.7  & 549.98 & 549.98 & 0.00\% & 4.4    & 194.3  & 200.8  & 550.0 & 549.98 & 0.00\% \\
									& Avg   & 379.8 & 4.1    & 15.5 & 10.9   & 29.7   &        &        & 0.00\% &        & 29.4   & 30.6   &       &        & 0.00\% \\
									\multirow{14}{*}{10\%} & a2-16 & 38    & 0.1    & 5.2  & 1.8    & 7.3    & 237.44 & 237.44 & 0.00\% & 0.3    & 0.6    & 0.9    & 237.4 & 237.44 & 0.00\% \\
									& a2-20 & 54    & 0.1    & 14.1 & 3.2    & 17.6   & 279.42 & 279.42 & 0.00\% & 0.4    & 0.8    & 1.2    & 279.4 & 279.42 & 0.00\% \\
									& a2-24 & 69    & 0.2    & 13.0 & 6.8    & 20.4   & 343.53 & 343.53 & 0.00\% & 0.5    & 1.5    & 2.0    & 343.5 & 343.53 & 0.00\% \\
									& a3-18 & 77    & 0.2    & 4.6  & 1.9    & 6.8    & 226.18 & 226.18 & 0.00\% & 0.3    & 0.9    & 1.2    & 226.2 & 226.18 & 0.00\% \\
									& a3-24 & 132   & 0.5    & 13.2 & 5.8    & 19.8   & 274.84 & 274.84 & 0.00\% & 0.6    & 1.5    & 2.2    & 274.8 & 274.82 & 0.00\% \\
									& a3-30 & 101   & 0.3    & 19.4 & 5.3    & 25.5   & 414.10 & 414.10 & 0.00\% & 0.8    & 4.9    & 5.8    & 414.1 & 414.10 & 0.00\% \\
									& a3-36 & 188   & 0.7    & 34.1 & 9.1    & 44.5   & 440.71 & 440.71 & 0.00\% & 1.1    & 2.2    & 3.4    & 440.7 & 440.71 & 0.00\% \\
									& a4-16 & 145   & 0.4    & 2.9  & 1.1    & 4.5    & 222.34 & 222.34 & 0.00\% & 0.3    & 0.8    & 1.1    & 222.3 & 222.34 & 0.00\% \\
									& a4-24 & 124   & 0.3    & 11.8 & 4.4    & 16.8   & 308.67 & 308.67 & 0.00\% & 0.6    & 2.4    & 3.1    & 308.7 & 308.67 & 0.00\% \\
									& a4-32 & 1107  & 10.7   & 21.1 & 12.3   & 44.6   & 390.14 & 390.14 & 0.00\% & 1.0    & 9.0    & 10.2   & 390.1 & 390.14 & 0.00\% \\
									& a4-40 & 426   & 2.7    & 33.4 & 15.7   & 52.9   & 452.30 & 452.30 & 0.00\% & 1.5    & 7.0    & 9.0    & 452.3 & 452.30 & 0.00\% \\
									& a4-48 & 557   & 3.2    & 57.1 & 25.0   & 87.1   & 553.46 & 553.46 & 0.00\% & 3.5    & 47.5   & 52.8   & 553.5 & 553.46 & 0.00\% \\
									& a5-40 & 436   & 4.3    & 34.4 & 29.6   & 70.1   & 416.38 & 416.38 & 0.00\% & 3.9    & 15.5   & 20.0   & 416.4 & 416.38 & 0.00\% \\
									& a5-50 & 2125  & 21.7   & 69.6 & 30.7   & 123.8  & 548.66 & 548.66 & 0.00\% & 4.3    & 63.9   & 69.3   & 548.7 & 548.66 & 0.00\% \\
									& Avg   & 379.8 & 3.9    & 15.6 & 8.4    & 27.3   &        &        & 0.00\% &        & 13.3   & 14.7   &       &        & 0.00\% \\ \bottomrule
								\end{tabular}
								\end{threeparttable}
								\label{table:detailDEADARPa2}
							\end{sidewaystable}

							\begin{sidewaystable}[htbp]
								\caption{Computational performance of different formulations on D-E-ADARP for type-a instances with 1-minute granularity}
								\scriptsize  
								\centering
								\setlength{\tabcolsep}{1.0 pt}
								\renewcommand{\arraystretch}{0.5}  
								\begin{threeparttable}
									\begin{tabular}{@{}cccccccccccccccc@{}}
										\toprule
										&       & \multicolumn{1}{l}{} & \multicolumn{7}{c}{TSFFCS}                                  & \multicolumn{6}{c}{IEBF}                         \\ \midrule
										$\gamma$              & Name  & $|F|$                & F-time & Net   & CPU    & Time   & OBJ    & LB     & Gap    & E-time & CPU   & Time  & OBJ   & LB     & Gap    \\
										\multirow{14}{*}{70\%} & a2-16 & 36                   & 0.1    & 25.9  & 13.1   & 39.1   & 237.02 & 237.01 & 0.00\% & 0.2    & 1.3   & 1.5   & 237.0 & 237.01 & 0.00\% \\
										& a2-20 & 51                   & 0.1    & 30.1  & 1768.9 & 1800.0 & 285.45 & 284.75 & 0.25\% & 0.3    & 4.0   & 4.3   & 285.5 & 285.45 & 0.00\% \\
										& a2-24 & 65                   & 0.2    & 57.7  & 53.3   & 111.5  & 346.70 & 346.70 & 0.00\% & 0.4    & 3.5   & 3.9   & 346.7 & 346.70 & 0.00\% \\
										& a3-18 & 68                   & 0.2    & 17.7  & 27.2   & 45.2   & 237.53 & 237.53 & 0.00\% & 0.3    & 2.5   & 2.8   & 237.5 & 237.53 & 0.00\% \\
										& a3-24 & 109                  & 0.3    & 54.0  & 31.9   & 86.5   & 274.99 & 274.99 & 0.00\% & 0.4    & 2.1   & 2.6   & 275.0 & 274.99 & 0.00\% \\
										& a3-30 & 91                   & 0.3    & 85.9  & 42.7   & 129.5  & 413.14 & 413.11 & 0.00\% & 0.7    & 3.7   & 4.5   & 413.1 & 413.11 & 0.00\% \\
										& a3-36 & 131                  & 0.5    & 112.0 & 76.7   & 190.0  & 443.21 & 443.21 & 0.00\% & 0.9    & 3.6   & 4.7   & 443.2 & 443.21 & 0.00\% \\
										& a4-16 & 94                   & 0.2    & 11.0  & 4.7    & 16.2   & 222.59 & 222.59 & 0.00\% & 0.2    & 1.0   & 1.2   & 222.6 & 222.59 & 0.00\% \\
										& a4-24 & 103                  & 0.3    & 48.7  & 90.1   & 139.5  & 311.78 & 311.78 & 0.00\% & 0.5    & 4.8   & 5.4   & 311.8 & 311.78 & 0.00\% \\
										& a4-32 & 297                  & 1.7    & 95.9  & 89.9   & 188.2  & 395.49 & 395.49 & 0.00\% & 0.9    & 11.1  & 12.1  & 395.5 & 395.49 & 0.00\% \\
										& a4-40 & 308                  & 2.0    & 190.2 & 401.1  & 594.7  & 458.50 & 458.48 & 0.00\% & 2.1    & 28.9  & 31.5  & 458.5 & 458.50 & 0.00\% \\
										& a4-48 & 372                  & 1.6    & 216.4 & 1270.2 & 1490.0 & 557.71 & 557.71 & 0.00\% & 3.0    & 56.2  & 60.3  & 557.7 & 557.71 & 0.00\% \\
										& a5-40 & 345                  & 2.3    & 138.3 & 139.2  & 281.2  & 414.97 & 414.97 & 0.00\% & 2.1    & 21.5  & 24.4  & 415.0 & 414.97 & 0.00\% \\
										& a5-50 & 874                  & 5.7    & 234.7 & 1558.3 & 1800.0 & 566.85 & 557.64 & 1.62\% & 2.6    & 584.5 & 588.0 & 562.0 & 561.96 & 0.00\% \\
										& Avg   & 163.5                & 1.1    & 60.3  & 486.8  & 530.0  &        &        & 0.23\% &        & 76.7  & 77.4  &       &        & 0.00\% \\
										\multirow{14}{*}{40\%} & a2-16 & 36                   & 0.1    & 22.1  & 6.8    & 29.1   & 236.94 & 236.94 & 0.00\% & 0.1    & 0.5   & 0.7   & 236.9 & 236.94 & 0.00\% \\
										& a2-20 & 51                   & 0.1    & 43.9  & 16.0   & 60.2   & 280.17 & 280.15 & 0.00\% & 0.4    & 1.1   & 1.5   & 280.2 & 280.16 & 0.00\% \\
										& a2-24 & 65                   & 0.2    & 55.5  & 44.7   & 100.6  & 346.04 & 346.04 & 0.00\% & 0.4    & 1.9   & 2.4   & 346.0 & 346.04 & 0.00\% \\
										& a3-18 & 68                   & 0.2    & 17.6  & 7.1    & 25.0   & 235.51 & 235.51 & 0.00\% & 0.3    & 0.8   & 1.1   & 235.5 & 235.51 & 0.00\% \\
										& a3-24 & 109                  & 0.3    & 50.9  & 20.4   & 71.9   & 274.59 & 274.59 & 0.00\% & 0.5    & 1.3   & 1.9   & 274.6 & 274.59 & 0.00\% \\
										& a3-30 & 91                   & 0.3    & 73.6  & 24.5   & 98.9   & 412.97 & 412.97 & 0.00\% & 0.7    & 2.8   & 3.6   & 413.0 & 412.97 & 0.00\% \\
										& a3-36 & 131                  & 0.5    & 114.1 & 53.2   & 168.5  & 442.94 & 442.94 & 0.00\% & 1.1    & 5.5   & 6.7   & 442.9 & 442.91 & 0.00\% \\
										& a4-16 & 94                   & 0.2    & 16.2  & 5.2    & 21.8   & 222.59 & 222.59 & 0.00\% & 0.4    & 1.2   & 1.7   & 222.6 & 222.59 & 0.00\% \\
										& a4-24 & 103                  & 0.3    & 46.7  & 23.7   & 71.2   & 310.24 & 310.24 & 0.00\% & 0.8    & 2.5   & 3.5   & 310.2 & 310.24 & 0.00\% \\
										& a4-32 & 297                  & 1.3    & 77.5  & 44.6   & 124.1  & 394.81 & 394.81 & 0.00\% & 1.1    & 10.8  & 12.2  & 394.8 & 394.81 & 0.00\% \\
										& a4-40 & 308                  & 2.7    & 136.3 & 55.2   & 196.4  & 454.58 & 454.58 & 0.00\% & 1.4    & 9.6   & 11.4  & 454.6 & 454.58 & 0.00\% \\
										& a4-48 & 372                  & 1.6    & 221.0 & 111.1  & 335.5  & 555.39 & 555.39 & 0.00\% & 2.2    & 15.7  & 18.8  & 555.4 & 555.39 & 0.00\% \\
										& a5-40 & 345                  & 2.5    & 138.4 & 78.9   & 221.4  & 414.06 & 414.06 & 0.00\% & 2.2    & 17.5  & 20.1  & 414.1 & 414.06 & 0.00\% \\
										& a5-50 & 874                  & 6.2    & 249.1 & 263.8  & 521.0  & 557.74 & 557.74 & 0.00\% & 2.6    & 66.5  & 70.0  & 557.7 & 557.74 & 0.00\% \\
										& Avg   & 163.5                & 1.2    & 58.3  & 41.9   & 101.7  &        &        & 0.00\% &        & 10.2  & 11.0  &       &        & 0.00\% \\
										\multirow{14}{*}{10\%} & a2-16 & 36                   & 0.1    & 25.1  & 7.2    & 32.4   & 236.94 & 236.94 & 0.00\% & 0.4    & 0.7   & 1.1   & 236.9 & 236.94 & 0.00\% \\
										& a2-20 & 51                   & 0.1    & 41.8  & 9.3    & 51.3   & 278.67 & 278.67 & 0.00\% & 0.5    & 1.2   & 1.7   & 278.7 & 278.67 & 0.00\% \\
										& a2-24 & 65                   & 0.2    & 42.8  & 43.7   & 86.9   & 346.04 & 346.04 & 0.00\% & 0.5    & 2.2   & 2.8   & 346.0 & 346.04 & 0.00\% \\
										& a3-18 & 68                   & 0.2    & 17.0  & 7.1    & 24.5   & 235.51 & 235.51 & 0.00\% & 0.4    & 1.2   & 1.6   & 235.5 & 235.51 & 0.00\% \\
										& a3-24 & 109                  & 0.3    & 55.4  & 24.7   & 80.6   & 274.59 & 274.59 & 0.00\% & 0.6    & 1.6   & 2.4   & 274.6 & 274.59 & 0.00\% \\
										& a3-30 & 91                   & 0.2    & 78.7  & 24.9   & 104.4  & 412.96 & 412.96 & 0.00\% & 0.9    & 2.6   & 3.6   & 413.0 & 412.96 & 0.00\% \\
										& a3-36 & 131                  & 0.5    & 103.2 & 52.8   & 157.3  & 442.71 & 442.67 & 0.00\% & 1.0    & 2.5   & 3.7   & 442.7 & 442.71 & 0.00\% \\
										& a4-16 & 94                   & 0.4    & 13.1  & 4.1    & 17.7   & 222.59 & 222.59 & 0.00\% & 0.2    & 0.5   & 0.8   & 222.6 & 222.59 & 0.00\% \\
										& a4-24 & 103                  & 0.3    & 50.9  & 19.4   & 70.9   & 310.19 & 310.19 & 0.00\% & 0.5    & 1.7   & 2.2   & 310.2 & 310.19 & 0.00\% \\
										& a4-32 & 297                  & 1.2    & 74.7  & 27.9   & 104.5  & 394.50 & 394.50 & 0.00\% & 0.8    & 7.9   & 8.8   & 394.5 & 394.50 & 0.00\% \\
										& a4-40 & 308                  & 1.6    & 137.4 & 52.6   & 192.7  & 454.58 & 454.58 & 0.00\% & 2.0    & 12.6  & 15.0  & 454.6 & 454.55 & 0.00\% \\
										& a4-48 & 372                  & 1.9    & 235.2 & 101.8  & 340.7  & 555.39 & 555.39 & 0.00\% & 2.2    & 13.2  & 16.3  & 555.4 & 555.39 & 0.00\% \\
										& a5-40 & 345                  & 2.6    & 148.8 & 109.1  & 261.9  & 414.06 & 414.06 & 0.00\% & 2.8    & 19.0  & 22.4  & 414.1 & 414.06 & 0.00\% \\
										& a5-50 & 874                  & 6.2    & 237.5 & 151.7  & 397.2  & 556.13 & 556.13 & 0.00\% & 2.7    & 43.7  & 47.3  & 556.1 & 556.13 & 0.00\% \\
										& Avg   & 163.5                & 1.1    & 58.8  & 34.4   & 94.7   &        &        & 0.00\% &        & 8.1   & 9.1   &       &        & 0.00\% \\ \bottomrule
									\end{tabular}
									\end{threeparttable}
									\label{table:detailDEADARPa1}
								\end{sidewaystable}

								\begin{sidewaystable}[htbp]
									\caption{Computational performance of different formulations on D-E-ADARP for type-u instances with 2-minute granularity}
									\scriptsize  
									\centering
									\setlength{\tabcolsep}{1.0 pt}
									\renewcommand{\arraystretch}{0.5}  
									\begin{threeparttable}
										\begin{tabular}{@{}cccccccccccccccccccccc@{}}
											\toprule
											&       & \multicolumn{8}{c}{TSFFCS}                                         & \multicolumn{6}{c}{IEBF}                            & \multicolumn{6}{c}{BTSFF}                                                                                                               \\ \midrule
											$\gamma$              & Name  & $|F|$ & F-time & Net  & CPU    & Time   & OBJ    & LB     & Gap    & E-time & CPU    & Time   & OBJ    & LB     & Gap    & Net                  & CPU                  & Time                 & OBJ                  & LB                   & Gap                  \\
											\multirow{14}{*}{0.7} & u2-16 & 79    & 0.2    & 2.1  & 4.4    & 6.9    & 56.77  & 56.77  & 0.00\% & 0.6    & 4.9    & 5.6    & 56.77  & 56.77  & 0.00\% & 17.8                 & 6.7                  & 24.8                 & 56.77                & 56.77                & 0.00\%               \\
											& u2-20 & 226   & 2.4    & 17.0 & 39.1   & 58.8   & 58.64  & 58.64  & 0.00\% & 0.9    & 25.0   & 26.0   & 58.64  & 58.64  & 0.00\% & 28.2                 & 144.0                & 173.7                & 58.64                & 58.64                & 0.00\%               \\
											& u2-24 & 81    & 0.2    & 3.9  & 11.1   & 15.4   & 92.47  & 92.47  & 0.00\% & 1.2    & 18.0   & 19.3   & 92.47  & 92.47  & 0.00\% & 57.5                 & 95.6                 & 153.6                & 92.47                & 92.47                & 0.00\%               \\
											& u3-18 & 108   & 0.2    & 2.8  & 3.9    & 7.1    & 50.50  & 50.50  & 0.00\% & 0.9    & 8.8    & 9.7    & 50.50  & 50.50  & 0.00\% & 31.5                 & 16.9                 & 49.4                 & 50.50                & 50.50                & 0.00\%               \\
											& u3-24 & 157   & 0.4    & 3.9  & 122.0  & 126.6  & 70.31  & 70.30  & 0.00\% & 1.4    & 103.7  & 105.3  & 70.31  & 70.31  & 0.00\% & 61.4                 & 192.5                & 255.1                & 70.31                & 70.31                & 0.00\%               \\
											& u3-30 & 200   & 1.7    & 7.0  & 14.7   & 24.0   & 77.41  & 77.41  & 0.00\% & 2.2    & 33.1   & 35.7   & 77.41  & 77.41  & 0.00\% & 64.4                 & 151.7                & 218.5                & 77.41                & 77.41                & 0.00\%               \\
											& u3-36 & 241   & 1.1    & 8.2  & 157.3  & 167.3  & 105.27 & 105.27 & 0.00\% & 7.1    & 203.5  & 211.2  & 105.27 & 105.27 & 0.00\% & 109.9                & 395.7                & 508.5                & 105.27               & 105.27               & 0.00\%               \\
											& u4-16 & 86    & 0.2    & 2.3  & 9.3    & 12.0   & 54.73  & 54.73  & 0.00\% & 1.6    & 16.2   & 17.9   & 54.73  & 54.73  & 0.00\% & 18.3                 & 21.7                 & 40.3                 & 54.73                & 54.73                & 0.00\%               \\
											& u4-24 & 70    & 0.1    & 7.9  & 3.7    & 12.1   & 88.19  & 88.19  & 0.00\% & 2.4    & 3.6    & 6.1    & 88.19  & 88.19  & 0.00\% & 58.4                 & 10.7                 & 69.5                 & 88.19                & 88.19                & 0.00\%               \\
											& u4-32 & 233   & 1.1    & 7.8  & 109.2  & 118.9  & 100.55 & 100.55 & 0.00\% & 4.8    & 57.7   & 63.5   & 100.55 & 100.55 & 0.00\% & 95.2                 & 431.7                & 528.2                & 100.55               & 100.54               & 0.00\%               \\
											& u4-40 & 182   & 1.2    & 12.3 & 35.5   & 50.4   & 133.47 & 133.47 & 0.00\% & 7.2    & 104.6  & 112.1  & 133.47 & 133.47 & 0.00\% & 144.0                & 135.0                & 280.7                & 133.47               & 133.46               & 0.00\%               \\
											& u4-48 & 1371  & 27.9   & 15.2 & 1753.5 & 1800.0 & 151.92 & 150.65 & 0.83\% & 7.2    & 1791.4 & 1800.0 & 153.02 & 144.92 & 5.30\% & 197.0                & 1583.1               & 1800.0               & 151.76               & 150.56               & 0.79\%               \\
											& u5-40 & 675   & 6.5    & 20.7 & 1770.2 & 1800.0 & 125.50 & 124.48 & 0.81\% & 6.6    & 1790.9 & 1800.0 & 125.18 & 122.49 & 2.15\% & 156.2                & 544.0                & 707.4                & 125.29               & 125.29               & 0.00\%               \\
											& u5-50 & 1202  & 101.8  & 16.7 & 1679.3 & 1800.0 & 143.37 & 142.76 & 0.42\% & 6.8    & 1791.7 & 1800.0 & 143.23 & 142.32 & 0.63\% & 226.6                & 1181.5               & 1509.5               & 143.23               & 143.23               & 0.00\%               \\
											& Avg   & 313.7 & 15.6   & 5.2  & 568.4  & 587.8  &        &        & 0.23\% & 2.6    & 585.5  & 587.4  &        &        & 0.91\% & 55.2                 & 340.3                & 399.5                & \multicolumn{1}{l}{} & \multicolumn{1}{l}{} & 0.10\%               \\
											\multirow{14}{*}{0.4} & u2-16 & 79    & 0.2    & 3.1  & 7.1    & 10.6   & 56.23  & 56.23  & 0.00\% & 0.8    & 4.1    & 5.0    & 56.23  & 56.23  & 0.00\% & 13.9                 & 16.8                 & 31.0                 & 56.23                & 56.23                & 0.00\%               \\
											& u2-20 & 226   & 1.6    & 4.3  & 9.4    & 15.8   & 57.46  & 57.46  & 0.00\% & 1.0    & 13.5   & 14.6   & 57.46  & 57.46  & 0.00\% & 24.7                 & 179.7                & 205.4                & 57.46                & 57.46                & 0.00\%               \\
											& u2-24 & 81    & 0.2    & 4.8  & 10.8   & 16.1   & 91.63  & 91.63  & 0.00\% & 1.0    & 15.9   & 16.9   & 91.63  & 91.63  & 0.00\% & 52.8                 & 45.5                 & 98.9                 & 91.38               & 91.38                & 0.00\%               \\
											& u3-18 & 108   & 0.3    & 3.3  & 1.8    & 5.6    & 50.04  & 50.04  & 0.00\% & 0.9    & 3.2    & 4.1    & 50.04  & 50.04  & 0.00\% & 30.3                 & 14.5                 & 45.2                 & 50.04                & 50.04                & 0.00\%               \\
											& u3-24 & 157   & 0.6    & 5.1  & 6.7    & 12.6   & 69.15  & 69.15  & 0.00\% & 2.3    & 26.0   & 28.5   & 69.15  & 69.15  & 0.00\% & 53.3                 & 112.2                & 166.3                & 69.15                & 69.15                & 0.00\%               \\
											& u3-30 & 200   & 1.8    & 10.3 & 16.9   & 29.5   & 76.91  & 76.91  & 0.00\% & 2.1    & 33.5   & 36.0   & 76.91  & 76.91  & 0.00\% & 77.5                 & 129.2                & 212.7                & 76.91                & 76.91                & 0.00\%               \\
											& u3-36 & 241   & 1.8    & 9.8  & 50.0   & 62.5   & 103.87 & 103.87 & 0.00\% & 3.2    & 50.6   & 54.3   & 103.87 & 103.87 & 0.00\% & 119.9                & 1017.8               & 1144.7               & 103.87               & 103.87               & 0.00\%               \\
											& u4-16 & 86    & 0.3    & 2.5  & 1.4    & 4.4    & 53.41  & 53.41  & 0.00\% & 0.7    & 3.7    & 4.4    & 53.41  & 53.41  & 0.00\% & 13.3                 & 5.3                  & 19.0                 & 53.41                & 53.41                & 0.00\%               \\
											& u4-24 & 70    & 0.2    & 5.3  & 2.9    & 8.9    & 88.19  & 88.19  & 0.00\% & 1.2    & 2.8    & 4.0    & 88.19  & 88.19  & 0.00\% & 54.4                 & 15.1                 & 69.9                 & 88.19                & 88.19                & 0.00\%               \\
											& u4-32 & 233   & 1.2    & 8.1  & 14.2   & 24.4   & 99.46  & 99.46  & 0.00\% & 2.4    & 20.5   & 23.5   & 99.46  & 99.46  & 0.00\% & 77.5                 & 148.5                & 227.5                & 99.46                & 99.46                & 0.00\%               \\
											& u4-40 & 182   & 0.7    & 11.8 & 39.1   & 52.9   & 133.05 & 133.05 & 0.00\% & 3.1    & 64.0   & 67.4   & 133.05 & 133.05 & 0.00\% & 121.7                & 340.3                & 466.1                & 133.05               & 133.05               & 0.00\%               \\
											& u4-48 & 1371  & 45.7   & 16.6 & 684.3  & 749.0  & 150.29 & 150.29 & 0.00\% & 6.0    & 1792.8 & 1800.0 & 150.71 & 144.65 & 4.02\% & 190.7                & 1577.0               & 1800.0               & 153.16               & 149.36               & 2.48\%               \\
											& u5-40 & 675   & 4.8    & 17.8 & 39.2   & 63.0   & 124.02 & 124.02 & 0.00\% & 6.1    & 1197.5 & 1204.8 & 124.02 & 124.02 & 0.00\% & 115.4                & 597.9                & 736.9                & 124.17              & 124.17               & 0.00\%               \\
											& u5-50 & 1202  & 95.1   & 21.1 & 188.5  & 306.5  & 142.17 & 142.17 & 0.00\% & 10.6   & 624.2  & 639.7  & 142.17 & 142.17 & 0.00\% & 195.1                & 1538.8               & 1800.0               & 273.65               & 141.77               & 48.19\%              \\
											& Avg   & 313.7 & 17.0   & 4.9  & 102.8  & 123.0  &        &        & 0.00\% & 2.0    & 398.4  & 401.2  &        &        & 0.53\% & 47.9                 & 441.7                & 496.4                & \multicolumn{1}{l}{} & \multicolumn{1}{l}{} & 6.37\%               \\
											\multirow{14}{*}{0.1} & u2-16 & 79    & 0.2    & 3.8  & 3.8    & 7.8    & 55.49  & 55.49  & 0.00\% & 0.8    & 3.0    & 3.8    & 55.49  & 55.49  & 0.00\% & \multicolumn{1}{l}{} & \multicolumn{1}{l}{} & \multicolumn{1}{l}{} & \multicolumn{1}{l}{} & \multicolumn{1}{l}{} & \multicolumn{1}{l}{} \\
											& u2-20 & 226   & 1.8    & 4.5  & 7.9    & 14.7   & 56.86  & 56.86  & 0.00\% & 1.1    & 11.7   & 12.9   & 56.86  & 56.86  & 0.00\% & \multicolumn{1}{l}{} & \multicolumn{1}{l}{} & \multicolumn{1}{l}{} & \multicolumn{1}{l}{} & \multicolumn{1}{l}{} & \multicolumn{1}{l}{} \\
											& u2-24 & 81    & 0.2    & 5.6  & 10.8   & 17.1   & 91.38  & 91.38  & 0.00\% & 1.2    & 8.8    & 10.1   & 91.38  & 91.38  & 0.00\% & \multicolumn{1}{l}{} & \multicolumn{1}{l}{} & \multicolumn{1}{l}{} & \multicolumn{1}{l}{} & \multicolumn{1}{l}{} & \multicolumn{1}{l}{} \\
											& u3-18 & 108   & 0.3    & 4.0  & 3.3    & 7.7    & 50.04  & 50.04  & 0.00\% & 1.2    & 3.6    & 4.9    & 50.04  & 50.04  & 0.00\% & \multicolumn{1}{l}{} & \multicolumn{1}{l}{} & \multicolumn{1}{l}{} & \multicolumn{1}{l}{} & \multicolumn{1}{l}{} & \multicolumn{1}{l}{} \\
											& u3-24 & 157   & 0.9    & 5.7  & 7.2    & 14.3   & 68.99  & 68.99  & 0.00\% & 1.7    & 26.9   & 28.8   & 68.99  & 68.99  & 0.00\% & \multicolumn{1}{l}{} & \multicolumn{1}{l}{} & \multicolumn{1}{l}{} & \multicolumn{1}{l}{} & \multicolumn{1}{l}{} & \multicolumn{1}{l}{} \\
											& u3-30 & 200   & 1.7    & 7.4  & 10.2   & 19.8   & 76.86  & 76.86  & 0.00\% & 2.9    & 23.1   & 26.2   & 76.86  & 76.86  & 0.00\% & \multicolumn{1}{l}{} & \multicolumn{1}{l}{} & \multicolumn{1}{l}{} & \multicolumn{1}{l}{} & \multicolumn{1}{l}{} & \multicolumn{1}{l}{} \\
											& u3-36 & 241   & 2.5    & 12.9 & 36.0   & 53.0   & 103.34 & 103.34 & 0.00\% & 4.0    & 32.5   & 37.1   & 103.34 & 103.34 & 0.00\% & \multicolumn{1}{l}{} & \multicolumn{1}{l}{} & \multicolumn{1}{l}{} & \multicolumn{1}{l}{} & \multicolumn{1}{l}{} & \multicolumn{1}{l}{} \\
											& u4-16 & 86    & 0.2    & 2.6  & 1.5    & 4.4    & 53.41  & 53.41  & 0.00\% & 0.7    & 3.0    & 3.8    & 53.41  & 53.41  & 0.00\% & \multicolumn{1}{l}{} & \multicolumn{1}{l}{} & \multicolumn{1}{l}{} & \multicolumn{1}{l}{} & \multicolumn{1}{l}{} & \multicolumn{1}{l}{} \\
											& u4-24 & 70    & 0.2    & 5.0  & 2.2    & 7.8    & 88.19  & 88.19  & 0.00\% & 1.3    & 1.9    & 3.2    & 88.19  & 88.19  & 0.00\% & \multicolumn{1}{l}{} & \multicolumn{1}{l}{} & \multicolumn{1}{l}{} & \multicolumn{1}{l}{} & \multicolumn{1}{l}{} & \multicolumn{1}{l}{} \\
											& u4-32 & 233   & 1.2    & 9.5  & 18.0   & 29.3   & 99.46  & 99.46  & 0.00\% & 2.7    & 11.0   & 14.2   & 99.46  & 99.46  & 0.00\% & \multicolumn{1}{l}{} & \multicolumn{1}{l}{} & \multicolumn{1}{l}{} & \multicolumn{1}{l}{} & \multicolumn{1}{l}{} & \multicolumn{1}{l}{} \\
											& u4-40 & 182   & 0.7    & 11.5 & 20.1   & 33.7   & 132.95 & 132.95 & 0.00\% & 3.4    & 47.7   & 51.5   & 132.95 & 132.95 & 0.00\% & \multicolumn{1}{l}{} & \multicolumn{1}{l}{} & \multicolumn{1}{l}{} & \multicolumn{1}{l}{} & \multicolumn{1}{l}{} & \multicolumn{1}{l}{} \\
											& u4-48 & 1371  & 43.7   & 22.2 & 245.6  & 314.1  & 149.73 & 149.73 & 0.00\% & 7.2    & 1791.7 & 1800.0 & 149.73 & 144.61 & 3.42\% & \multicolumn{1}{l}{} & \multicolumn{1}{l}{} & \multicolumn{1}{l}{} & \multicolumn{1}{l}{} & \multicolumn{1}{l}{} & \multicolumn{1}{l}{} \\
											& u5-40 & 675   & 6.4    & 14.0 & 22.2   & 44.6   & 123.64 & 123.64 & 0.00\% & 4.4    & 580.3  & 586.0  & 123.64 & 123.64 & 0.00\% & \multicolumn{1}{l}{} & \multicolumn{1}{l}{} & \multicolumn{1}{l}{} & \multicolumn{1}{l}{} & \multicolumn{1}{l}{} & \multicolumn{1}{l}{} \\
											& u5-50 & 1202  & 114.8  & 24.6 & 72.7   & 214.3  & 142.10 & 142.10 & 0.00\% & 22.0   & 436.3  & 460.4  & 142.10 & 142.10 & 0.00\% & \multicolumn{1}{l}{} & \multicolumn{1}{l}{} & \multicolumn{1}{l}{} & \multicolumn{1}{l}{} & \multicolumn{1}{l}{} & \multicolumn{1}{l}{} \\
											& Avg   & 313.7 & 19.1   & 5.4  & 36.5   & 59.5   &        &        & 0.00\% & 3.1    & 309.9  & 313.5  &        &        & 0.45\% & \multicolumn{1}{l}{} & \multicolumn{1}{l}{} & \multicolumn{1}{l}{} & \multicolumn{1}{l}{} & \multicolumn{1}{l}{} & \multicolumn{1}{l}{} \\ \bottomrule
										\end{tabular}
								\end{threeparttable}
								\label{table:detailDEADARPu2}
							\end{sidewaystable}

							\begin{sidewaystable}[htbp]
								\caption{Computational performance of different formulations on D-E-ADARP for type-u instances with 1-minute granularity}
								\scriptsize  
								\centering
								\setlength{\tabcolsep}{1.0 pt}
								\renewcommand{\arraystretch}{0.5}  
								\begin{threeparttable}
									\begin{tabular}{@{}ccccccccccccccccllllll@{}}
										\toprule
										&       & \multicolumn{8}{c}{TSFFCS}                                         & \multicolumn{6}{c}{IEBF}                            & \multicolumn{6}{c}{BTSFF}                                                                                                                                                  \\ \midrule
										$\gamma$              & Name  & $|F|$ & F-time & Net  & CPU    & Time   & OBJ    & LB     & Gap    & E-time & CPU    & Time   & OBJ    & LB     & Gap    & \multicolumn{1}{c}{Net}    & \multicolumn{1}{c}{CPU}    & \multicolumn{1}{c}{Time}   & \multicolumn{1}{c}{OBJ}   & \multicolumn{1}{c}{LB}     & \multicolumn{1}{c}{Gap}    \\
										\multirow{14}{*}{0.7} & u2-16 & 79    & 0.3    & 12.7 & 24.3   & 37.5   & 58.08  & 58.08  & 0.00\% & 1.0    & 11.3   & 12.4   & 58.08  & 58.08  & 0.00\% & \multicolumn{1}{c}{116.3}  & \multicolumn{1}{c}{515.5}  & \multicolumn{1}{c}{632.2}  & \multicolumn{1}{c}{58.08} & \multicolumn{1}{c}{58.08}  & \multicolumn{1}{c}{0.00\%} \\
										& u2-20 & 273   & 2.1    & 16.7 & 28.4   & 47.4   & 56.55  & 56.55  & 0.00\% & 0.9    & 17.9   & 18.9   & 56.55  & 56.55  & 0.00\% & \multicolumn{1}{c}{217.1}  & \multicolumn{1}{c}{1596.9} & \multicolumn{1}{c}{1817.1} & \multicolumn{1}{c}{60.11} & \multicolumn{1}{c}{55.87}  & \multicolumn{1}{c}{7.05\%} \\
										& u2-24 & 85    & 0.2    & 18.2 & 236.4  & 255.1  & 88.46  & 88.46  & 0.00\% & 1.6    & 19.4   & 21.1   & 88.46  & 88.46  & 0.00\% & \multicolumn{1}{c}{311.5}  & \multicolumn{1}{c}{866.6}  & \multicolumn{1}{c}{1178.7} & \multicolumn{1}{c}{88.46} & \multicolumn{1}{c}{88.46}  & \multicolumn{1}{c}{0.00\%} \\
										& u3-18 & 126   & 0.4    & 12.4 & 21.4   & 34.4   & 50.75  & 50.75  & 0.00\% & 1.1    & 16.7   & 17.9   & 50.75  & 50.75  & 0.00\% & \multicolumn{1}{c}{175.4}  & \multicolumn{1}{c}{513.9}  & \multicolumn{1}{c}{690.1}  & \multicolumn{1}{c}{50.75} & \multicolumn{1}{c}{50.75}  & \multicolumn{1}{c}{0.00\%} \\
										& u3-24 & 198   & 1.4    & 20.6 & 45.0   & 67.3   & 68.08  & 68.08  & 0.00\% & 2.3    & 62.7   & 65.2   & 68.08  & 68.08  & 0.00\% & \multicolumn{1}{c}{334.2}  & \multicolumn{1}{c}{1359.6} & \multicolumn{1}{c}{1694.8} & \multicolumn{1}{c}{68.08} & \multicolumn{1}{c}{68.08}  & \multicolumn{1}{c}{0.00\%} \\
										& u3-30 & 269   & 6.3    & 25.6 & 83.7   & 116.2  & 75.92  & 75.92  & 0.00\% & 2.0    & 137.9  & 140.3  & 75.92  & 75.92  & 0.00\% & \multicolumn{1}{c}{487.2}  & \multicolumn{1}{c}{1302.5} & \multicolumn{1}{c}{1800.0} & \multicolumn{1}{c}{NA}    & \multicolumn{1}{c}{75.22}  & \multicolumn{1}{c}{NA}     \\
										& u3-36 & 391   & 3.2    & 39.3 & 215.5  & 259.1  & 105.27 & 105.27 & 0.00\% & 2.7    & 234.1  & 237.4  & 105.27 & 105.27 & 0.00\% & \multicolumn{1}{c}{670.5}  & \multicolumn{1}{c}{1120.1} & \multicolumn{1}{c}{1800.0} & \multicolumn{1}{c}{NA}    & \multicolumn{1}{c}{-21.25} & \multicolumn{1}{c}{NA}     \\
										& u4-16 & 157   & 0.6    & 11.5 & 11.9   & 24.3   & 53.85  & 53.85  & 0.00\% & 1.6    & 19.0   & 20.8   & 53.85  & 53.85  & 0.00\% & \multicolumn{1}{c}{140.8}  & \multicolumn{1}{c}{144.0}  & \multicolumn{1}{c}{285.8}  & \multicolumn{1}{c}{53.85} & \multicolumn{1}{c}{53.85}  & \multicolumn{1}{c}{0.00\%} \\
										& u4-24 & 78    & 0.3    & 23.8 & 15.2   & 39.8   & 89.91  & 89.91  & 0.00\% & 1.2    & 3.0    & 4.3    & 89.91  & 89.91  & 0.00\% & \multicolumn{1}{c}{323.2}  & \multicolumn{1}{c}{258.3}  & \multicolumn{1}{c}{582.2}  & \multicolumn{1}{c}{89.91} & \multicolumn{1}{c}{89.91}  & \multicolumn{1}{c}{0.00\%} \\
										& u4-32 & 334   & 2.3    & 36.0 & 184.2  & 223.4  & 99.33  & 99.33  & 0.00\% & 10.6   & 43.5   & 54.7   & 99.33  & 99.33  & 0.00\% & \multicolumn{1}{c}{491.9}  & \multicolumn{1}{c}{1301.2} & \multicolumn{1}{c}{1800.0} & \multicolumn{1}{c}{NA}    & \multicolumn{1}{c}{-36.74} & \multicolumn{1}{c}{NA}     \\
										& u4-40 & 211   & 2.3    & 51.0 & 120.2  & 175.4  & 134.60 & 134.60 & 0.00\% & 3.3    & 163.7  & 167.3  & 134.60 & 134.60 & 0.00\% & \multicolumn{1}{c}{813.7}  & \multicolumn{1}{c}{980.8}  & \multicolumn{1}{c}{1800.0} & \multicolumn{1}{c}{NA}    & \multicolumn{1}{c}{133.97} & \multicolumn{1}{c}{NA}     \\
										& u4-48 & 3154  & 132.9  & 74.7 & 1588.5 & 1800.0 & 148.58 & 147.77 & 0.55\% & 8.0    & 1790.9 & 1800.0 & 148.78 & 139.80 & 6.03\% & \multicolumn{1}{c}{1194.2} & \multicolumn{1}{c}{507.8}  & \multicolumn{1}{c}{1800.0} & \multicolumn{1}{c}{NA}    & \multicolumn{1}{c}{NA}     & \multicolumn{1}{c}{NA}     \\
										& u5-40 & 745   & 7.7    & 57.1 & 910.0  & 976.5  & 121.64 & 121.63 & 0.00\% & 9.6    & 1047.0 & 1057.7 & 121.64 & 121.64 & 0.00\% & \multicolumn{1}{c}{913.5}  & \multicolumn{1}{c}{880.6}  & \multicolumn{1}{c}{1800.0} & \multicolumn{1}{c}{NA}    & \multicolumn{1}{c}{-25.38} & \multicolumn{1}{c}{NA}     \\
										& u5-50 & 1350  & 349.0  & 69.9 & 183.0  & 606.7  & 140.30 & 140.30 & 0.00\% & 9.1    & 773.9  & 785.7  & 140.30 & 140.30 & 0.00\% & \multicolumn{1}{c}{1560.5} & \multicolumn{1}{c}{53.4}   & \multicolumn{1}{c}{1800.0} & \multicolumn{1}{c}{NA}    & \multicolumn{1}{c}{NA}     & \multicolumn{1}{c}{NA}     \\
										& Avg   & 521.8 & 58.5   & 18.1 & 282.1  & 340.6  &        &        & 0.07\% & 3.1    & 383.1  & 385.7  &        &        & 0.80\% & \multicolumn{1}{c}{340.6}  & \multicolumn{1}{c}{413.3}  & \multicolumn{1}{c}{512.6}  &                           &                            & \multicolumn{1}{c}{1.73\%} \\
										\multirow{14}{*}{0.4} & u2-16 & 79    & 0.2    & 8.1  & 14.0   & 22.4   & 57.69  & 57.69  & 0.00\% & 0.5    & 4.6    & 5.2    & 57.69  & 57.69  & 0.00\% &                            &                            &                            &                           &                            &                            \\
										& u2-20 & 273   & 2.6    & 13.2 & 29.4   & 45.5   & 55.97  & 55.97  & 0.00\% & 1.1    & 7.9    & 9.1    & 55.97  & 55.97  & 0.00\% &                            &                            &                            &                           &                            &                            \\
										& u2-24 & 85    & 0.3    & 16.8 & 29.0   & 46.4   & 87.58  & 87.58  & 0.00\% & 1.0    & 9.3    & 10.4   & 87.58  & 87.58  & 0.00\% &                            &                            &                            &                           &                            &                            \\
										& u3-18 & 126   & 0.4    & 11.6 & 7.7    & 19.8   & 50.29  & 50.29  & 0.00\% & 0.7    & 10.1   & 10.8   & 50.29  & 50.29  & 0.00\% &                            &                            &                            &                           &                            &                            \\
										& u3-24 & 198   & 0.8    & 21.9 & 23.1   & 46.1   & 67.56  & 67.56  & 0.00\% & 1.6    & 34.3   & 36.2   & 67.56  & 67.56  & 0.00\% &                            &                            &                            &                           &                            &                            \\
										& u3-30 & 269   & 6.2    & 28.2 & 72.7   & 107.5  & 75.53  & 75.53  & 0.00\% & 2.7    & 32.5   & 35.5   & 75.53  & 75.53  & 0.00\% &                            &                            &                            &                           &                            &                            \\
										& u3-36 & 391   & 4.0    & 42.5 & 91.8   & 139.7  & 104.18 & 104.18 & 0.00\% & 4.0    & 159.6  & 164.3  & 104.18 & 104.18 & 0.00\% &                            &                            &                            &                           &                            &                            \\
										& u4-16 & 157   & 0.4    & 8.1  & 5.5    & 14.1   & 53.57  & 53.57  & 0.00\% & 0.7    & 26.9   & 27.7   & 53.57  & 53.57  & 0.00\% &                            &                            &                            &                           &                            &                            \\
										& u4-24 & 78    & 0.2    & 15.6 & 7.9    & 24.0   & 89.91  & 89.91  & 0.00\% & 2.0    & 3.5    & 5.7    & 89.91  & 89.91  & 0.00\% &                            &                            &                            &                           &                            &                            \\
										& u4-32 & 334   & 2.0    & 33.1 & 194.3  & 230.0  & 98.90  & 98.90  & 0.00\% & 3.2    & 38.6   & 42.9   & 98.90  & 98.90  & 0.00\% &                            &                            &                            &                           &                            &                            \\
										& u4-40 & 211   & 0.9    & 43.7 & 61.4   & 107.6  & 133.87 & 133.87 & 0.00\% & 7.6    & 37.4   & 45.3   & 133.87 & 133.87 & 0.00\% &                            &                            &                            &                           &                            &                            \\
										& u4-48 & 3154  & 127.4  & 75.9 & 250.7  & 456.0  & 147.54 & 147.54 & 0.00\% & 7.3    & 1790.9 & 1800.0 & 147.87 & 140.69 & 4.86\% &                            &                            &                            &                           &                            &                            \\
										& u5-40 & 745   & 8.3    & 57.7 & 63.8   & 132.6  & 120.57 & 120.57 & 0.00\% & 7.5    & 311.7  & 320.6  & 120.57 & 120.57 & 0.00\% &                            &                            &                            &                           &                            &                            \\
										& u5-50 & 1350  & 383.1  & 87.0 & 109.9  & 583.4  & 140.23 & 140.23 & 0.00\% & 6.7    & 591.6  & 600.3  & 140.23 & 140.23 & 0.00\% &                            &                            &                            &                           &                            &                            \\
										& Avg   & 521.8 & 62.0   & 20.2 & 53.7   & 120.9  &        &        & 0.00\% & 2.3    & 291.2  & 293.4  &        &        & 0.64\% &                            &                            &                            &                           &                            &                            \\
										\multirow{14}{*}{0.1} & u2-16 & 79    & 0.7    & 29.8 & 8.3    & 39.3   & 57.53  & 57.53  & 0.00\% & 0.7    & 4.8    & 5.6    & 57.53  & 57.53  & 0.00\% &                            &                            &                            &                           &                            &                            \\
										& u2-20 & 273   & 1.6    & 10.4 & 16.5   & 28.7   & 55.27  & 55.27  & 0.00\% & 0.8    & 4.9    & 5.8    & 55.27  & 55.27  & 0.00\% &                            &                            &                            &                           &                            &                            \\
										& u2-24 & 85    & 0.2    & 14.2 & 19.0   & 33.7   & 87.31  & 87.31  & 0.00\% & 1.0    & 8.8    & 9.9    & 87.31  & 87.31  & 0.00\% &                            &                            &                            &                           &                            &                            \\
										& u3-18 & 126   & 0.3    & 12.2 & 6.8    & 19.5   & 50.29  & 50.29  & 0.00\% & 0.7    & 9.3    & 10.1   & 50.29  & 50.29  & 0.00\% &                            &                            &                            &                           &                            &                            \\
										& u3-24 & 198   & 0.9    & 15.1 & 19.1   & 35.4   & 67.56  & 67.56  & 0.00\% & 1.5    & 24.2   & 25.9   & 67.56  & 67.56  & 0.00\% &                            &                            &                            &                           &                            &                            \\
										& u3-30 & 269   & 3.6    & 38.8 & 74.5   & 117.3  & 75.53  & 75.53  & 0.00\% & 1.9    & 18.4   & 20.6   & 75.53  & 75.53  & 0.00\% &                            &                            &                            &                           &                            &                            \\
										& u3-36 & 391   & 2.6    & 33.3 & 69.2   & 105.9  & 104.18 & 104.18 & 0.00\% & 3.0    & 61.6   & 65.1   & 104.18 & 104.18 & 0.00\% &                            &                            &                            &                           &                            &                            \\
										& u4-16 & 157   & 0.5    & 8.5  & 5.4    & 14.7   & 53.57  & 53.57  & 0.00\% & 1.0    & 11.5   & 12.6   & 53.57  & 53.57  & 0.00\% &                            &                            &                            &                           &                            &                            \\
										& u4-24 & 78    & 0.2    & 14.6 & 6.1    & 21.2   & 89.91  & 89.91  & 0.00\% & 1.3    & 2.2    & 3.5    & 89.91  & 89.91  & 0.00\% &                            &                            &                            &                           &                            &                            \\
										& u4-32 & 334   & 1.8    & 25.8 & 68.1   & 96.4   & 98.90  & 98.90  & 0.00\% & 2.7    & 17.6   & 20.9   & 98.90  & 98.90  & 0.00\% &                            &                            &                            &                           &                            &                            \\
										& u4-40 & 211   & 1.4    & 36.6 & 19.2   & 58.8   & 133.49 & 133.49 & 0.00\% & 3.0    & 25.9   & 29.2   & 133.49 & 133.49 & 0.00\% &                            &                            &                            &                           &                            &                            \\
										& u4-48 & 3154  & 117.0  & 65.8 & 211.3  & 396.1  & 147.22 & 147.22 & 0.00\% & 10.0   & 1787.5 & 1800.0 & 147.41 & 140.27 & 4.84\% &                            &                            &                            &                           &                            &                            \\
										& u5-40 & 745   & 5.4    & 43.4 & 29.3   & 79.5   & 120.47 & 120.47 & 0.00\% & 4.3    & 222.3  & 227.5  & 120.47 & 120.47 & 0.00\% &                            &                            &                            &                           &                            &                            \\
										& u5-50 & 1350  & 341.0  & 69.2 & 87.7   & 500.4  & 140.23 & 140.23 & 0.00\% & 8.1    & 429.0  & 440.1  & 140.23 & 140.23 & 0.00\% &                            &                            &                            &                           &                            &                            \\
										& Avg   & 521.8 & 55.7   & 15.4 & 40.3   & 97.5   &        &        & 0.00\% & 2.0    & 268.0  & 270.6  &        &        & 0.64\% &                            &                            &                            &                           &                            &                            \\ \bottomrule
									\end{tabular}
									\end{threeparttable}
									\label{table:detailDEADARPu1}
								\end{sidewaystable}
								
								\clearpage

								\section{Computational performance of different formulations on D-E-ADARP for type‑a instances with limited battery capacity}\label{DEADARP025}
								
								Intuitively, we believe that BTSFF may exhibit advantages in scenarios with limited battery capacity. To test this, we consider a setting where the battery size is reduced to 25\% of the original capacity. This experiment focuses solely on assessing whether BTSFF demonstrates improved performance for D-E-ADARP under low‑battery conditions. Here, we choose type-a instances with 2-minute granularity. We evaluate three formulations, TSFFCS, IEBF, and BTSFF, on D-E-ADARP with $N_{\max}=\infty$, $\gamma=0.7$, and a battery size of 25\% of the baseline. The computational results are shown in Table~\ref{table:detailDEADARP025}.
								
								\begin{sidewaystable}[htbp]
									\caption{Computational performance of different formulations on D-E-ADARP for type-a instances with $N_{\max}=\infty$, $\gamma=70\%$, and 25\% battery size}
									\scriptsize  
									\centering
									\setlength{\tabcolsep}{1.0 pt}
									\renewcommand{\arraystretch}{0.5}  
									\begin{threeparttable}
										\begin{tabular}{ccccccccccccccccccccc}
											\hline
											\multicolumn{1}{l}{} & \multicolumn{8}{c}{TSFFCS}                                         & \multicolumn{6}{c}{IEBF}                            & \multicolumn{6}{c}{BTSFF}                          \\ \hline
											Name                 & $|F|$ & F-time & Net  & CPU    & Time   & OBJ    & LB     & Gap    & E-time & CPU    & Time   & OBJ    & LB     & Gap    & Net   & CPU    & Time   & OBJ    & LB     & Gap    \\
											a2-16                & 38    & 0.1    & 3.6  & 111.2  & 115.1  & 251.87 & 251.87 & 0.00\% & 0.2    & 3.6    & 3.7    & 251.87 & 251.87 & 0.00\% & 123.6 & 66.5   & 190.3  & 251.87 & 251.87 & 0.00\% \\
											a2-24                & 69    & 0.6    & 4.3  & 8.0    & 12.7   & 360.14 & 360.14 & 0.00\% & 0.5    & 3.3    & 3.8    & 360.14 & 360.14 & 0.00\% & 157.7 & 82.6   & 241.7  & 360.14 & 360.14 & 0.00\% \\
											a3-18                & 77    & 0.4    & 2.3  & 12.1   & 14.7   & 240.53 & 240.53 & 0.00\% & 0.2    & 4.4    & 4.7    & 240.53 & 240.53 & 0.00\% & 58.1  & 37.1   & 95.9   & 240.53 & 240.53 & 0.00\% \\
											a3-24                & 132   & 0.6    & 6.5  & 327.1  & 334.3  & 289.70 & 289.70 & 0.00\% & 0.5    & 16.9   & 17.4   & 289.70 & 289.70 & 0.00\% & 106.5 & 124.9  & 232.4  & 289.70 & 289.70 & 0.00\% \\
											a3-30                & 101   & 0.4    & 8.1  & 135.3  & 144.3  & 427.46 & 427.42 & 0.01\% & 0.6    & 11.7   & 12.5   & 427.46 & 427.46 & 0.00\% & 139.8 & 113.7  & 254.3  & 427.46 & 427.46 & 0.00\% \\
											a3-36                & 188   & 0.8    & 27.0 & 1770.8 & 1800.0 & 469.91 & 468.99 & 0.20\% & 1.0    & 77.5   & 78.6   & 469.91 & 469.91 & 0.00\% & 443.8 & 711.4  & 1156.6 & 469.91 & 469.91 & 0.00\% \\
											a4-16                & 145   & 0.5    & 2.6  & 1797.0 & 1800.0 & 235.61 & 230.10 & 2.34\% & 0.2    & 28.0   & 28.2   & 235.61 & 235.61 & 0.00\% & 34.5  & 48.0   & 83.1   & 235.61 & 235.61 & 0.00\% \\
											a4-32                & 1107  & 7.6    & 18.6 & 1774.0 & 1800.0 & 421.42 & 402.40 & 4.52\% & 1.3    & 1334.9 & 1336.4 & 417.67 & 417.67 & 0.00\% & 144.2 & 320.3  & 472.9  & 417.67 & 417.67 & 0.00\% \\
											a4-40                & 426   & 2.6    & 12.9 & 757.1  & 776.0  & 472.26 & 472.26 & 0.00\% & 3.1    & 152.4  & 156.7  & 472.26 & 472.26 & 0.00\% & 537.8 & 1170.9 & 1712.2 & 472.26 & 472.26 & 0.00\% \\
											a4-48                & 557   & 14.5   & 17.5 & 1767.5 & 1800.0 & 577.61 & 567.98 & 1.67\% & 9.8    & 1785.3 & 1800.0 & 575.97 & 570.71 & 0.91\% & 603.9 & 604.8  & 1230.3 & 575.97 & 575.97 & 0.00\% \\
											a5-40                & 436   & 15.9   & 17.1 & 1766.5 & 1800.0 & 449.56 & 431.47 & 4.03\% & 2.2    & 994.0  & 997.6  & 448.25 & 448.25 & 0.00\% & 676.7 & 857.0  & 1554.2 & 448.25 & 448.25 & 0.00\% \\
											a5-50                & 2125  & 16.8   & 28.6 & 1749.7 & 1800.0 & NA     & 572.05 & NA     & 3.7    & 1795.3 & 1800.0 & 602.35 & 577.69 & 4.09\% & 738.9 & 1059.5 & 1800.0 & 613.48 & 593.85 & 3.20\% \\
											\multicolumn{1}{l}{} &       &        &      & 768.6  & 777.1  &        &        & 1.44\% &        &        & 531.5  &        &        & 0.15\% &       & 334.5  & 550.3  &        &        & 0.00\% \\ \hline
										\end{tabular}	\begin{tablenotes}
											\item[1] For a more informative comparison, the last row is excluded from the calculation of average values. \end{tablenotes}
									\end{threeparttable}
									\label{table:detailDEADARP025}
								\end{sidewaystable}
								
								\clearpage
								\section{The counts of events, event arcs, and fragments in (D-)E-ADARP}\label{eventEADARP}

								Table~\ref{eventcount} summarizes the counts of events, event arcs, and fragments for (D-)E-ADARP instances, which are generated with varying levels of time granularity. The counts of events and event arcs are denoted as $|V_E|$ and $|A_E|$, respectively.

								\begin{sidewaystable}[htbp]
									\caption{The counts of events, event arcs, and fragments in (D-)E-ADARP}
									\scriptsize  
									\centering
									\setlength{\tabcolsep}{0.5 pt}
									\renewcommand{\arraystretch}{0.2}  
									\begin{threeparttable}
										\begin{tabular}{@{}cccccccccccccccccccc@{}}
											\toprule
											& \multicolumn{9}{c}{Type a}                                                                    & \multicolumn{1}{l}{} & \multicolumn{9}{c}{Type u}                                                                    \\ \midrule
											Unit: & 2     & 1     & C     & \multicolumn{2}{c}{2} & \multicolumn{2}{c}{1} & \multicolumn{2}{c}{C} & \multicolumn{1}{l}{} & 2     & 1     & C     & \multicolumn{2}{c}{2} & \multicolumn{2}{c}{1} & \multicolumn{2}{c}{C} \\\midrule
											Name        & $|F|$ & $|F|$ & $|F|$ & $|V_E|$   & $|A_E|$   & $|V_E|$   & $|A_E|$   & $|V_E|$   & $|A_E|$   & Name                 & $|F|$ & $|F|$ & $|F|$ & $|V_E|$   & $|A_E|$   & $|V_E|$   & $|A_E|$   & $|V_E|$   & $|A_E|$   \\\midrule
											a2-16       & 38    & 36    & 32    & 72        & 371       & 65        & 361       & 62        & 358       & u2-16                & 79    & 79    & 60    & 156       & 780       & 163       & 816       & 131       & 697       \\
											a2-20       & 54    & 51    & 50    & 108       & 544       & 97        & 526       & 97        & 525       & u2-20                & 226   & 273   & 168   & 248       & 1214      & 270       & 1276      & 232       & 1171      \\
											a2-24       & 69    & 65    & 64    & 115       & 671       & 108       & 659       & 106       & 655       & u2-24                & 81    & 85    & 66    & 199       & 1096      & 188       & 1074      & 158       & 1013      \\
											a3-18       & 77    & 68    & 65    & 131       & 561       & 111       & 530       & 102       & 518       & u3-18                & 108   & 126   & 78    & 227       & 1092      & 214       & 1053      & 155       & 889       \\
											a3-24       & 132   & 109   & 106   & 200       & 906       & 182       & 865       & 182       & 857       & u3-24                & 157   & 198   & 130   & 296       & 1462      & 310       & 1545      & 253       & 1313      \\
											a3-30       & 101   & 91    & 89    & 213       & 1030      & 201       & 1009      & 200       & 1003      & u3-30                & 200   & 269   & 182   & 386       & 1963      & 405       & 2040      & 330       & 1790      \\
											a3-36       & 188   & 131   & 116   & 268       & 1408      & 245       & 1347      & 217       & 1307      & u3-36                & 241   & 391   & 223   & 468       & 2386      & 487       & 2485      & 394       & 2160      \\
											a4-16       & 145   & 94    & 82    & 177       & 615       & 167       & 599       & 157       & 575       & u4-16                & 86    & 157   & 74    & 195       & 891       & 244       & 1057      & 170       & 845       \\
											a4-24       & 124   & 103   & 95    & 215       & 949       & 190       & 908       & 169       & 878       & u4-24                & 70    & 78    & 57    & 163       & 1086      & 169       & 1103      & 123       & 1024      \\
											a4-32       & 1107  & 297   & 239   & 384       & 1548      & 317       & 1381      & 288       & 1340      & u4-32                & 233   & 334   & 174   & 580       & 2510      & 661       & 2882      & 443       & 2101      \\
											a4-40       & 426   & 308   & 239   & 565       & 2318      & 480       & 2143      & 434       & 2050      & u4-40                & 182   & 211   & 144   & 327       & 2311      & 355       & 2408      & 250       & 2095      \\
											a4-48       & 557   & 372   & 346   & 991       & 3721      & 726       & 2947      & 685       & 2872      & u4-48                & 1371  & 3154  & 922   & 983       & 4819      & 1198      & 5668      & 807       & 4233      \\
											a5-40       & 436   & 345   & 319   & 566       & 2379      & 455       & 2148      & 437       & 2095      & u5-40                & 675   & 745   & 330   & 721       & 3342      & 815       & 3601      & 548       & 2887      \\
											a5-50       & 2125  & 874   & 692   & 970       & 3742      & 811       & 3288      & 717       & 3111      & u5-50                & 1202  & 1350  & 548   & 1108      & 5218      & 1140      & 5419      & 823       & 4325      \\\midrule
											Avg         & 379.8 & 163.5 & 132.9 & 242.7     & 898.9     & 186.4     & 747.9     & 169.3     & 714.1     &                      & 313.7 & 521.8 & 160.5 & 242.4     & 1093.7    & 276.7     & 1223.7    & 184.9     & 917.9     \\ \bottomrule
										\end{tabular}
										\begin{tablenotes}
											\item[1] Unit: The numbers 2 and 1 represent the 2-minute and 1-minute granularity in  time and SoC, respectively. ``C" indicates that the instance uses continuous parameters for both time and SoC.
										\end{tablenotes}
									\end{threeparttable}
									\label{eventcount}
								\end{sidewaystable}

	\end{document}